\documentclass[%
	,
	submission,
	]{IEEEtran}
\usepackage{epsfig,amssymb}
\usepackage[cmex10]{amsmath}
\usepackage{array}
\usepackage{graphicx,times}
\usepackage{tabularx}
\usepackage{cite}
\usepackage{subfigure}
\usepackage{algorithm, algorithmic}

\usepackage[top=1in, bottom=1in, left=0.5in, right=0.5in]{geometry}\usepackage{array} 
\usepackage{booktabs} 
\usepackage{multirow} 
\usepackage{tabularx}
\usepackage{setspace}
\usepackage{cite}
\usepackage{url}
\usepackage{scrextend}
\usepackage{hyperref}
\usepackage{xcolor}
\usepackage{blindtext}

\begin{document}

\title{Signal Denoising Using the Minimum-Probability-of-Error Criterion}
\author{Jishnu Sadasivan, Subhadip Mukherjee, and Chandra Sekhar Seelamantula, \emph{Senior member, IEEE }

\thanks{ J. Sadasivan  is with the Department of Electrical Communication Engineering, Indian Institute of Science, Bangalore, India.  Phone: +91 80 2293 2276. Fax: +91 80 2360 0563. E-mail: jishnus@ece.iisc.ernet.in.}

\thanks{ S. Mukherjee and C. S. Seelamantula are with the Department of Electrical Engineering, Indian Institute of Science, Bangalore, India.  Phone: +91 80 2293 2695. Fax: +91 80 2360 0444. E-mails: \{subhadip, chandra.sekhar\}@ee.iisc.ernet.in.}


\markboth{IEEE Transactions on Signal Processing}}

\maketitle

\begin{abstract}
We address the problem of signal denoising via transform-domain shrinkage based on a novel \textit{risk} criterion called  the minimum probability of error (MPE), which measures  the probability that the estimated parameter lies outside an $\epsilon$-neighborhood of the actual value. However, the MPE,  similar to the mean-squared error (MSE), depends on the ground-truth parameter, and has to be estimated from the noisy observations. We consider linear shrinkage-based denoising functions, wherein the optimum shrinkage parameter is obtained by minimizing an estimate of the MPE. When the probability of error is integrated over  $\epsilon$, it leads to  the expected $\ell_1$ distortion. The proposed MPE and $\ell_1$ distortion formulations are applicable to various noise distributions by invoking a Gaussian mixture model approximation. Within the realm of MPE, we also develop an extension of the transform-domain shrinkage by grouping transform coefficients, resulting in \textit{subband shrinkage}. The denoising performance obtained within the proposed framework is shown to be better than that obtained using the minimum MSE-based approaches formulated within \textbf{\textit {Stein's unbiased risk estimation}} (SURE) framework, especially in the low measurement signal-to-noise ratio (SNR) regime. Performance comparison with three state-of-the-art denoising algorithms, carried out on electrocardiogram signals and two test signals taken from the \textit{Wavelab} toolbox, exhibits that the MPE  framework results in consistent SNR gains for input SNRs below $5$ dB.
\end{abstract}
\section{Introduction}
\IEEEPARstart{S}{ignal} denoising  algorithms are often developed with the objective of minimizing the mean-squared error (MSE) between an estimate and the ground-truth, which may be deterministic or stochastic with a known prior. The latter formalism leads to Bayesian estimators. Within the deterministic signal estimation paradigm, which is also the formalism considered in this paper, one typically desires that the estimator has minimum variance and is unbiased (MVU) \cite{poor,kay}. An MVU estimator may not always exist, and if it does, it can be obtained using the theory of sufficient statistics. Eldar and Kay \cite{kay} showed that, when it comes to minimizing the MSE, biased estimates may outperform the MVU estimate. For example, one could shrink the MVU estimate and optimize for the shrinkage parameter so that the MSE is minimum.\\
\indent In this paper, we consider the problem of estimating a deterministic signal corrupted by additive white noise. The noise distribution is assumed to be known, but not restricted to be Gaussian. We propose a new distortion metric based on the probability of error and develop estimators using a transform-domain shrinkage approach. Before proceeding with the developments, we review some important literature related to the problem at hand. 
\subsection{Prior Art} 
\indent The MSE is by far the most widely used metric for obtaining the optimum shrinkage parameter. Since the MSE is a function of the parameter to be estimated, direct minimization might result in an unrealizable estimate, in the sense that it might depend on the unknown parameter. However, in some  cases, it is possible to find the optimum shrinkage parameter, for example, using a min-max approach \cite{kay}, where the parameter is constrained to a known set.  An optimum shrinkage estimator, when the  variance of the unbiased  estimate (or MVU) is a  scaled version of the square of the parameter, with a known scaling, is proposed in \cite{kay}.\\
 \indent Optimum shrinkage estimators have also been computed based on \textit{risk estimation}, where an unbiased estimate of the MSE that depends only on the noisy observations is obtained and subsequently minimized over the shrinkage parameter.  Under the assumption of Gaussian noise, an unbiased estimate of the MSE, namely Stein's unbiased risk estimator (SURE), was developed based on Stein's lemma \cite{Stein}, and has been successfully employed in numerous denoising applications. In his seminal work \cite{Stein}, Stein  proved that the  shrinkage estimator of the mean of a multivariate Gaussian distribution, obtained from its independent and identically distributed (i.i.d.) samples by minimizing SURE, dominates the classical least-squares estimate when the number of samples exceeds three \cite{Jamesproof}.\\
\indent A risk minimization approach for denoising using a linear expansion of elementary thresholding functions has been addressed in \cite{Luisier,Bluv,Blu3,Blu,Zheng}, wherein the combining weights are chosen optimally to minimize the SURE objective. SURE-optimized wavelet-domain thresholding techniques have been developed in \cite{Donoho,Benazza,Zhang}. Atto et al. \cite{Atto1, Atto2} have investigated the problem of signal denoising based on optimally selecting the parameters of a wavelet-domain smooth sigmoidal shrinkage function by minimizing the SURE criterion. The use of SURE objective is not restricted to denoising; it has found applications in image deconvolution as well \cite{Pesquet}.\\
\indent Ramani et al. \cite{Ramani} developed a Monte-Carlo technique to select the parameters of a generic denoising operator based on SURE. An image denoising algorithm based on non-local means (NLM) is proposed in \cite{Dimitri2}, where parameters of NLM are optimized using SURE. Notable denoising algorithms that aim to optimize the SURE objective include wavelet-domain multivariate shrinkage \cite{Hsung}, local affine transform for image denoising \cite{Qiu}, optimal basis selection for denoising \cite{Krim}, raised-cosine-based fast bilateral filtering for image denoising \cite{Harini}, SURE-optimized Savitzky-Golay filter \cite{Skrishnan}, etc..\\  
\indent The original formulation of SURE, which assumed independent Gaussian noise was extended to certain distributions in continuous and discrete exponential families in \cite{Hudson} and \cite{Hwang}, respectively, with the assumption of independence left unchanged. Eldar generalized SURE (GSURE) for distributions  belonging to the  non-i.i.d. multivariate exponential family  \cite{Eldar}. Giryes et al. \cite{Giryes} used a projected version of GSURE for selecting parameters in the context of solving inverse problems. An unbiased estimate of the  Itakura-Saito (IS) distortion  and corresponding pointwise shrinkage was developed  in \cite{Nag} and \cite{Krishnan}, and successfully applied to speech denoising. A detailed discussion of Gaussian parameter estimation using shrinkage estimators, together with a performance comparison of SURE with the maximum-likelihood (ML) and soft-thresholding-based estimators, can be found in \cite{Johnstone} (Chapter $2$). 
It is shown in \cite{Johnstone} that the soft-thresholding-based estimator dominates the James-Stein shrinkage estimator in terms of MSE if the parameter vector to be estimated is  sparse. On the other hand, shrinkage estimator dominates if all coordinates of the parameter to be estimated are nearly equal.
\subsection{This Paper}
\indent We address the problem of signal denoising based on the minimum probability of error (MPE), which
we first considered in \cite{icassp2014}. The MPE quantifies the probability of the estimate lying outside an $\epsilon$-neighborhood of the true value. Since the MPE risk depends on the ground truth, we consider a surrogate, which may be  biased, and optimize it to obtain the shrinkage parameter (Section~\ref{sec_prob_form}). The optimization is carried out in the discrete cosine transform (DCT) domain, either in a pointwise fashion or on a subband basis. We derive the MPE risk for Gaussian, Laplacian, and Student's-$t$ noise distributions (Sections~\ref{MPE_risk_est_sec_point} and \ref{mpe_vector_shrink_define}). In practical applications, where the noise distribution may be multimodal and not known explicitly, we propose to use a Gaussian mixture model (GMM) approximation \cite{Plataniotis,Sorenson} (Section~\ref{mpe_sec_gmm_approx}).  We show the performance of the MPE-based denoising technique on the \textit{Piece-Regular} signals  taken from the \textit{Wavelab} toolbox in Gaussian, Student's-$t$, and Laplacian noise contaminations (Section~\ref{exp_results_sec}). Proceeding further,  we also consider the probability of error accumulated over $0 < \epsilon < \infty$ (Section~\ref{sec_integrated_error}), which results in the expected $\ell_1$ distortion between the parameter and its estimate.  The estimators for the expected $\ell_1$ distortion are also derived by invoking the GMM approximation (Section~\ref{GMM_l1_risk}). We also assess the denoising performance of the shrinkage estimator obtained by minimizing the $\ell_1$ distortion for different input SNRs and for different number of noisy realizations (Section~\ref{L1_sec_results}). \\
\indent To further boost the denoising performance of the $\ell_1$ distortion-based estimator, we develop an iterative algorithm to successively refine the cost function and the resulting estimate, starting with the noisy signal as the initialization (Section~\ref{L1_sec_results}). The iterations lead to an improvement of $2$--$3$ dB in output signal-to-noise ratio (SNR) (Section~\ref{Iterative_versus_Noniterative}).\\
\indent Performance comparison of the MPE and $\ell_1$ distortion-based estimators is carried out on the \textit{Piece-Regular} and the \textit{HeaviSine} signals from the \textit{Wavelab} toolbox \cite{wavelab}, and electrocardiogram (ECG) signals from the \textit{PhysioBank} database \cite{database}, with three benchmark techniques: (i) wavelet-domain soft-thresholding \cite{donoho}, (ii) SURE-based orthonormal wavelet thresholding using a linear expansion of thresholds (SURE-LET) \cite{Luisier}, and (iii) SURE-based smooth sigmoid shrinkage (SS) in wavelet domain\cite{Atto1}; all assuming Gaussian noise contamination (Section~\ref{sec_State-Of-The-Art}) for fair comparison. 
 
\section{The MPE Risk}
\label{sec_prob_form}
\indent Consider  the observation model $\bf x=s+w$ in $\mathbb{R}^n$, where ${\bf x}$ and ${\bf s}$ denote the noisy and clean signals, respectively. The noise vector ${\bf w}$ is assumed to have i.i.d. entries with zero mean and variance $\sigma^2$. The goal is to estimate  $\bf s$  from $\bf x$ by minimizing a suitable risk function. The signal model is  considered in an appropriate transform domain, where the signal admits a parsimonious representation, but noise does not. We consider two types of shrinkage estimators: (i) pointwise, where a shrinkage factor $a_i \in [0,1]$ is applied to $x_i$ to obtain an estimate $\widehat{s}_i = a_i x_i$; and (ii) subband-based, wherein a single shrinkage factor $a_J$ is applied to a group of coefficients \{$x_i$, $i\in J$\} in subband $J\subset \{1,2,\cdots,n\}$. Shrinkage estimators may also be interpreted as premultiplication of $\bold x$ with a diagonal matrix.
\subsection{MPE Risk for Pointwise Shrinkage}
\label{MPE_risk_est_sec_point}
\indent Assuming that the estimate of $s_i$ does not depend on $x_j$, for $j \neq i$, we drop the index $i$ for brevity of notation. The MPE risk is defined as 
\begin{eqnarray}
\mathcal{R} &=& \mathbb{P}\left( \left| \widehat{s}-s \right|>\epsilon \right),
\end{eqnarray}
where $\epsilon>0$ is a predefined tolerance parameter. The risk $\mathcal{R}$ quantifies the estimation error using the probability measure and takes into account the noise distribution in its entirety. On the contrary, the MSE relies only on the first- and second-order statistics of noise for linear shrinkage estimators. Substituting $\widehat{s}=ax=a(s+w)$, the risk $\mathcal{R}$ evaluates to
\begin{eqnarray}   
\mathcal{R}\left(s;a\right) &=& \mathbb{P}\left( \left| a(s+w) -s \right| > \epsilon  \right)\nonumber \\
&=&1-F  \left(  \frac{\epsilon-(a-1)s}{a}  \right) + F\left( - \frac{\epsilon+(a-1)s}{a}  \right),\nonumber\\
\label{basic_MPE_expr}
\end{eqnarray}
where $F\left(\cdot\right)$ is the cumulative distribution function (c.d.f.) of the additive noise. Since $\mathcal{R}$ depends on $s$, which is the parameter to be estimated, it is impractical to optimize it directly over $a$. To circumvent the problem, we minimize an estimate of $\mathcal{R}$, which is obtained by  replacing $s$ with an estimate $\tilde{s}$, which, for example, may be obtained using any baseline denoising algorithm, or can even be taken as $\tilde{s}=x$ (which is also the ML estimate of $s$). In the first instance, the proposed technique becomes an add-on to an existing denoising algorithm, and in the second, it is a denoising scheme in itself. Such an estimate $\widehat{ \mathcal{R}}=\mathcal{R}\left(\tilde{s};a\right)$ takes the form
\begin{eqnarray}   
\widehat{\mathcal{R}}=1-F  \left(  \frac{\epsilon-(a-1)\tilde{s}}{a}  \right) + F\left( - \frac{\epsilon+(a-1)\tilde{s}}{a}  \right),
\label{rhat}
\end{eqnarray}
and correspondingly, the optimal shrinkage parameter is obtained as $a_{\text{opt}} = \text{arg\,} \underset{0 \leq a \leq 1}{\min\,} \widehat{\mathcal{R}}$. A grid search is performed to optimize $\widehat{\mathcal{R}}$ over $a \in [0,1]$, and the clean signal is obtained as $\widehat{s}=a_{\text{opt}}x$. We next derive explicit formulae for the risk function for Gaussian, Laplacian, and Student's-$t$ noise distributions.\\ 
\noindent \textit{(i) Gaussian distribution}: In this case, the noisy observation $x$ also follows a Gaussian distribution, and therefore, $\widehat{s}-s=ax-s$ is distributed as $\mathcal{N}\left((a-1)s,a^2\sigma^2 \right)$. The MPE risk estimate is given as
\begin{eqnarray}
\widehat{ \mathcal{R}} &=& Q \left(  \frac{\epsilon-(a-1)\tilde{s}}{a\sigma} \right) + Q \left(  \frac{\epsilon+(a-1)\tilde{s}}{a\sigma} \right),
\label{MPE_gauss_eqn}
\end{eqnarray}
where $Q(u)=\displaystyle \frac{1}{\sqrt{2 \pi}}\int_{u}^{\infty}e^{-\frac{t^2}{2}}\mathrm{d}t$.\\
\noindent \textit{(ii) Student's-$t$ distribution:} Consider the case where the noise follow a Student's-$t$ distribution with parameter $\lambda>2$ and the probability density function (p.d.f.) of noise is given by
\begin{eqnarray*}
f(w) &=& \frac{\Gamma \left( \frac{\lambda+1}{2}  \right) }{\sqrt  {\lambda \pi  } \text{\,\,}\Gamma \left( \frac{\lambda}{2} \right) } \left( 1+\frac{w^2}{\lambda}  \right)^{-\frac{\lambda+1}{2}}.
\end{eqnarray*}
The variance of $w$ is  $\sigma^2=\frac{\lambda}{\lambda-2}$. The expression for $ \widehat{\mathcal{R}}$ is the one given in (\ref{rhat}) with 
\begin{eqnarray}
F(w) &=& \frac{1}{2}+ w\Gamma \left( \frac{\lambda+1}{2}  \right) \frac{G_1 \left( \frac{1}{2},\frac{\lambda+1}{2};\frac{3}{2};-\frac{w^2}{\lambda}  \right)} {\sqrt  {\lambda \pi  } \text{\,\,}\Gamma \left( \frac{\lambda}{2} \right) },
\label{student_t_cdf_eq}
\end{eqnarray}
where $G_1$ is the hypergeometric function defined as
\begin{eqnarray*}
 G_1 \left( a,b;c;z \right) &=& \sum_{k=0}^{\infty}\frac{(a)_k (b)_k}{(c)_k}\frac{z^k}{k!},
\end{eqnarray*}
and $(q)_k$ denotes the Pochhammer symbol:
\begin{eqnarray*}
  (q)_k &\stackrel{\Delta}{=}&\begin{cases}1\text{\,\,\,for\,\,\,} k=0,  \,\\
 q(q+1)(q+2)\cdots(q+k-1), \mbox{\,\,\,for\,\,\,}  k>0.
\end{cases}
\end{eqnarray*}
\noindent\textit{(iii) Laplacian distribution:} Considering the noise to be  i.i.d. Laplacian with zero-mean and parameter $b$ (variance $\sigma^2=2b^2$), with the p.d.f. $f(w)=\frac{1}{2b}\exp \left(- \frac{|w|}{b} \right)$, the MPE risk can be obtained by using the following expression for $F(w)$ in \eqref{rhat}: 
\begin{eqnarray}
F(w) &=& \frac{1}{2} + \frac{1}{2} \text{sgn}(w) \left( 1-\exp  \left(  -  \frac{|w|}{b}    \right) \right).
\label{laplacian_cdf_eq}
\end{eqnarray}
\subsubsection{Closeness of $\widehat{\mathcal{R}}$ to $\mathcal{R}$}
\begin{figure}[t]
\centering
$\begin{array}{cc}
\hspace{-0.21cm}\includegraphics[width=1.5in]{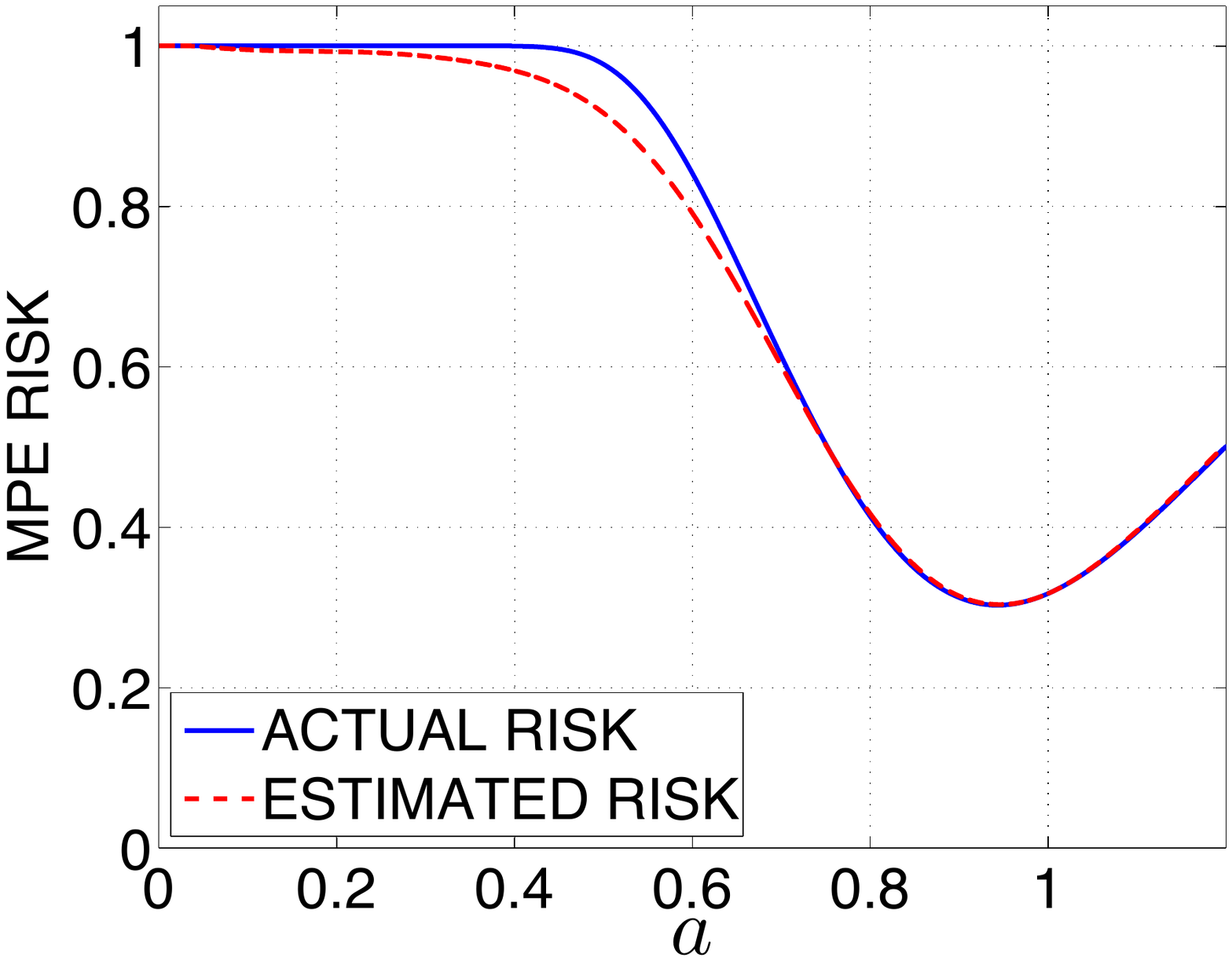}&
\hspace{-0.21cm}\includegraphics[width=1.5in]{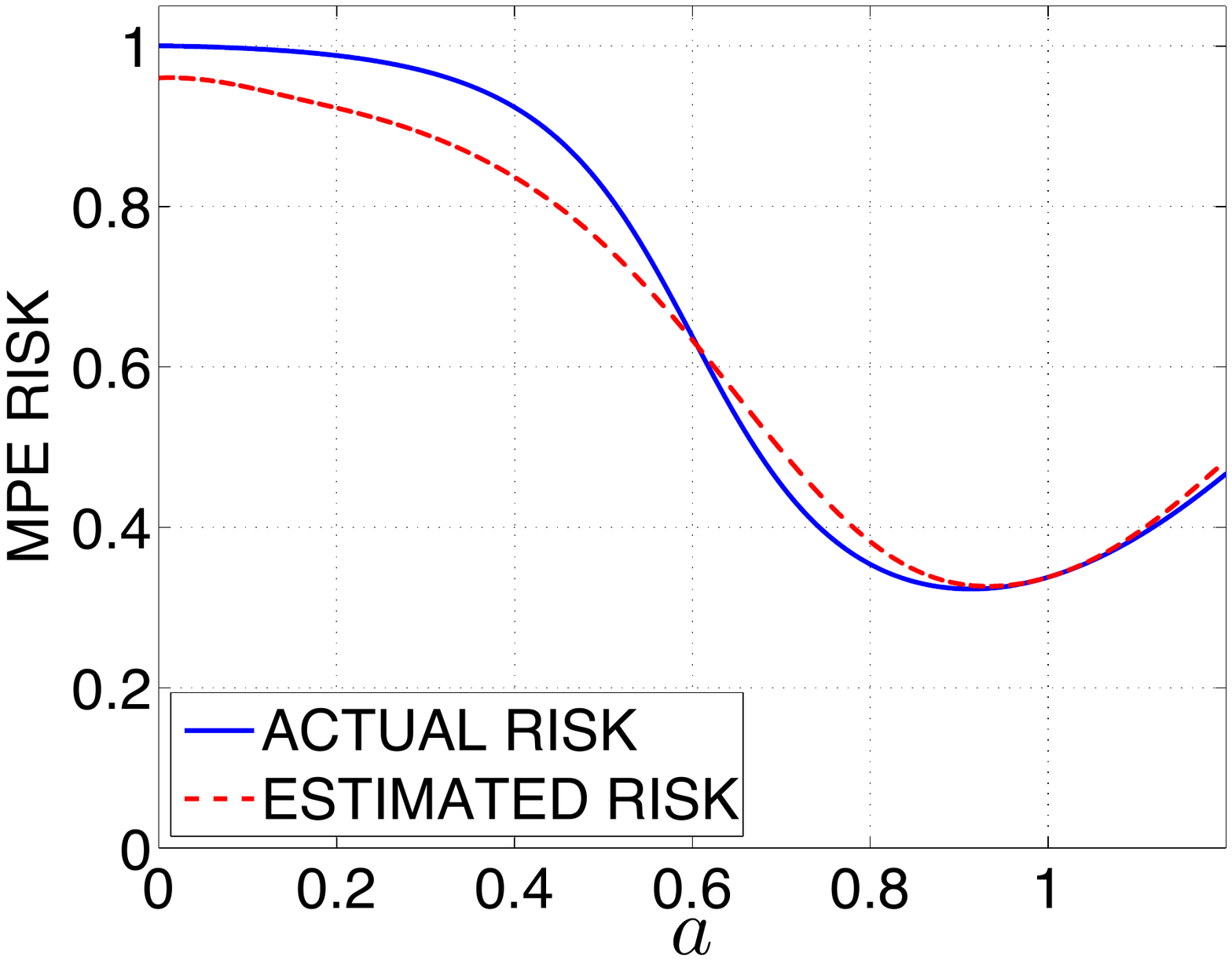}\\
 \text{\small{(a) Gaussian}} &  \text{\small{(b) Student's}-$t$} \\
\hspace{-0.21cm}\includegraphics[width=1.5in]{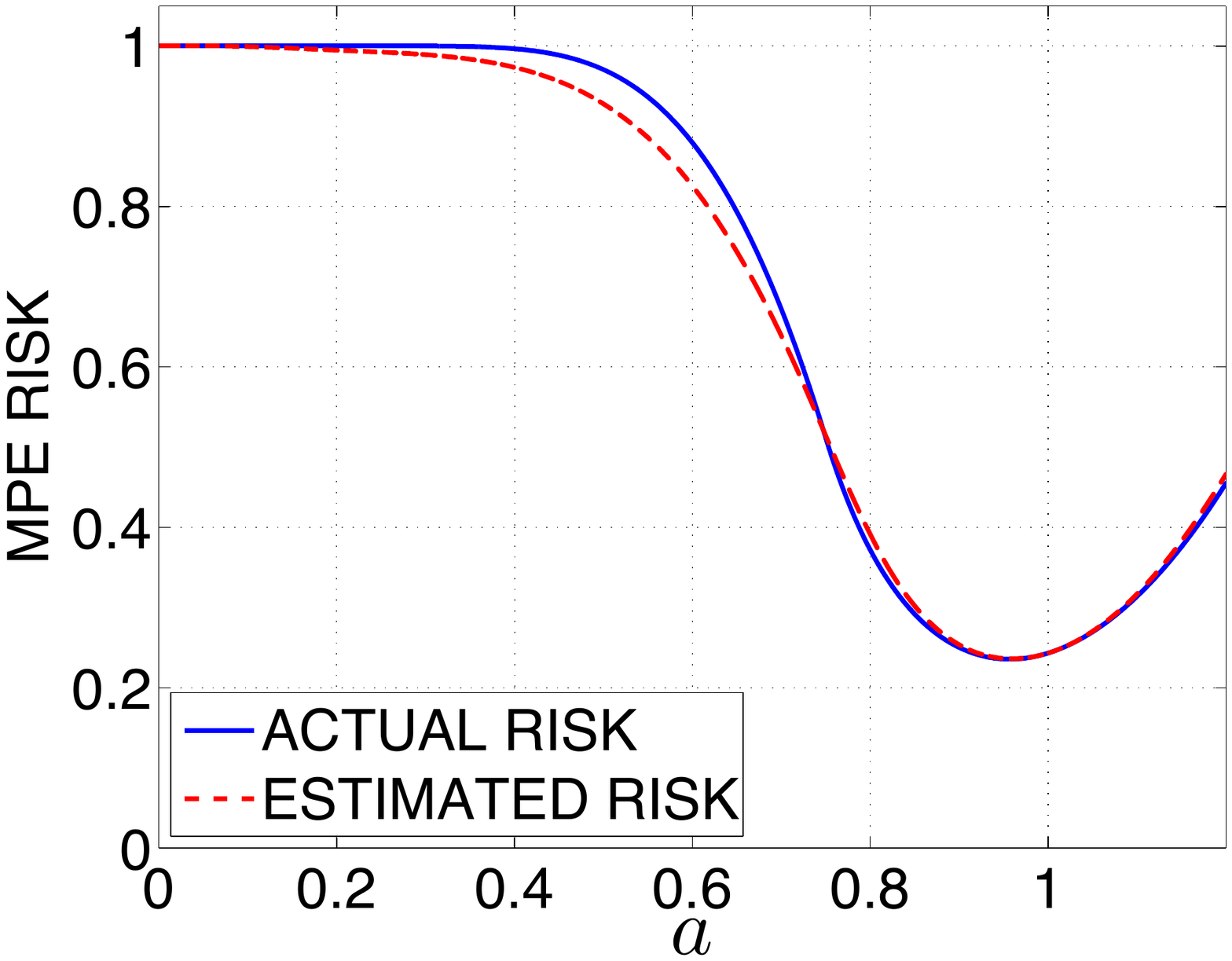}&
\hspace{-0.21cm}\includegraphics[width=1.5in]{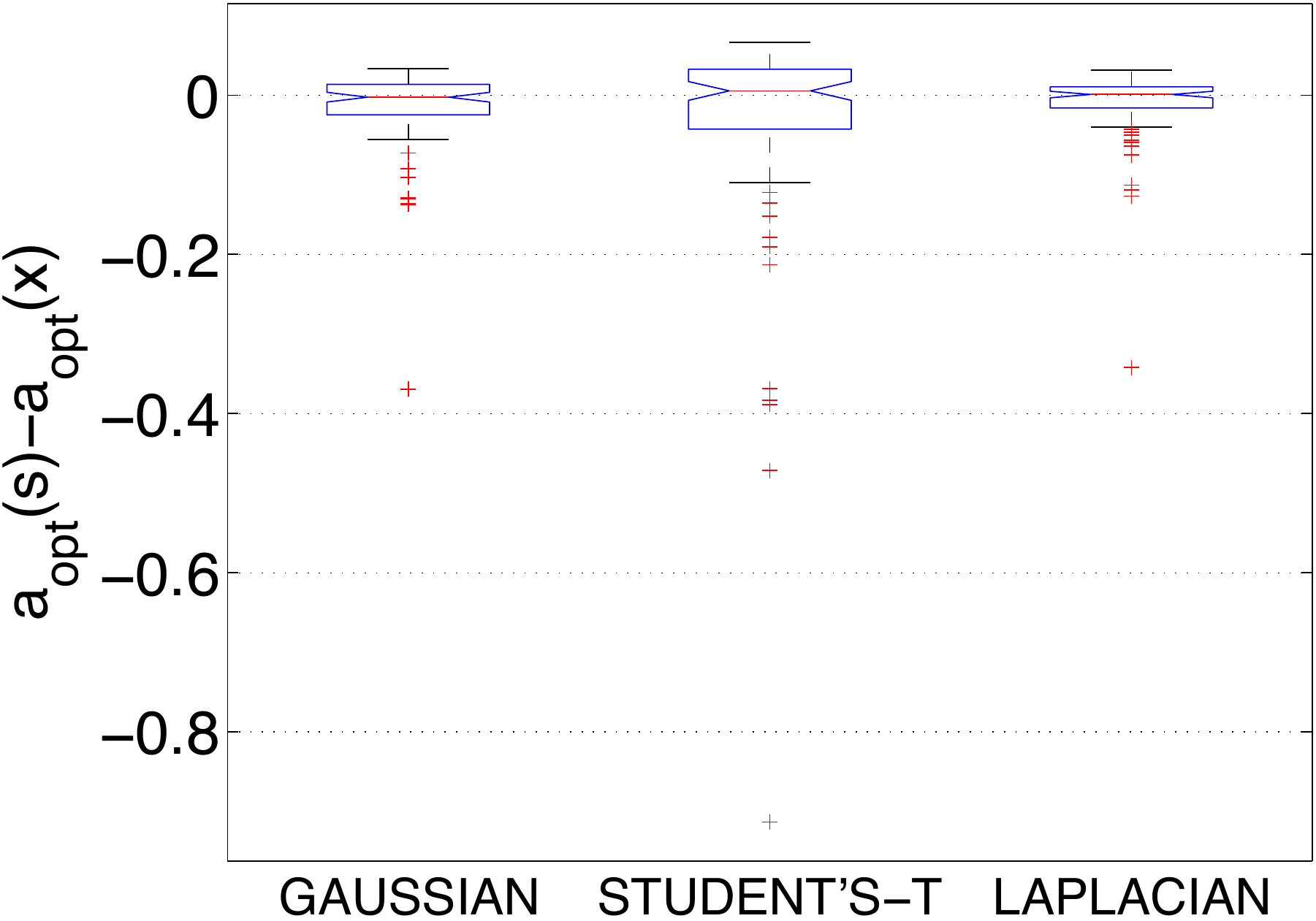}\\
 \text{\small{(c) Laplacian}} &  \text{\small{(d) Percentiles of error in minima}}\\
\end{array}$
\caption{\small  (Color online) The MPE risk averaged over 100 realizations for: (a) Gaussian, (b) Student's-$t$, and (c) Laplacian noise, versus the shrinkage parameter $a$; and (d) the percentiles of error in minima (obtained with $x$ versus $s$ (oracle)).}
\label{riskest}
\end{figure}

\indent To measure the closeness of $\widehat{\mathcal{R}}$ to $\mathcal{R}$, consider the example of estimating a scalar $s=4$ from a noisy observation $x$. The MPE risk estimate $\widehat{\mathcal{R}}$ is obtained by setting $\tilde{s}=x$. In Figures~\ref{riskest}(a),~\ref{riskest}(b), and~\ref{riskest}(c), we show the variation of the  actual risk $\mathcal{R}$ and its estimate $\widehat{\mathcal{R}}$ with $a$, averaged over $100$ independent trials, for Gaussian, Student's-$t$, and Laplacian noise distributions, respectively. The noise has zero mean, and the variance is taken as $\sigma^2=1$ for Gaussian and Laplacian models, whereas for Student's-$t$ model, the variance is $\sigma^2=2$. The value of $\epsilon$ is set equal to $\sigma$ while computing the MPE risk. We observe that $\widehat{\mathcal{R}}$ is a good approximation to $\mathcal{R}$, particularly in the vicinity of the minima. The deviation of the shrinkage parameter $a_{\text{opt}}(x)$, obtained by minimizing $\widehat{\mathcal{R}}$, with respect to its true value $a_{\text{opt}}(s)$ resulted from the minimization of $\mathcal{R}$, is shown in Figure~\ref{riskest}(d) for three noise models under consideration. The central red lines in Figure~\ref{riskest}(d) indicate the medians, whereas the black lines on the top and bottom denote the $25$ and the $75$ percentile points, respectively. We observe that $a_{\text{opt}}(x)$ is well concentrated around $a_{\text{opt}}(s)$, especially for Gaussian and Laplacian noise, barring a small number of outliers. 
\subsubsection{Perturbation Probability of the location of minimum}
\begin{figure}[t]
\centering
$\begin{array}{c}
\includegraphics[width=3.5 in]{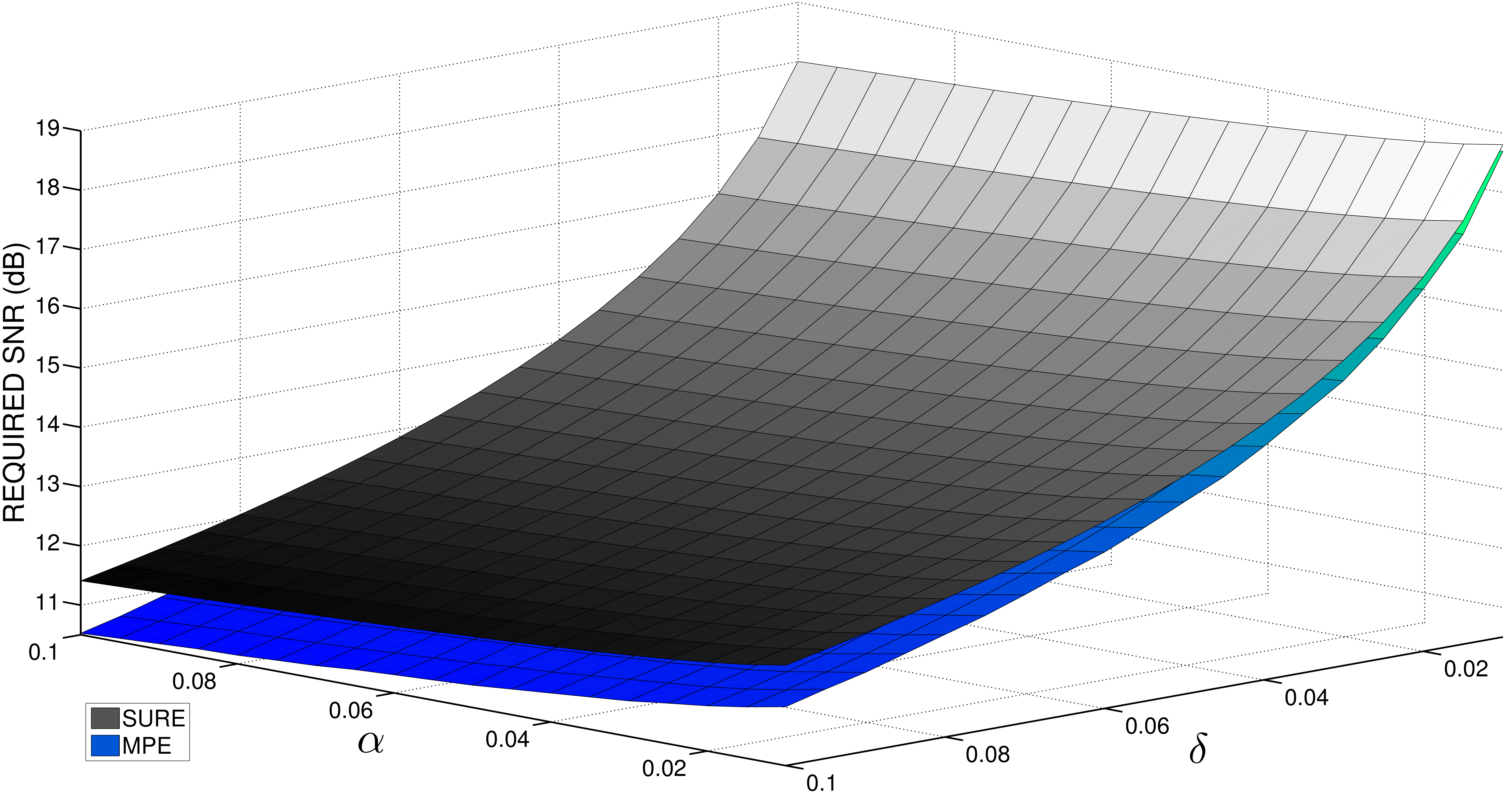}
\end{array}$
\caption{\small (Color online) Input SNR requirement for SURE (black) and MPE (blue) to ensure that the probability of $\delta$-perturbation of the minima is less than or equal to $\alpha$.}
\label{peturb_anal_deviation_mpe}
\end{figure}

\indent The location of the minimum of the MPE risk determines the shrinkage parameter. Therefore, one must ensure that it does not deviate too much from its actual value, with high probability, when $s$ is replaced by $x$ in the original risk $\mathcal{R}$. Let $a_{\text{opt}}\left(s\right)=\arg \underset{0\leq a \leq 1}{\min} \mathcal{R}\left(s;a\right)$ denote the argument that minimizes the true risk $\mathcal{R}$.  Consider the probability of deviation, given by
\begin{equation}
P_e^{\text{MPE}}=\mathbb{P}\left( \left|  a_{\text{opt}}\left(s\right) - a_{\text{opt}}\left(x\right)  \right| \geq \delta \right),
\label{devprob}
\end{equation}
for some $\delta>0$. Using a first-order Taylor series approximation of $a_{\text{opt}}\left(x\right)$ about $s$, and substituting $x=s+w$, we obtain $a_{\text{opt}}\left(x\right)\approx a_{\text{opt}}\left(s\right)+w a'_{\text{opt}}\left(s\right)$,
where $'$ denotes the derivative. 
The deviation probability $P_e^{\text{MPE}}$ in \eqref{devprob} simplifies to $P_e^{\text{MPE}}=\mathbb{P}\left( \left|  w  \right| \geq \frac{\delta}{\left| a'_{\text{opt}}\left(s\right) \right|} \right).$
For additive Gaussian noise $w$ with zero mean and variance $\sigma^2$, placing the Chernoff bound on $P_e^{\text{MPE}}$ leads to
\begin{equation*}
P_e^{\text{MPE}} \leq 2\exp\left( -\frac{\delta^2}{2\sigma^2\left| a'_{\text{opt}}\left(s\right) \right|^2}  \right).
\end{equation*}
To ensure that $P_e^{\text{MPE}}$ is less than $\alpha$, for a given $\alpha \in (0,1)$, it suffices to have
\begin{eqnarray}
\left| a'_{\text{opt}}\left(s\right) \right|^2 &\leq& \frac{\delta^2}{2\sigma^2 \log \left(\frac{2}{\alpha}\right)},
\label{MPE_aopt_robustness_criterion}
\end{eqnarray}  
which translates to a lower-bound on the input SNR. Since there is no closed-form expression available for $a'_{\text{opt}}\left(s\right)$ in the context of MPE risk, we empirically obtain the range of input SNR values $\displaystyle{\frac{s^2}{\sigma^2}}$, for which (\ref{MPE_aopt_robustness_criterion}) is satisfied.\\
\indent Analogously, to satisfy an upper bound on the deviation probability $P_e^{\text{SURE}}$ of the minimum in the case of  SURE, for a given deviation $\delta>0$, one must ensure that
\begin{equation}
  \frac{s^6}{8\sigma^6} \left( \delta - \frac{\sigma^4}{\left(  s^2+\sigma^2 \right)s^2}   \right)^2  \geq \log \left(\frac{2}{\alpha}\right).
\label{SURE_snr_requirement}
\end{equation}
The proof of \eqref{SURE_snr_requirement} is given in Appendix~\ref{sure_pert}.\\
\indent The minimum input SNR required to ensure $P_e\leq \alpha$ for both SURE- and MPE-based shrinkage estimators is shown in Figure~\ref{peturb_anal_deviation_mpe}, for different values of $\alpha$ and $\delta$. The MPE-risk estimate is obtained by replacing $s$ with $x$ and setting $\epsilon=\sigma$. We observe that reducing the amount of deviation $\delta$ for a given probability $\alpha$, or vice versa, leads to a higher input SNR requirement for both SURE and MPE. We also observe from Figure~\ref{peturb_anal_deviation_mpe} that, for given $\delta$ and $\alpha$, SURE requires a higher input SNR than MPE to keep the $\delta$-deviation probability under $\alpha$. Also, for a given input SNR, the $\delta$-deviation probability of the estimated shrinkage parameter $a_{\text{opt}}\left(x\right)$ from the optimum $a_{\text{opt}}\left(s\right)$ is smaller for MPE than SURE, thereby indicating that the MPE-based shrinkage is comparatively more reliable than the SURE-based one at lower input SNRs.
\subsubsection{Unknown noise distributions}
\label{mpe_sec_gmm_approx}
\begin{figure}[t]
\centering
$\begin{array}{cc}
\hspace{-0.2cm}\includegraphics[width=1.75in]{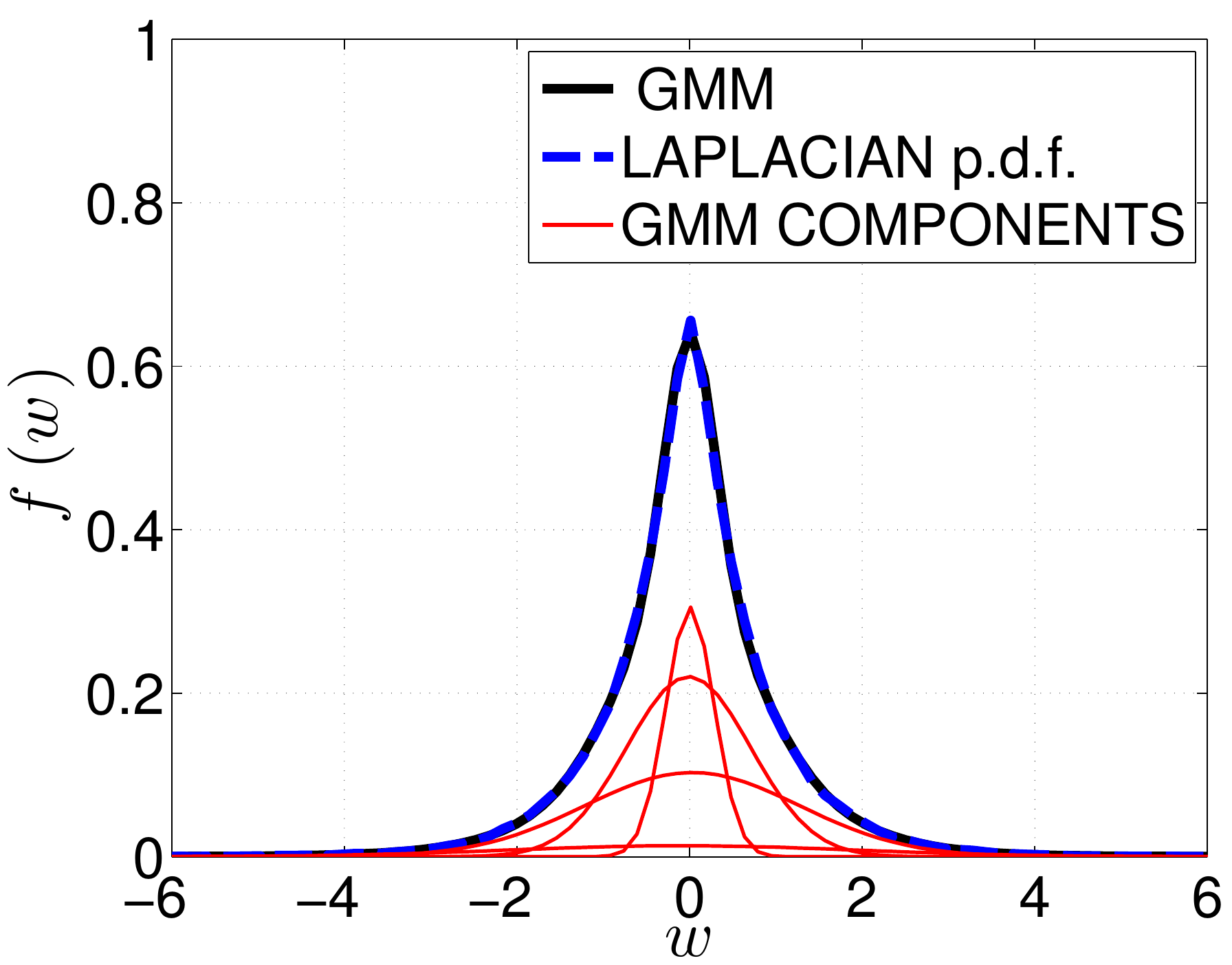}&
\hspace{-0.2cm}\includegraphics[width=1.75in]{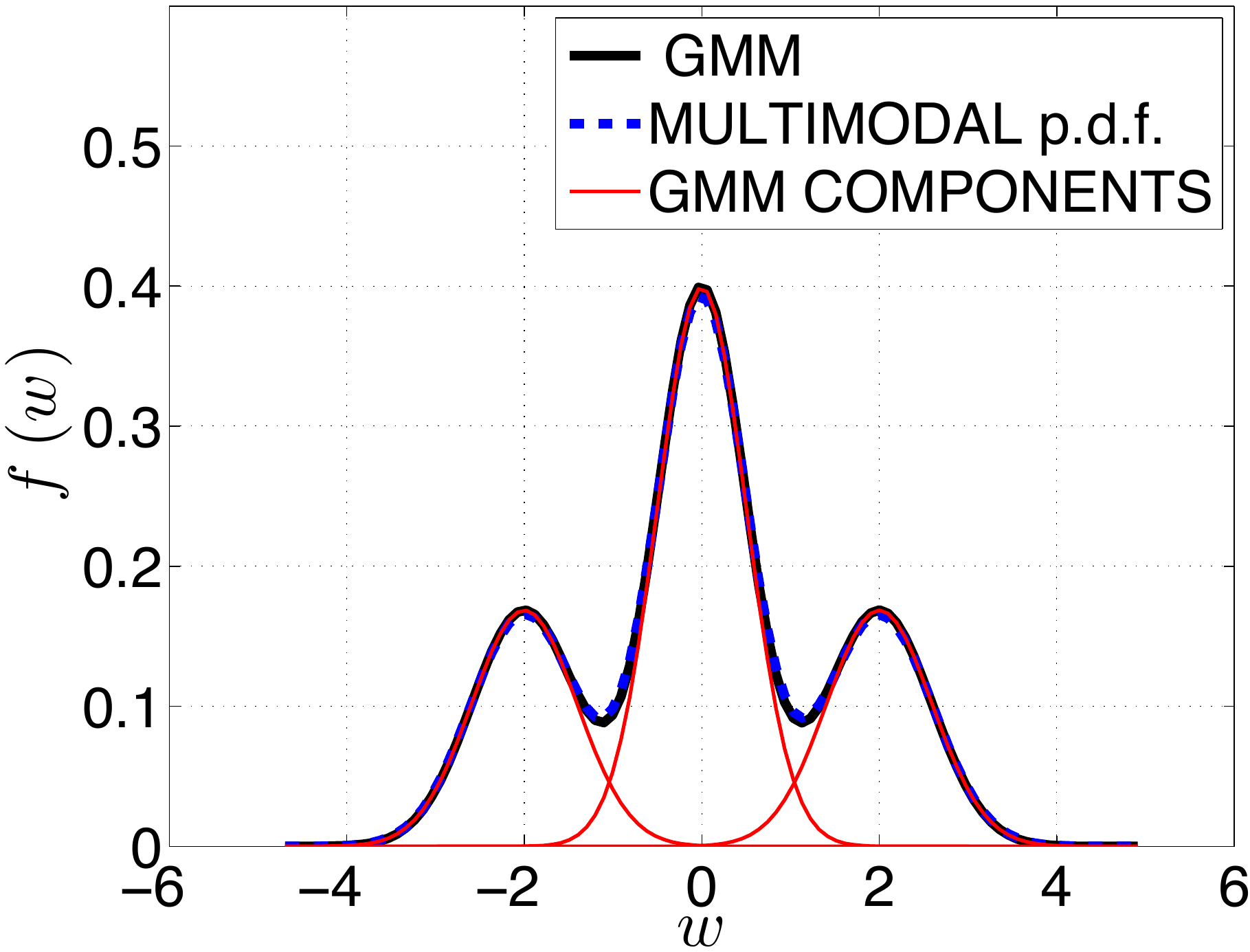}\\
\text{(a)} & \text{(b)}\\
\end{array}$
\caption{\small  (Color online) Original noise distribution and a GMM approximation: (a) Laplacian p.d.f. and its approximation using a four-component GMM; and (b) A multimodal p.d.f. and its three-component GMM approximation.}
\label{risk_gmm_fig}
\end{figure}

\begin{figure}[t]
\centering
$\begin{array}{ccc}
\hspace{-0.2cm}\includegraphics[width=1.2in]{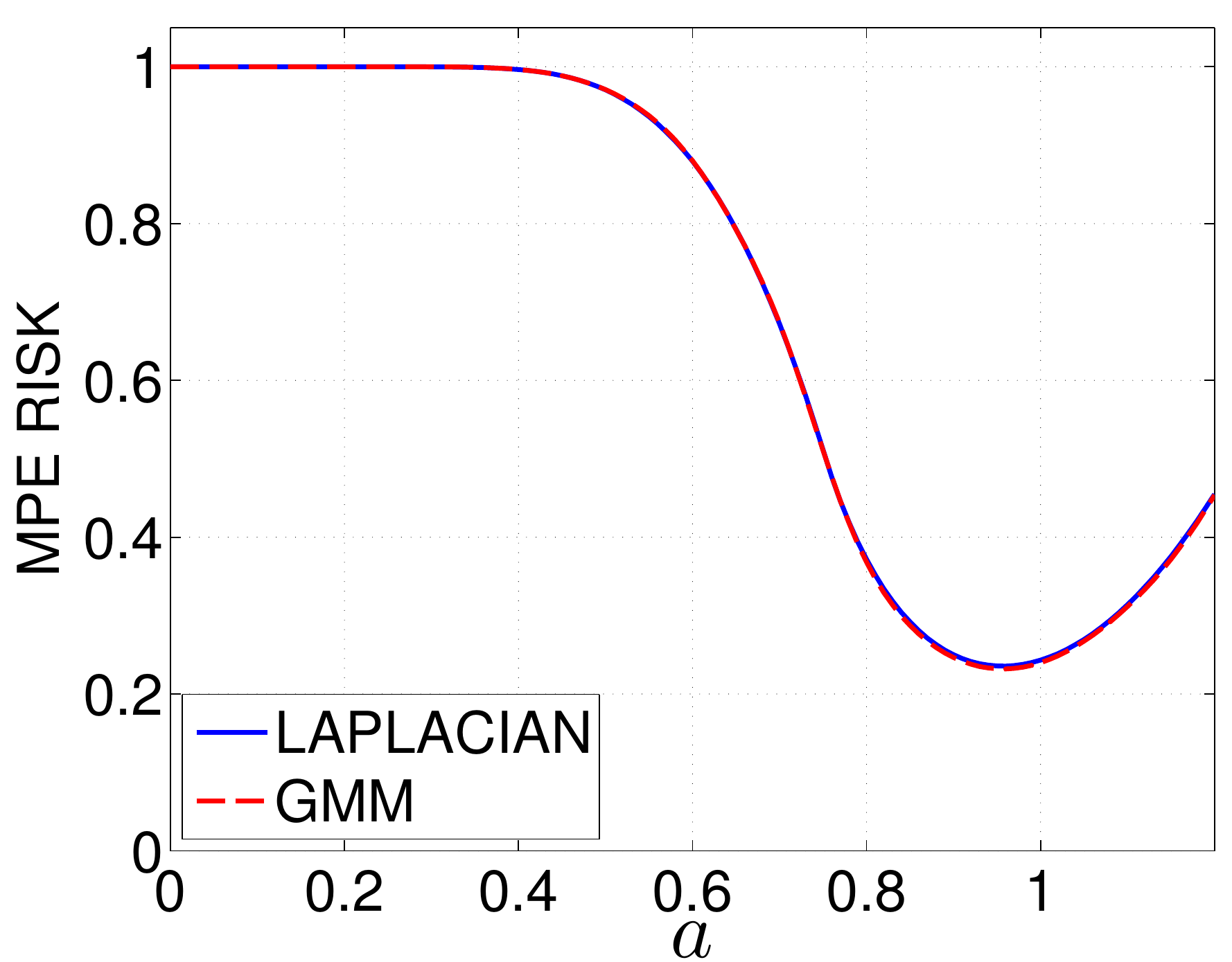}&
\hspace{-0.2cm}\includegraphics[width=1.2in]{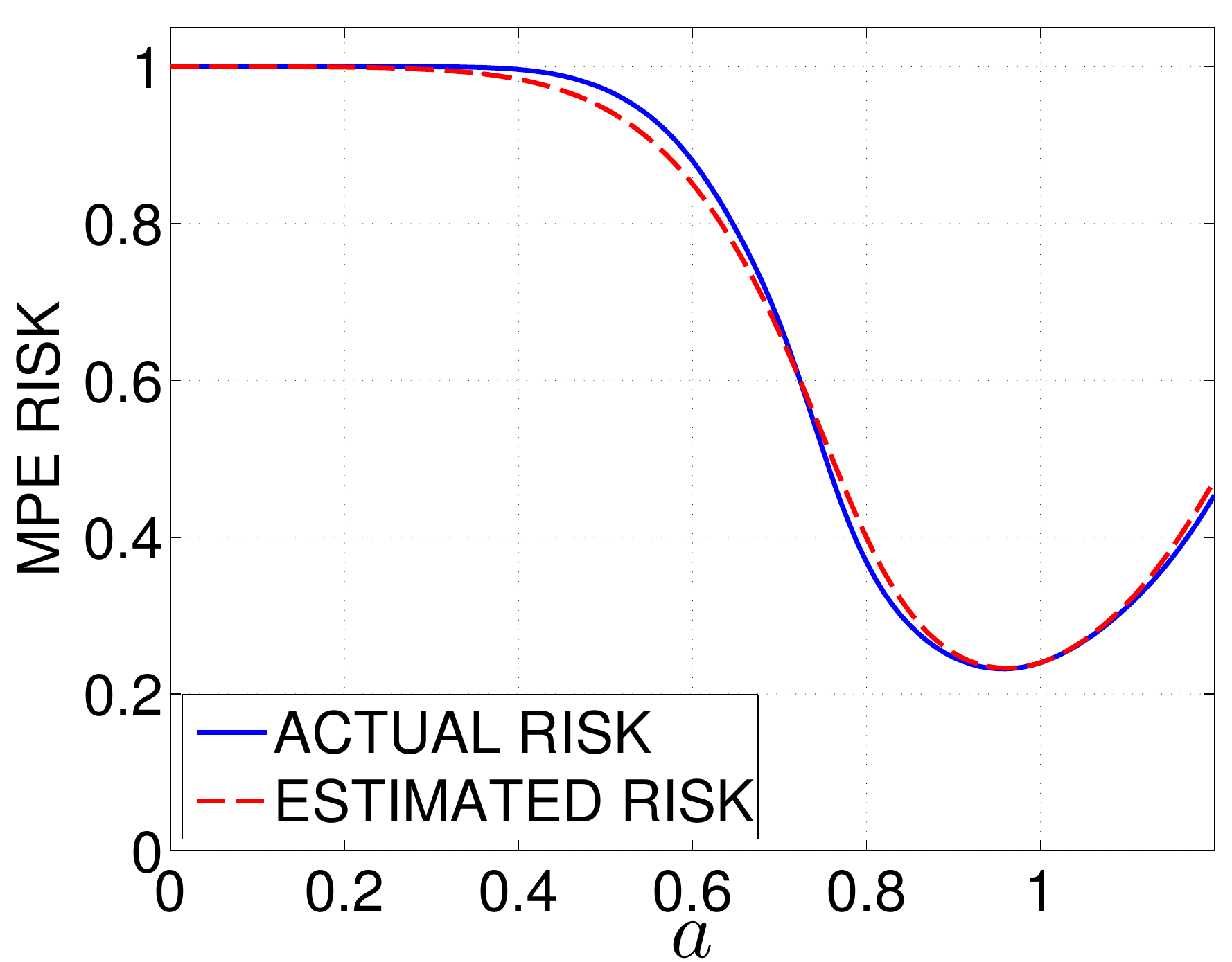}&
\hspace{-0.2cm}\includegraphics[width=1.2in]{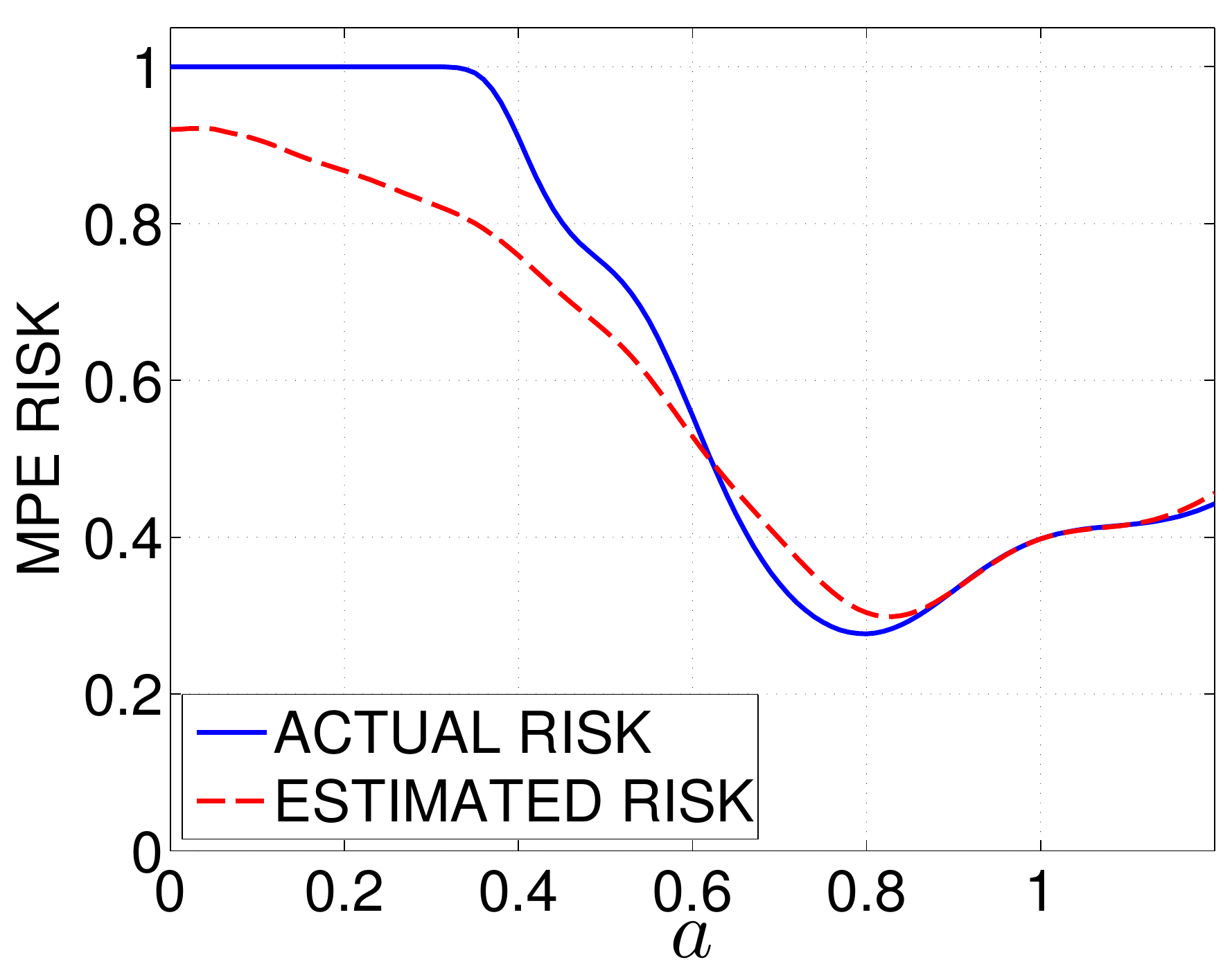}\\
\text{(a)} & \text{(b)} & \text{(c)}
\end{array}$
\caption{\small (Color online) The MPE risk estimate versus the shrinkage parameter $a$: (a) MPE risk for Laplacian noise, considering the Laplacian p.d.f. and its GMM approximation; (b) GMM-based MPE risk estimate for Laplacian noise; and  (c) MPE risk estimate for multimodal noise; for $\epsilon = \sigma$. The risk estimates are averaged over $100$ Monte Carlo realizations.}
\label{risk_gmm_fig_MPE}
\end{figure}

\indent In practical applications, the distribution of noise may not be known in a parametric form and may also be multimodal. At best, one would have access to realizations of the noise, from which the distribution has to be estimated. In such cases, approximation of the noise p.d.f. using a GMM is a viable alternative \cite{Plataniotis}, wherein  one can  estimate the parameters of GMM using the expectation-maximization algorithm \cite{Redner}. Gaussian mixture modeling is attractive as it comes with certain guarantees. For example, it is known  that a p.d.f. with a finite number of finite discontinuities can be approximated by a GMM to a desired accuracy except at the points of discontinuity \cite{Sorenson, Plataniotis}. The GMM approximation can be used even for non-Gaussian, unimodal distributions. For the GMM-based noise p.d.f.
\begin{eqnarray}
f(w)=\sum_{m=1}^{M}  \frac{\alpha_m}{\sigma_m \sqrt{2\pi}}\exp\left( -\frac{\left( w-\theta_m \right)^2}{2\sigma_m^2} \right),
\label{GMM_pdf_eqn}
\end{eqnarray}
the MPE risk turns out to be
\begin{eqnarray}
\widehat{\mathcal{R}}= \sum_{m=1}^{M}\alpha_m \left[ Q \left(  \frac{\epsilon-(a-1)\tilde{s}-\theta_m}{a\sigma_m} \right) +\right.\nonumber\\ \left. Q \left(  \frac{\epsilon+(a-1)\tilde{s}+\theta_m}{a\sigma_m} \right)\right],
\label{gmm_mpe_expr}
\end{eqnarray}
using \eqref{MPE_gauss_eqn}. For illustration, consider the estimation of a scalar $s=4$ in the transform domain from its noisy observation $x$. The additive noise is Laplacian distributed with zero mean and variance $\sigma^2=1$. The noise distribution is modeled using a GMM with $M=4$ components and the corresponding MPE risk estimate is obtained using (\ref{gmm_mpe_expr}) by setting $\tilde{s}=x$. In Figure~\ref{risk_gmm_fig}(a), we show a Laplacian p.d.f. and  its GMM approximation. Figure~\ref{risk_gmm_fig}(b) shows the GMM approximation to a multimodal distribution. Figure~\ref{risk_gmm_fig_MPE}(a) shows the MPE risk based on the original Laplacian distribution as well as  the GMM approximation, as a function of the shrinkage parameter $a$. The close match between the two indicates that the GMM is a viable alternative when the noise distribution is unknown or follows a complicated model. In Figure~\ref{risk_gmm_fig_MPE}(b),  we plot the GMM-based MPE risk and its estimate averaged over $100$ independent trials. We observe that the locations of minima of the actual risk and its estimate match closely, thereby justifying the minimization of $\widehat{\mathcal{R}}$. The MPE risk and its estimate are shown in  Figure~\ref{risk_gmm_fig_MPE}(c)  for the multimodal p.d.f. of Figure~\ref{risk_gmm_fig}(b). 
\subsection{MPE Risk for Subband Shrinkage}
\label{mpe_vector_shrink_define}
\begin{figure}[t]
\centering
$\begin{array}{cc}
\includegraphics[width=5.6cm]{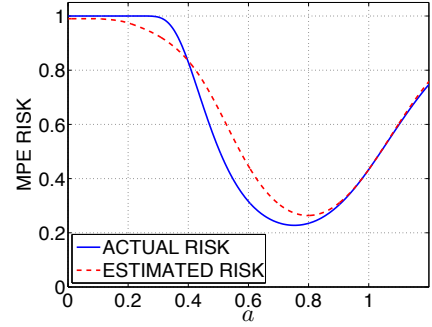}
\end{array}$
\caption{\small (Color online) The MPE risk and its estimate averaged over $100$ Monte Carlo trials for the subband shrinkage estimator versus $a$; where $\epsilon = \sqrt{k}\sigma$, with $k=8$. The additive noise is Gaussian with $\text{SNR}_{\text{in}} = 5$ dB. In each trial, $\mathbf{s}$ is generated by drawing samples from  $\mathcal{N}\left(2\times\bold 1_k, I_k\right)$.}
\label{risk_mpe_vector_fig}
\end{figure}

\indent Let $a_J$ be the shrinkage factor applied to the set of coefficients \{$x_i$,  $i\in J $\} in subband $J$. The estimate $\widehat{s}_J$ of the clean signal is obtained by $\widehat{\bold s}_J= a_J \bold x_J$, where $\bold x_J \in \mathbb{R}^{\left|J\right|}$ and $a_J \in \left[0,1\right]$. For notational brevity, we drop the subscript $J$, as we did for pointwise shrinkage, and express the estimator as $\widehat{\mathbf{s}} = a \mathbf{x}$, where boldface letters indicate vectors.\\
\indent Analogous to pointwise shrinkage, the MPE risk for subband shrinkage is defined as $\mathcal{R} = \mathbb{P}\left( \left\| \widehat{\mathbf{s}}-\mathbf{s} \right\|_2>\epsilon \right),$ which, for $\widehat{\mathbf{s}} = a \mathbf{x}$, becomes $\mathcal{R} = \mathbb{P}\left( \left\| a \mathbf{w} + (a-1)\mathbf{s} \right\|_2>\epsilon \right)$. For $\mathbf{w} \sim \mathcal{N}\left(0,\sigma^2 I\right)$, 
\begin{eqnarray}
\mathcal{R} &=& 1-F(\theta|k,\lambda),
\label{vector_MPE_eq3} 
\end{eqnarray}
where $k=\left|J\right|$, $\lambda = \sum_{j=1}^{k}\frac{\left(1-a\right)^2 s_j^2}{a^2\sigma^2}$, $\theta=\left(\frac{\epsilon}{a \sigma}\right)^2$, and $F(\theta|k,\lambda)$ is the c.d.f. of the non-central $\chi^2$ distribution, given by
\begin{eqnarray*}
F(\theta|k,\lambda)=\sum_{m=0}^{\infty} \frac{\lambda^m   e^{-\frac{\lambda}{2}} }{2^m\,m!}    \mathbb{P}\left[   \chi^{2}_{k+2m}  \leq \theta      \right],
\end{eqnarray*}
wherein $\chi^{2}_{v}$ denotes the central $\chi^{2}$ random variable having $v$ degrees of freedom.\\
\indent Similar to pointwise shrinkage, we propose to obtain an estimate $\widehat{\mathcal{R}}$ of $\mathcal{R}$ for subband shrinkage estimators either by replacing $s_j$ with $x_j$, or by an  estimate $\tilde{s}_j$ produced by any standard denoising algorithm. The optimum subband shrinkage factor is obtained by minimizing $\widehat{\mathcal{R}}$\\
\indent Figure~\ref{risk_mpe_vector_fig} shows the subband MPE risk and its estimate versus $a$, where the underlying clean signal $\mathbf{s}  \in \mathbb{R}^{\left|J\right|}$ is corrupted by Gaussian noise and the subband size is chosen to be $\left|J\right|=k=8$. The clean signal $\mathbf{s}$ is generated by drawing samples from  $\mathcal{N}\left(2\times\bold 1_k, I_k\right)$, where $\bold 1_k$ and $I_k$ denote a $k$-length vector of all ones and a $k\times k$ identity matrix, respectively. The observation $\mathbf{x}$ is obtained by adding zero-mean i.i.d. Gaussian noise  to $\mathbf{s}$, with an input SNR of $5$ dB, where the input SNR is defined as $\text{SNR}_{\text{in}}= 10 \log_{10}\left(  \displaystyle   \frac{1}{k\sigma^2} \sum_{n=1}^{k}s_n^2\right) \text{\,dB}$.  The MPE risk estimate is obtained by replacing $\mathbf{s}$ with $\mathbf{x}$ in (\ref{vector_MPE_eq3}), which does not significantly shift the location of the minimum (cf. Figure~\ref{risk_mpe_vector_fig}).
\section{Experimental Results for MPE-Based Denoising}
\label{exp_results_sec}
The performance of the MPE-based pointwise and subband shrinkage estimator is validated on a synthesized harmonic signal (of length $N=2048$) in Gaussian noise and the \textit{Piece-Regular} signal (of length $N=4096$) in Gaussian, Student's-$t$, and Laplacian noise. The \textit{Piece-Regular} signal has both smooth and rapidly-varying regions, making it a suitable candidate for the assessment of denoising performance.  
\subsection{Performance of Pointwise-Shrinkage Estimator}
\subsubsection{Harmonic signal denoising}
\indent Consider the signal 
\begin{equation}
s_n = \cos\left( \frac{5\pi n}{2048} \right)+ 2 \sin\left( \frac{10\pi n}{2048} \right), 0\leq n \leq 2047,
\label{harmonic_signal_def}
\end{equation}
in additive white Gaussian noise, with zero mean and variance $\sigma^2$. Since the denoising is carried out in the DCT \cite{KRRao} domain, the Gaussian noise statistics remain unaltered. For the purpose of illustration, we assume that $\sigma^2$ is known. In practice, $\sigma^2$ may not be known a priori and could be replaced by the robust median estimate \cite{mallat} or the trimmed estimate \cite{Socheleau}. The clean signal is estimated using inverse DCT after applying the optimum shrinkage. The denoising performance of the MPE and SURE-based approaches is compared in Table \ref{table_performance_on_synthesized_signal}. In case of the Wiener filter, the power spectrum of the clean signal is estimated using the standard spectral subtraction technique \cite{Boll,PLoizou}. We observe that MPE-based shrinkage with $\epsilon = 3.5\,\sigma$ is superior to SURE and Wiener filter (WF) by $8$--$12$ dB. The comparison also shows that the performance of the MPE depends critically on $\epsilon$. 
\begin{table}[t]
\centering
\caption{\small Comparison of  MPE,  SURE-based shrinkage estimator and Wiener filter (WF) for different input SNRs. The output SNR  values are averaged over $100$ noise realizations.}
\begin{tabular}{ p{1cm} |p{1cm}| p{1cm}|  p{1cm}| p{1cm} | p{1.cm}}
\hline
\hline
\small Input SNR &\multicolumn{5}{c}{ Output SNR (dB) } \\
\cline{2-6}
\small (dB)  &  \multicolumn{3}{c|}{MPE} &  SURE  & WF  \\
\cline{2-4}
&$\epsilon$$=$3.5$\sigma$&$\epsilon$$=$2.5$\sigma$&$\epsilon$$=$1.5$\sigma$& & \\
\hline
$ -5.0$&$11.67$&$5.99$&$-0.18$ &$-0.27$&$1.44 $\\ \hline
$ -2.5$&$14.42$&$8.62$&$2.34$&$2.23$&$3.96$ \\ \hline
$       0$&$17.02$&$10.96$&$4.80$&$4.71$&$6.35$ \\ \hline
$  2.5$&$19.08$&$13.36$&$7.31$&$7.21$&$8.79 $\\ \hline
$  5.0$&$21.25$&$15.52$&$9.72$&$9.64$&$11.09$ \\ \hline
$  7.5$&$22.93$&$18.26$&$12.32$&$12.23$&$13.60$ \\ \hline
$ 10.0$&$25.34$&$20.57$&$14.77$&$14.69$&$15.92$ \\ \hline
$  12.5$&$26.91$&$22.79$&$17.26$&$17.17$&$18.20$ \\ \hline
$  15.0$&$28.77$&$25.05$&$19.66$ &$19.59$&$20.33$\\ \hline
$  17.5$&$30.74$&$27.44$&$ 22.20$&$22.12$&$22.57$\\ \hline
$  20.0$&$32.65$&$29.61$&$ 24.61$&$24.54$&$24.60$ \\ \hline
 \hline
\end{tabular}
\label{table_performance_on_synthesized_signal}
\end{table}
\subsubsection{\textit{Piece-Regular} signal denoising}
\label{exp_results_ECG_pointwise}
\begin{figure*}[t]
\centering
$\begin{array}{ccc}
\includegraphics[width=5.6cm]{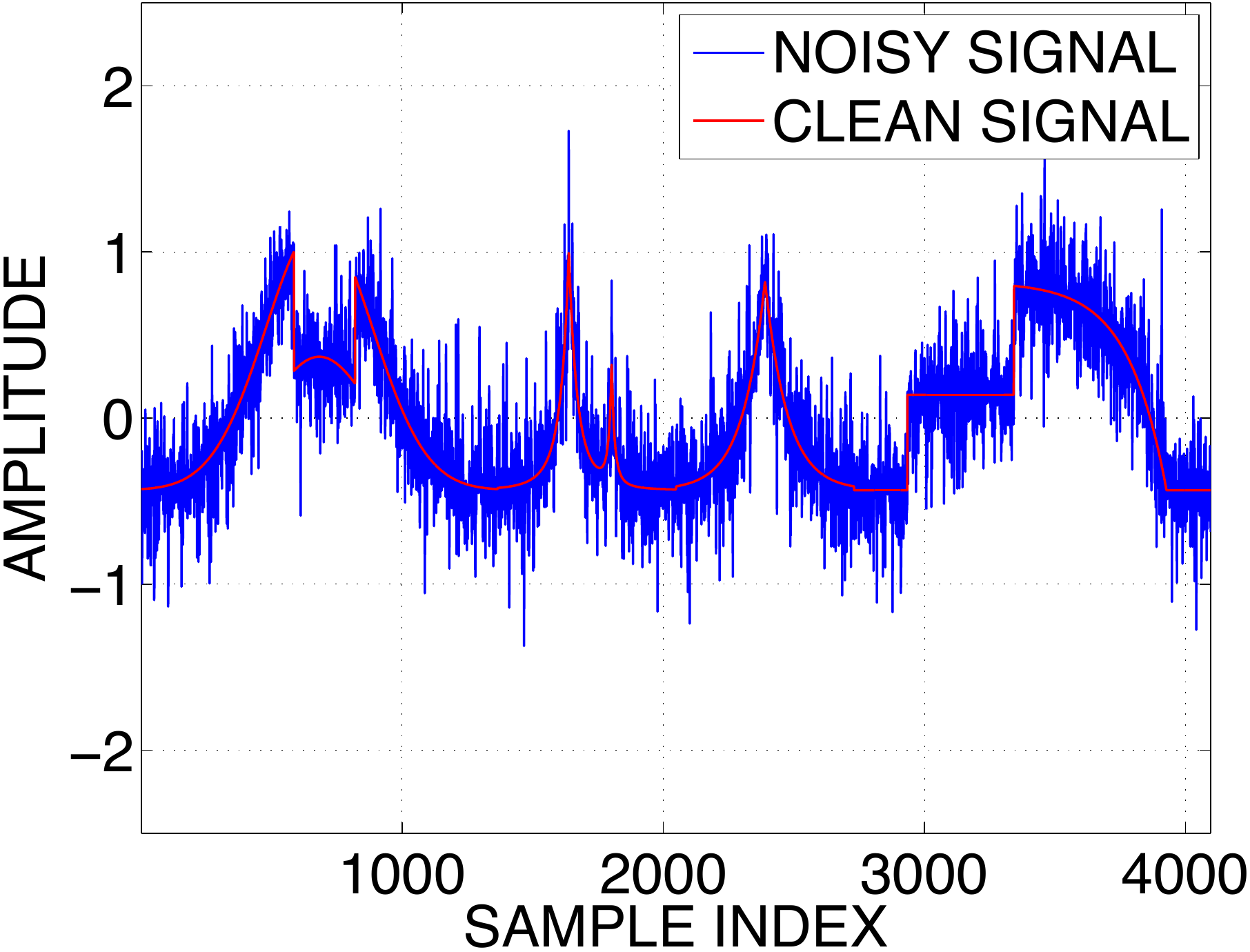}&
\includegraphics[width=5.6cm]{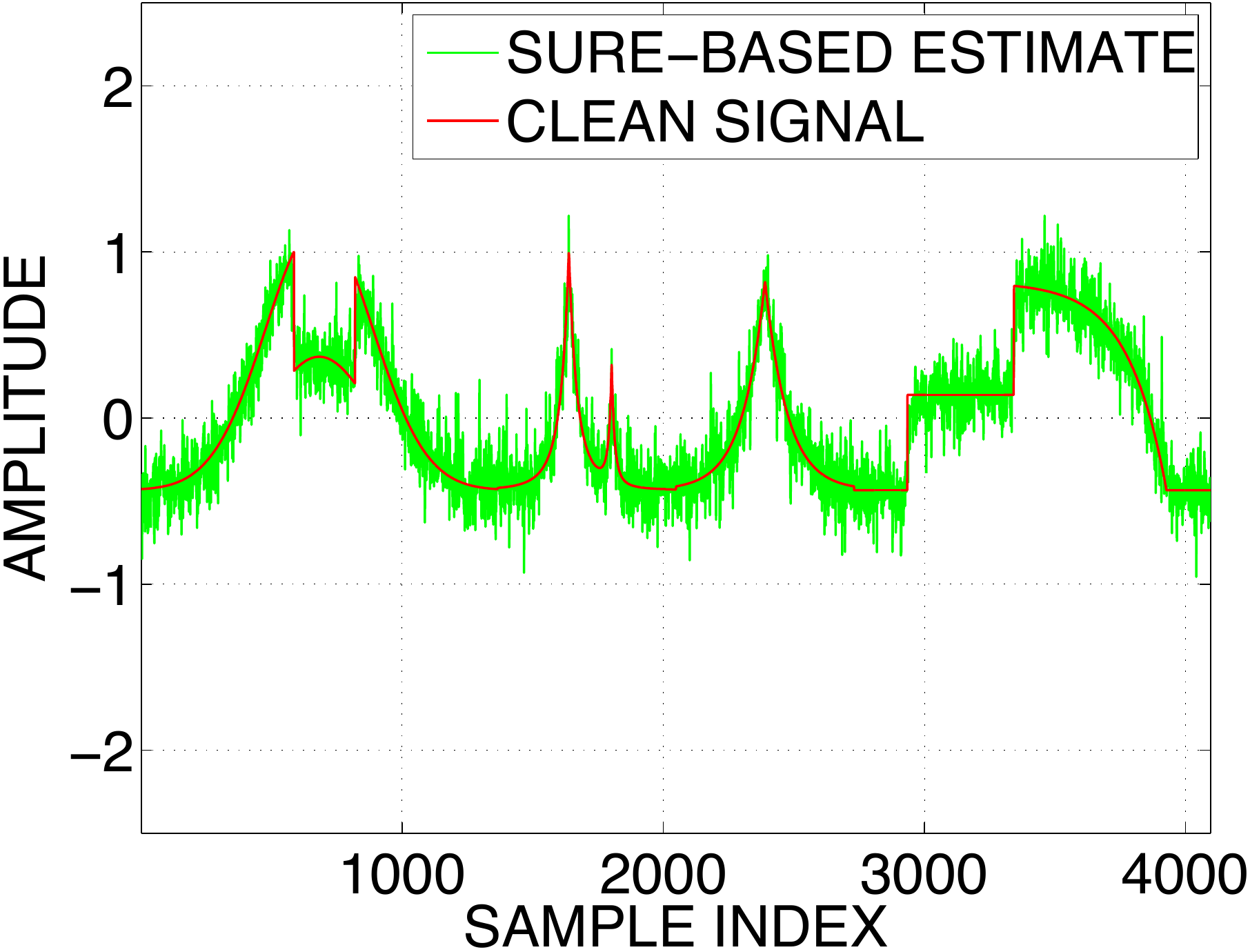}&
\includegraphics[width=5.6cm]{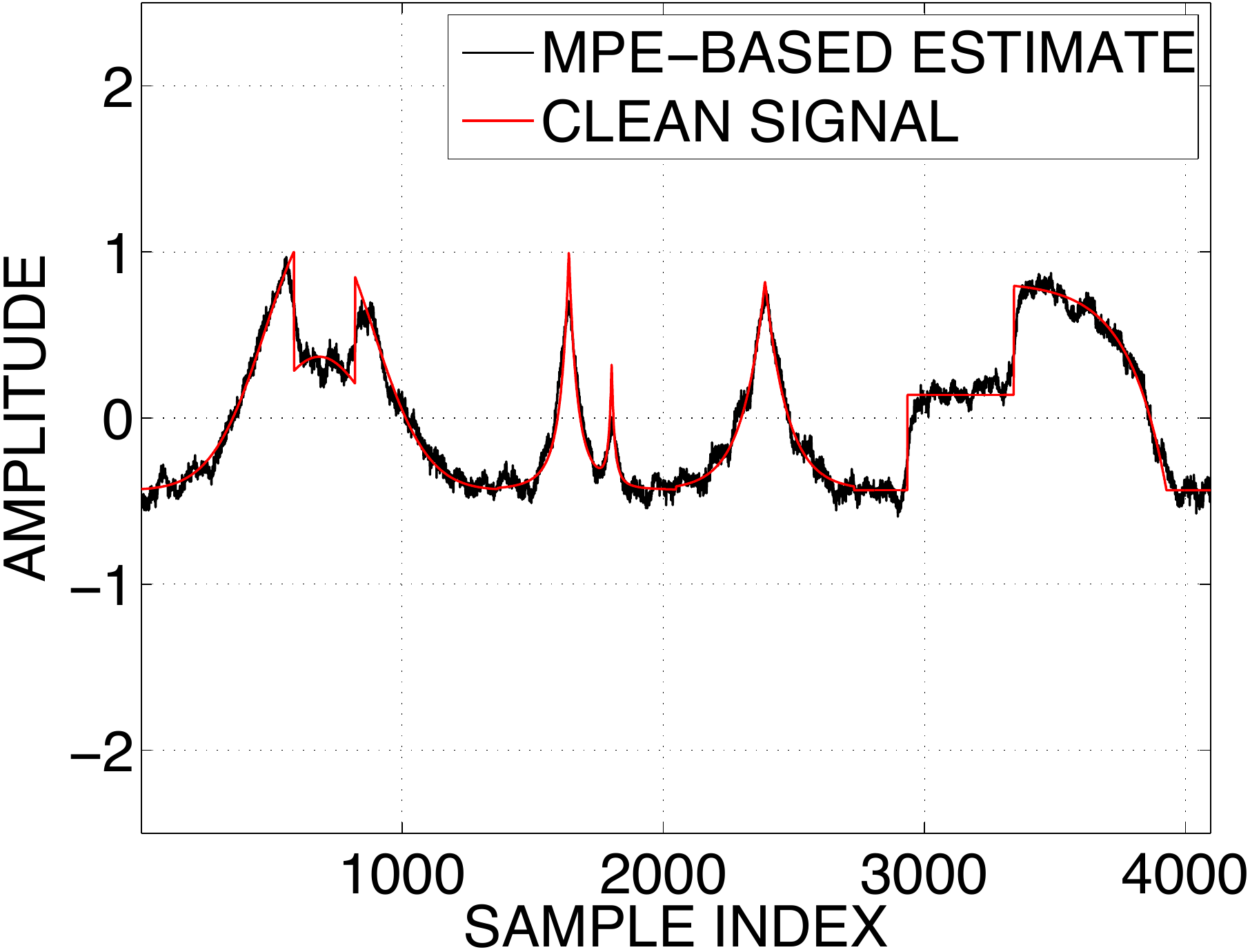}\\
\text{Input SNR = 5.47 dB} & \text{Output SNR = 9.71 dB} & \text{Output SNR = 14.88 dB}
\end{array}$
\caption{\small (Color online) Denoising performance of the MPE- and SURE-based pointwise shrinkage estimators for the \textit{Piece-Regular} signal corrupted by Laplacian noise. The MPE risk is calculated using a GMM approximation, and by setting $\epsilon=3\sigma$.}
\label{gmm_denoising_PR_laplacian_fig}
\end{figure*}

\begin{figure*}[t]
\centering
$\begin{array}{cccc}
\includegraphics[width=1.75in]{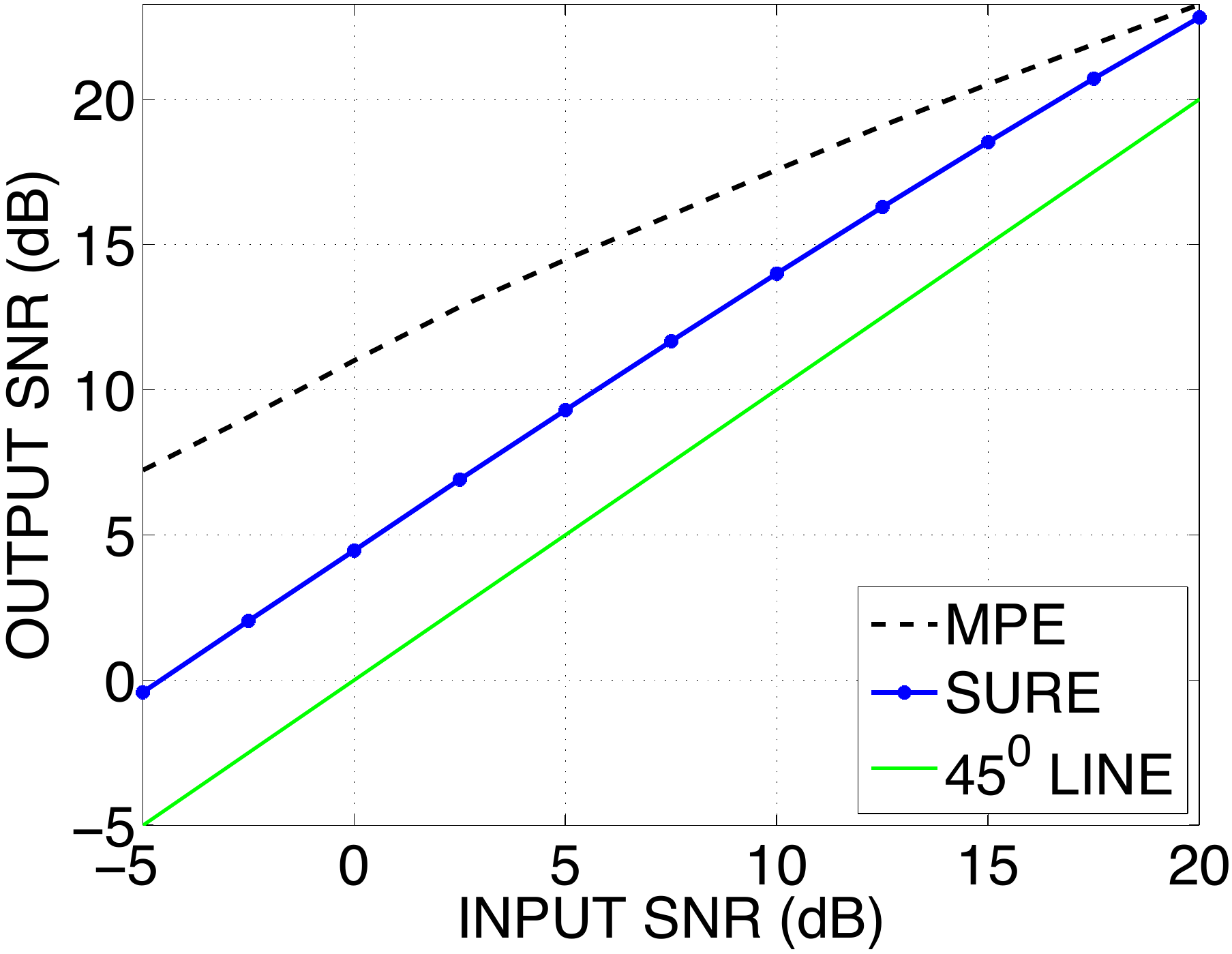}&
\includegraphics[width=1.75in]{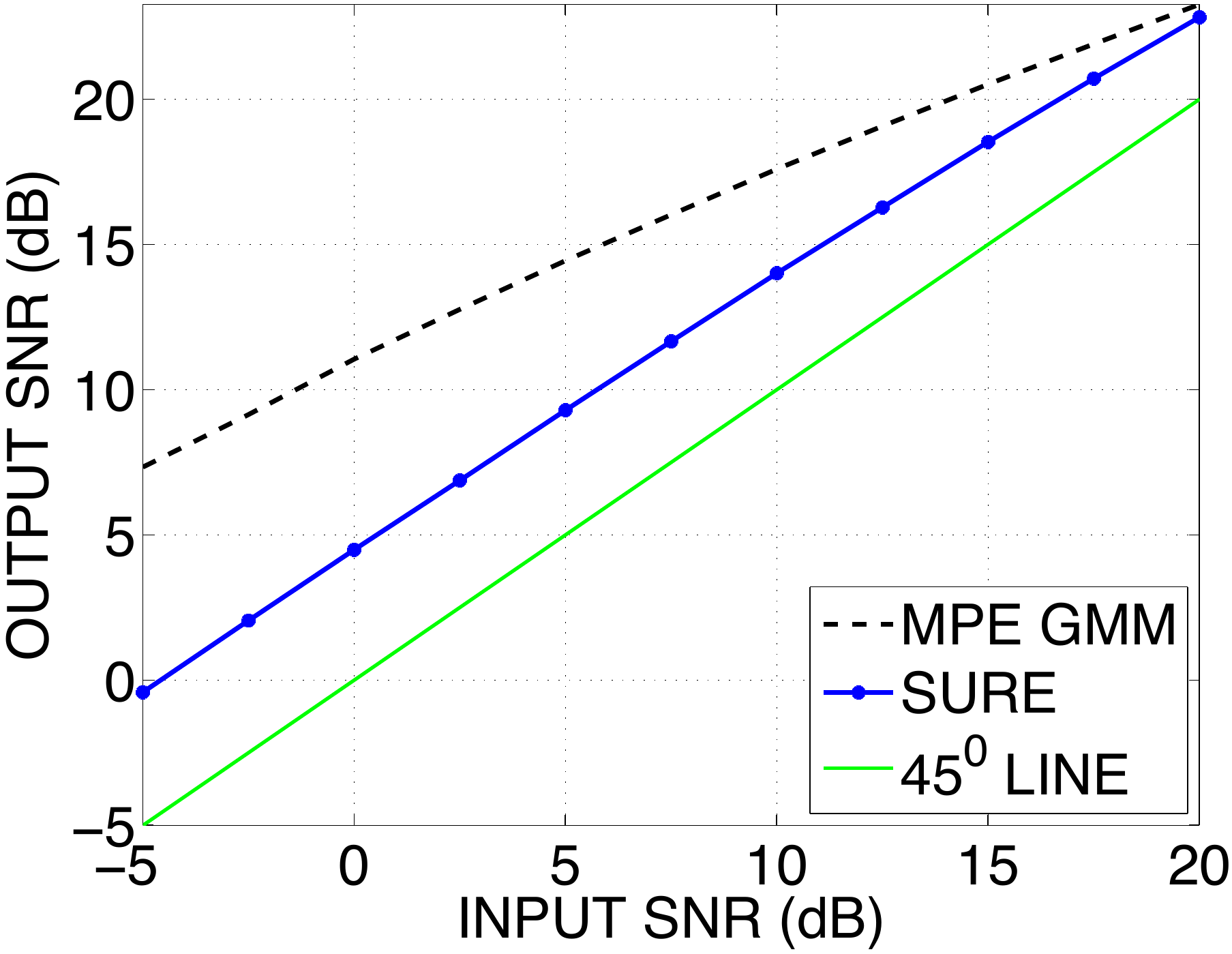}&
\includegraphics[width=1.75in]{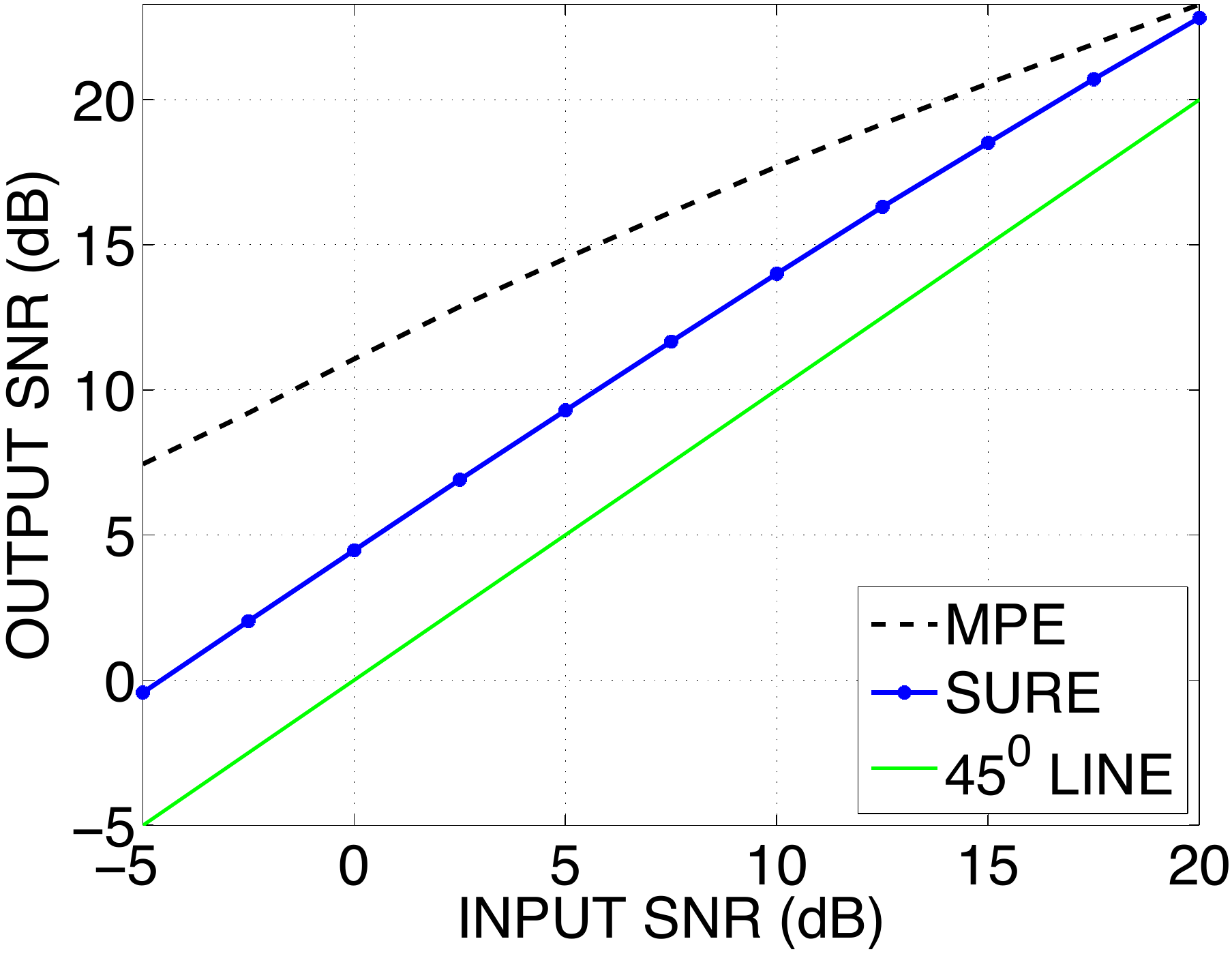}&
\includegraphics[width=1.75in]{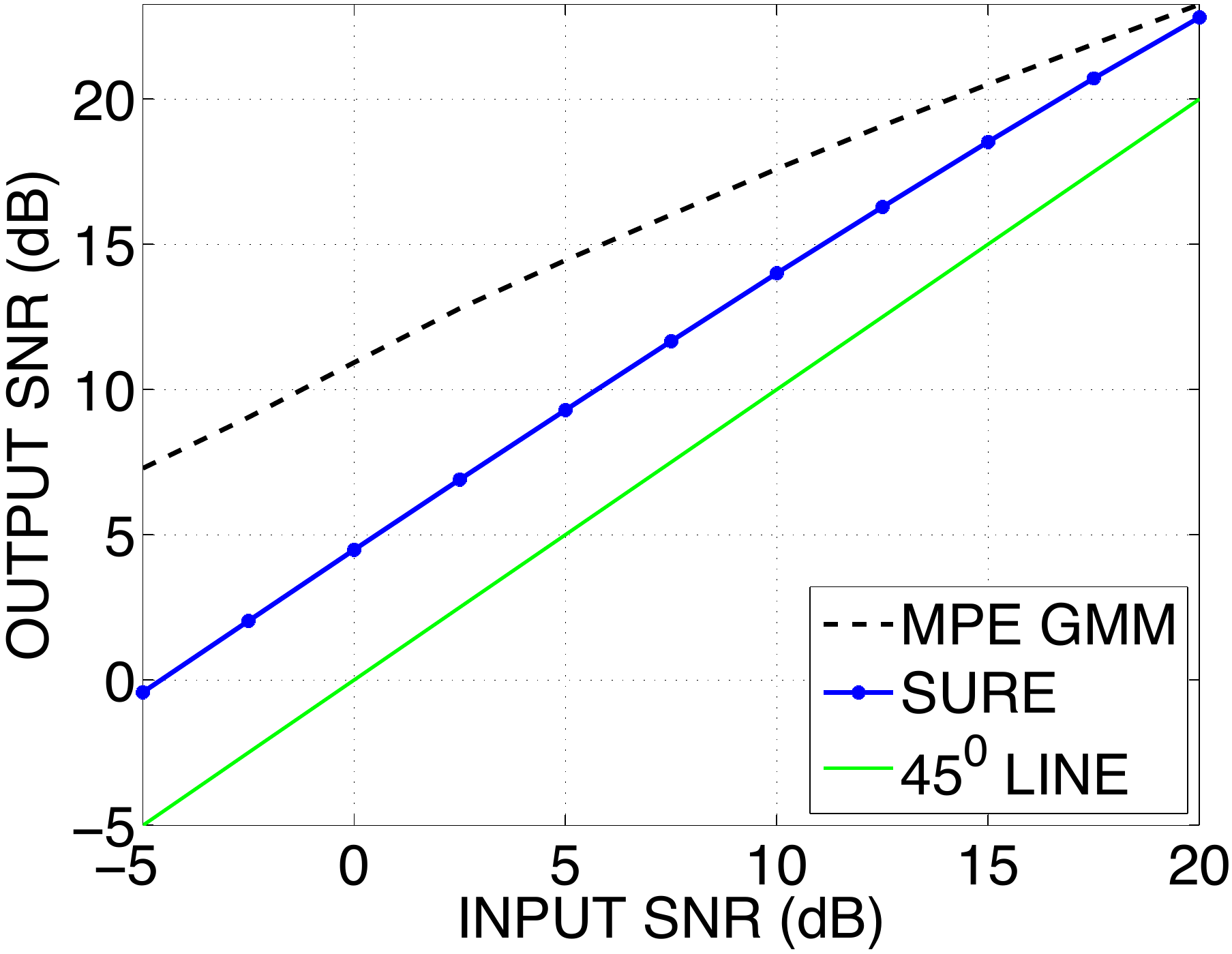}\\
\text{(a) Gaussian noise} & \text{(b) Laplacian noise} &\text{(c) Student's-$t$ noise} & \text{(d) GMM noise}\\ 
\end{array}$
\caption{\small (Color online) Output SNR-versus-input SNR corresponding to the MPE- and SURE-based pointwise shrinkages, under various noise distributions. The output SNR values are calculated by averaging over $100$ independent noise realizations.}
\label{PR_in_out_snr_fig_dif_noise}
\end{figure*}

\indent We consider noisy copies of the \textit{Piece-Regular} signal, taken from the \textit{Wavelab} toolbox \cite{wavelab}, under Gaussian, Student's-$t$, and Laplacian contaminations. The noise variance is assumed to be known. Notably, the Gaussian, GMM, and Student's-$t$ distributions of noise are preserved by an orthonormal transform \cite{roth}, unlike the Laplacian statistics. Therefore, the MPE estimate for Laplacian noise is computed based on a four-component GMM approximation in the DCT domain. The denoised output signal corresponding to Laplacian noise is shown in Figure~\ref{gmm_denoising_PR_laplacian_fig} for illustration. The MPE estimates are better than SURE estimates. The SNR plots in Figure~\ref{PR_in_out_snr_fig_dif_noise} indicate that the MPE outperforms SURE for the noise statistics under consideration and that the gains are particularly  high in the input SNR range of $-5$ to $20$ dB, and tend to reduce beyond $20$ dB.
\subsubsection{Effect of $\epsilon$ on the denoising performance of MPE}
\label{choice_epsilon}
\begin{figure}[t]
$\begin{array}{cc}
\textit{Piece-Regular}\,\text{signal}&\text{Harmonic signal}\\
\includegraphics[width=1.75in]{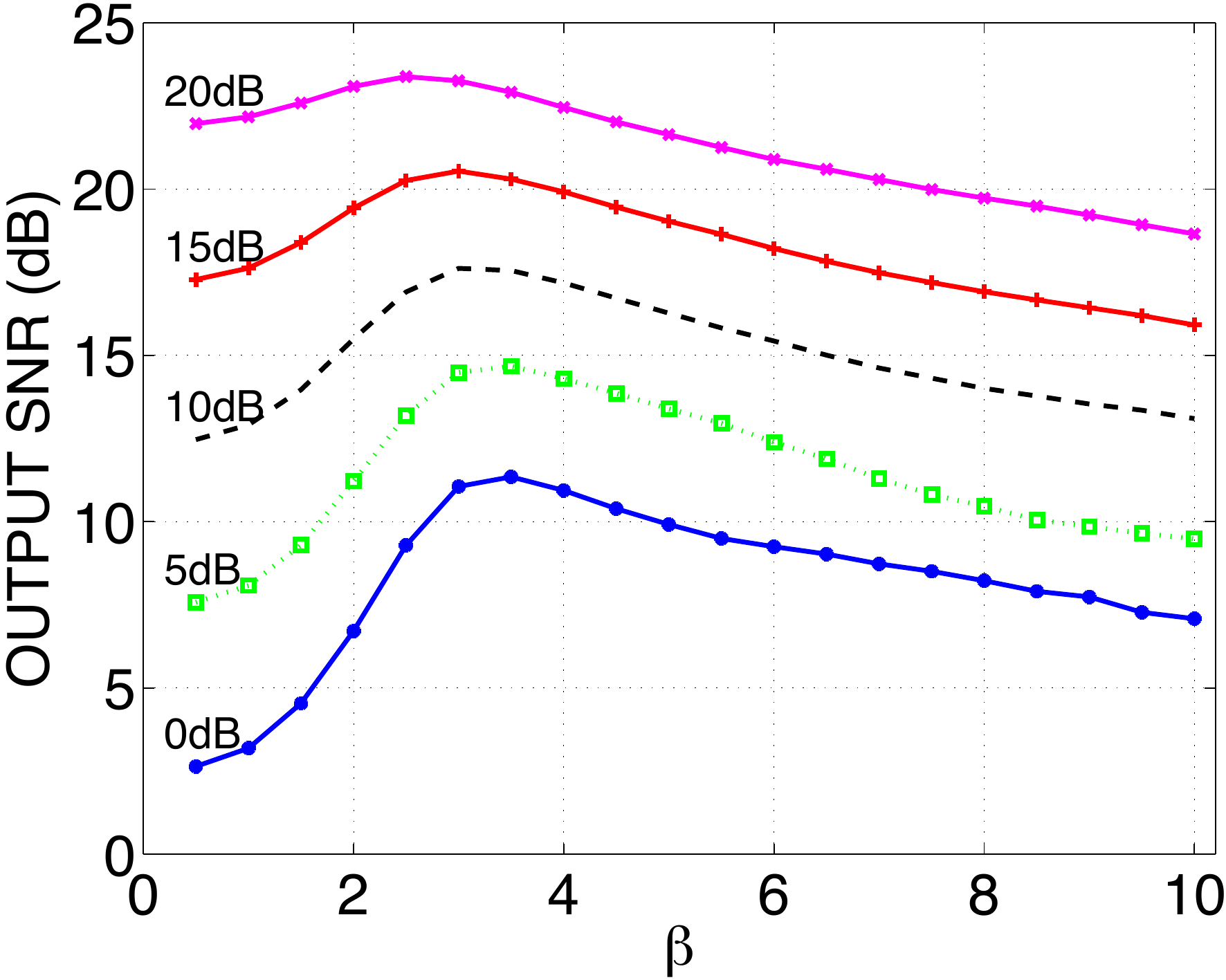}&
\includegraphics[width=1.75in]{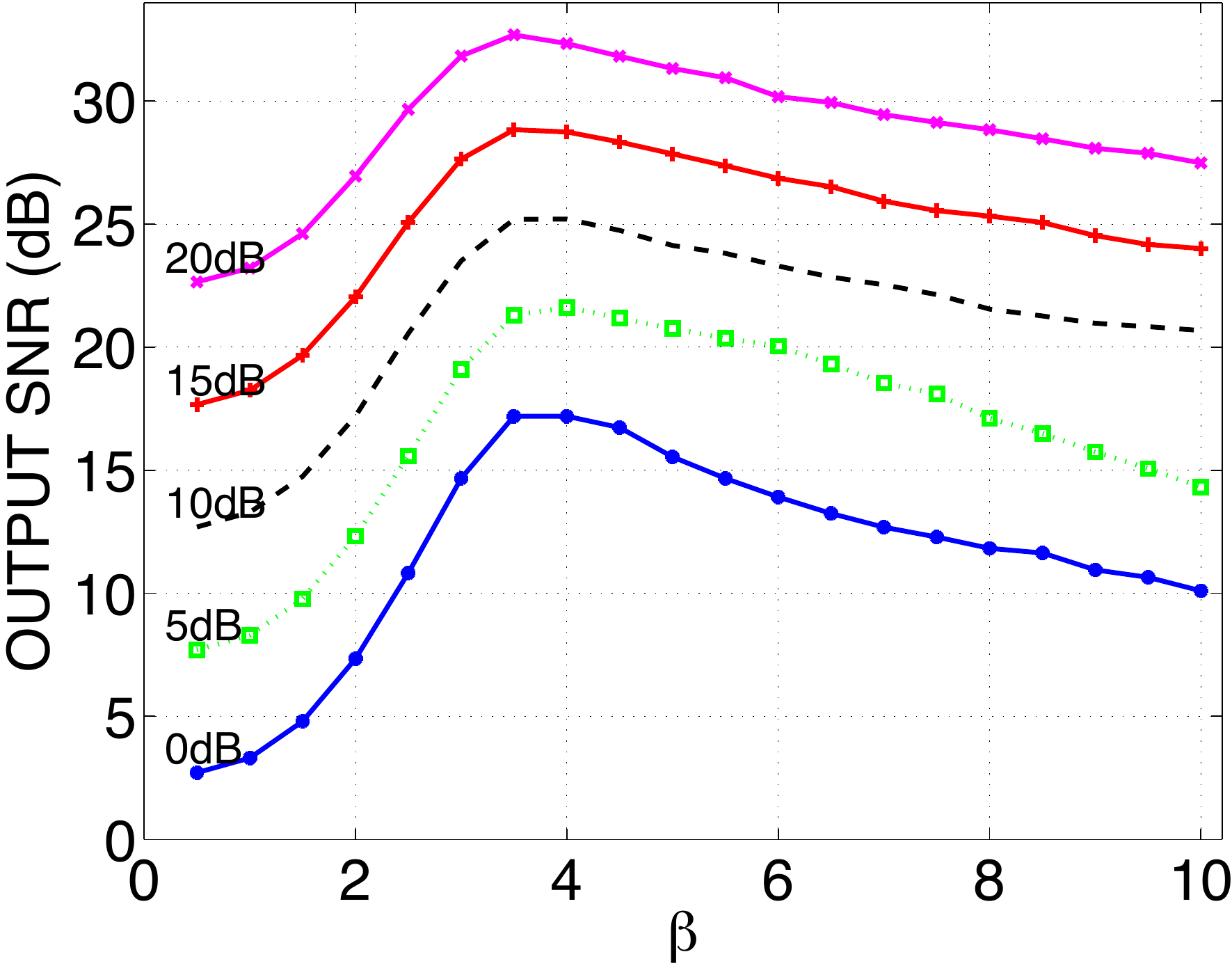}\\
\end{array}$
\caption{\small  (Color online) Average output SNR of the pointwise MPE shrinkage as a function of $\beta=\frac{\epsilon}{\sigma}$, for different values of input SNR. The output SNR curves attain peaks when $\beta \approx 3$.}
\label{in_out_snr_fig}
\end{figure}

\indent Obtaining a closed-form expression for the $\epsilon$  that maximizes the output SNR is not straightforward. We determine the optimum $\epsilon$ empirically by measuring the SNR gain as a function of $\epsilon$ (cf. Figure~\ref{in_out_snr_fig}), for i.i.d. Gaussian noise. We observe that the output SNR exhibits a peak approximately at $\beta = \frac{\epsilon}{\sigma}=3.5$ for the harmonic signal in \eqref{harmonic_signal_def} and at $\beta=3$ for the \textit{Piece-Regular} signal. As a rule of thumb, we recommend to choose $\epsilon=3\sigma$ for pointwise shrinkage estimators.
\subsection {Performance of Subband MPE Shrinkage}
\begin{figure*}[t]
\centering
$\begin{array}{ccc}
\text{Input SNR}=4.99 \text{\,dB} & \text{Output SNR}=15.19 \text{\,dB} & \text{Output SNR}=17.12 \text{\,dB}\\
\includegraphics[width=5.6cm]{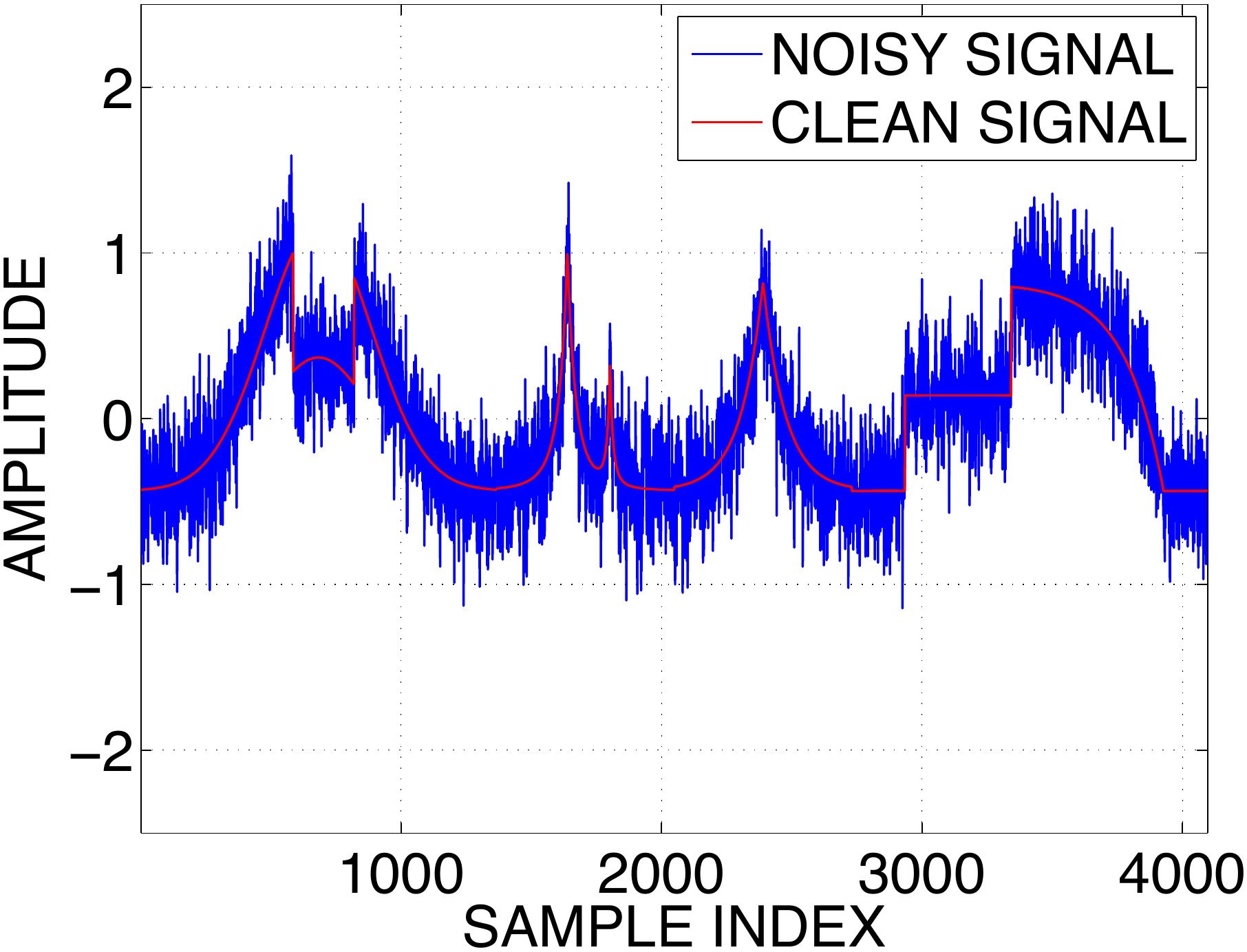}&
\includegraphics[width=5.6cm]{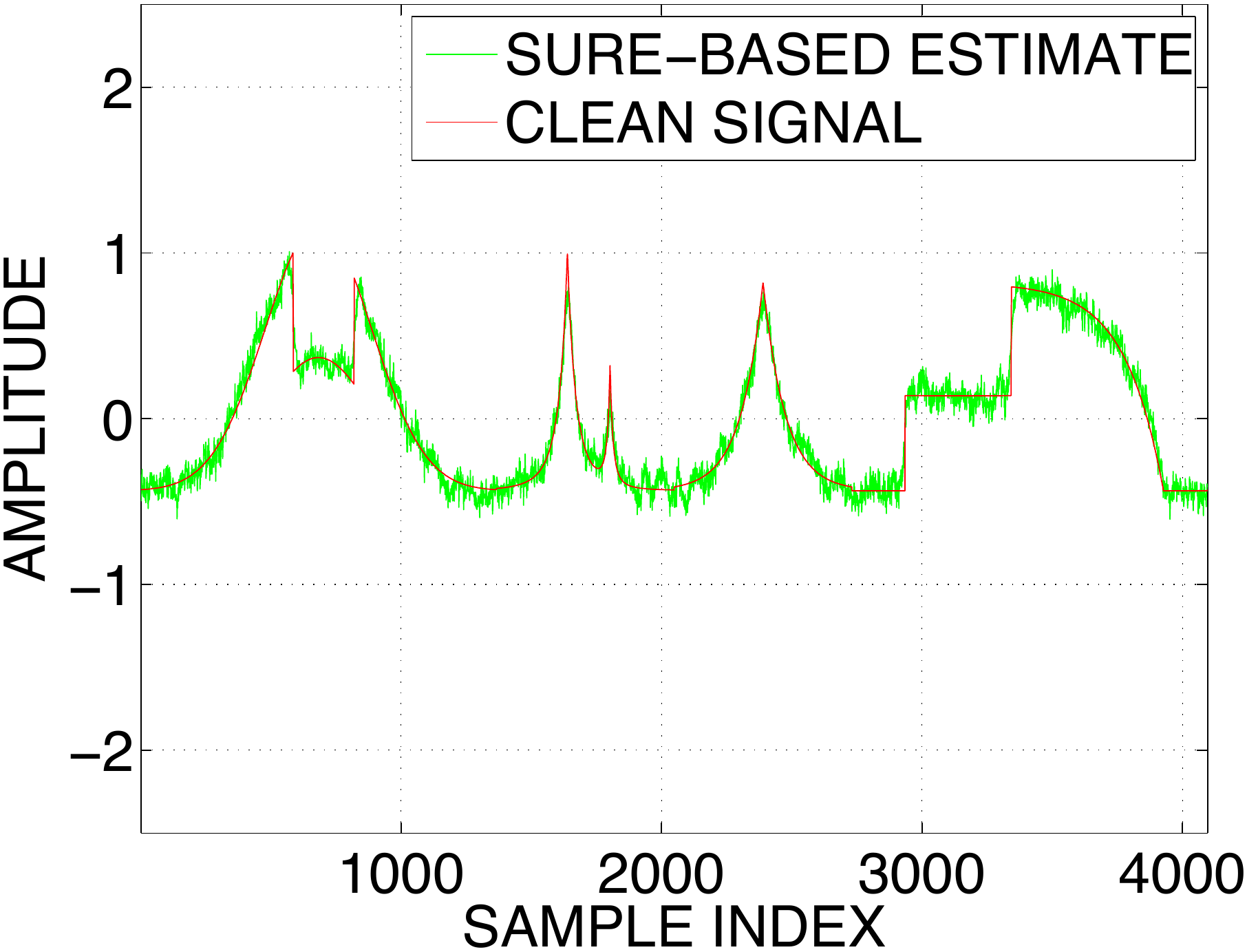}&
\includegraphics[width=5.6cm]{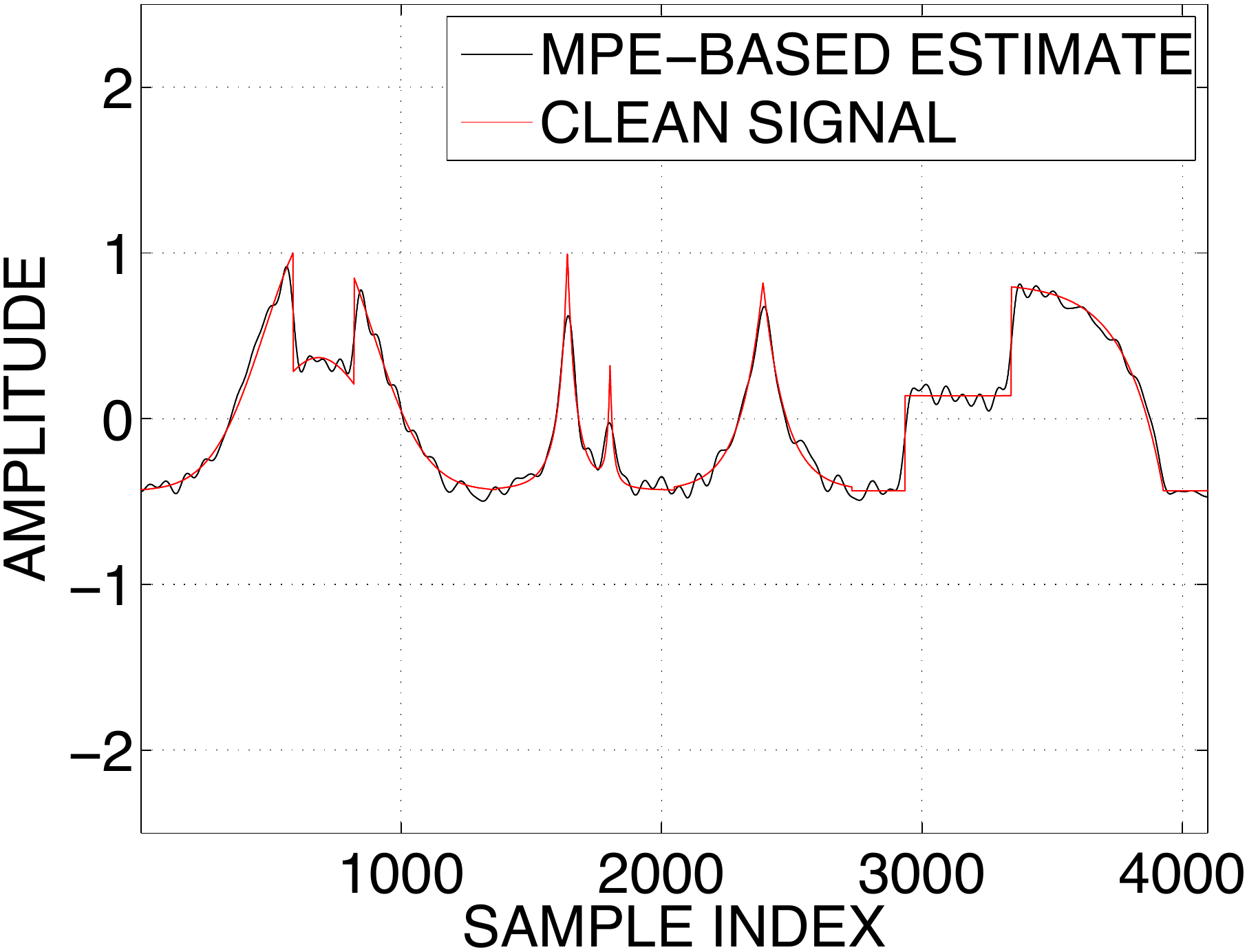}
\end{array}$
\caption{\small (Color online) Comparison of denoising performance of the subband shrinkage estimators using MPE and SURE, for the \textit{Piece-Regular} signal corrupted by additive Gaussian noise. The subband size is taken as $k=16$ and the value of $\epsilon$ is $1.75\sqrt{k}\sigma$.}
\label{MPE_vect_piece_regular}
\end{figure*}

\begin{figure}[t]
\centering
$\begin{array}{c}
\includegraphics[width=5.6cm]{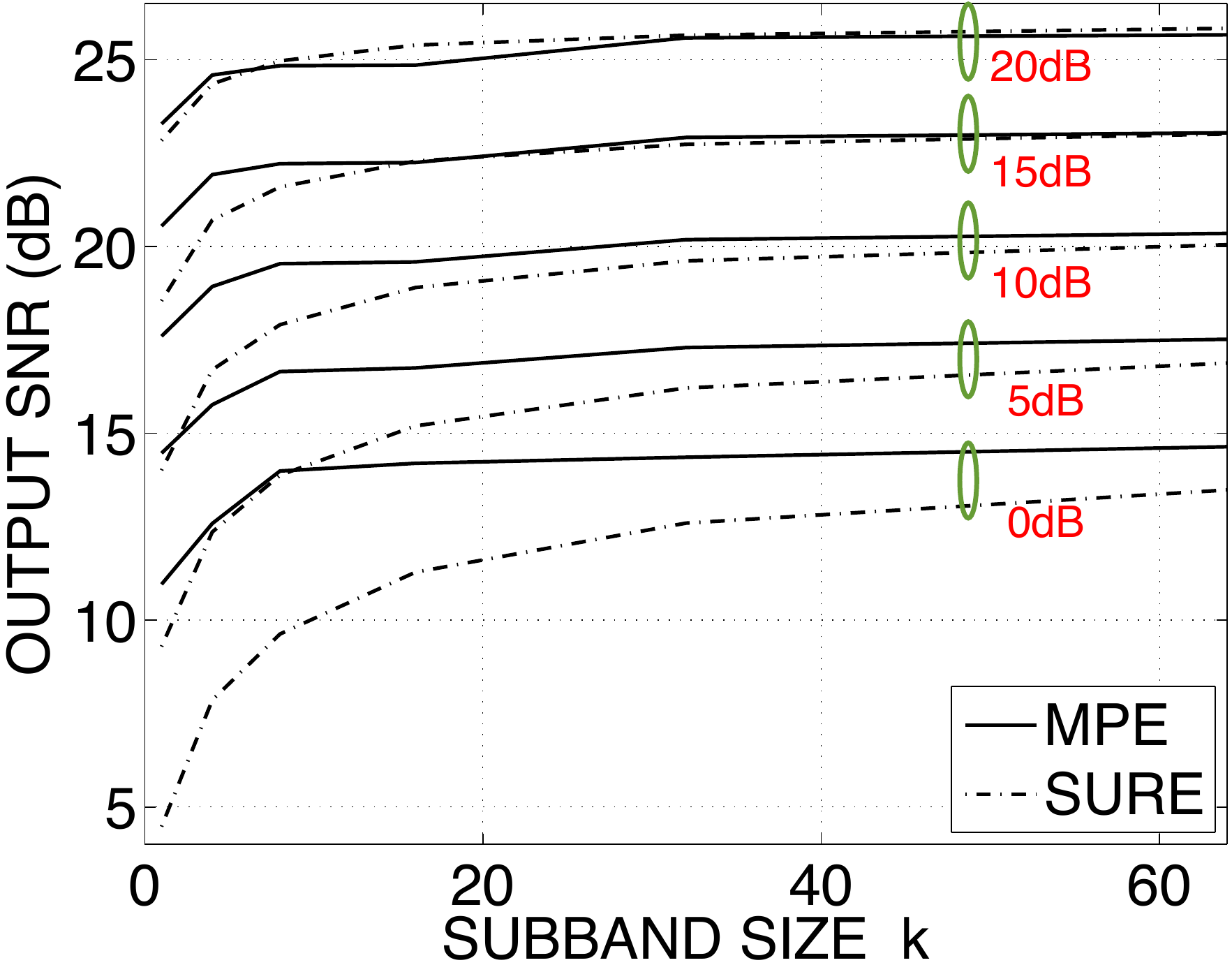}\\
\end{array}$
\caption{\small (Color online) Output SNR versus subband size $k$, averaged over $100$ noise realizations, for different input SNRs. The output SNR of MPE is consistently superior to that obtained using SURE, especially when $k\leq 40$ and the input SNR is below $15$ dB.}
\label{MPE_vectror_diffsnr_diflength}
\end{figure}

\indent To validate the performance of the MPE-based subband shrinkage estimator (cf. Section \ref{mpe_vector_shrink_define}), we consider denoising of  the \textit{Piece-Regular} signal in additive Gaussian noise. The clean signal and its noisy measurement are shown in Figure~\ref{MPE_vect_piece_regular}(a). Denoising is carried out by grouping $k$ adjacent DCT coefficients to form a subband. The denoised signals obtained using SURE and MPE are shown in Figures~\ref{MPE_vect_piece_regular}(b) and \ref{MPE_vect_piece_regular}(c), respectively.  The subband size $k$ is chosen to be $16$ and the parameter $\epsilon$ is set equal to $1.75\sqrt{k}\sigma$, a value that was determined experimentally and found to be nearly optimal. We observe that the MPE gives 1 dB improvement in SNR than the SURE approach.\\
\indent Variation of the output SNR is also studied as a function of $k$ (cf. Figure~\ref{MPE_vectror_diffsnr_diflength}). We experimented with  $\epsilon=3\sigma$, $\epsilon=1.75\sqrt{k}\sigma$, and $\epsilon=1.25\sqrt{k}\sigma$ corresponding to subband sizes $k=1$, $k\in[2,16]$, and $k>16$, respectively. For both SURE and MPE, as $k$ increases, the output SNR also increases and eventually saturates for $k \geq 40$. For input SNR below $15$ dB, MPE gives a comparatively higher SNR than SURE, and the margin diminishes with increase in input SNR or the subband size $k$. The degradation in performance of SURE for low SNRs is due to the large error in estimating the MSE at such SNRs. The SURE-based estimate of MSE becomes increasingly reliable as $k$ increases, thereby leading to superior performance. 
\section{Accumulated Probability of Error: MPE Meets  the Expected $\ell_1$ Distortion}
\label{sec_integrated_error}
\begin{figure}[t]
\centering
$\begin{array}{ccc}
\includegraphics[width=1.1in]{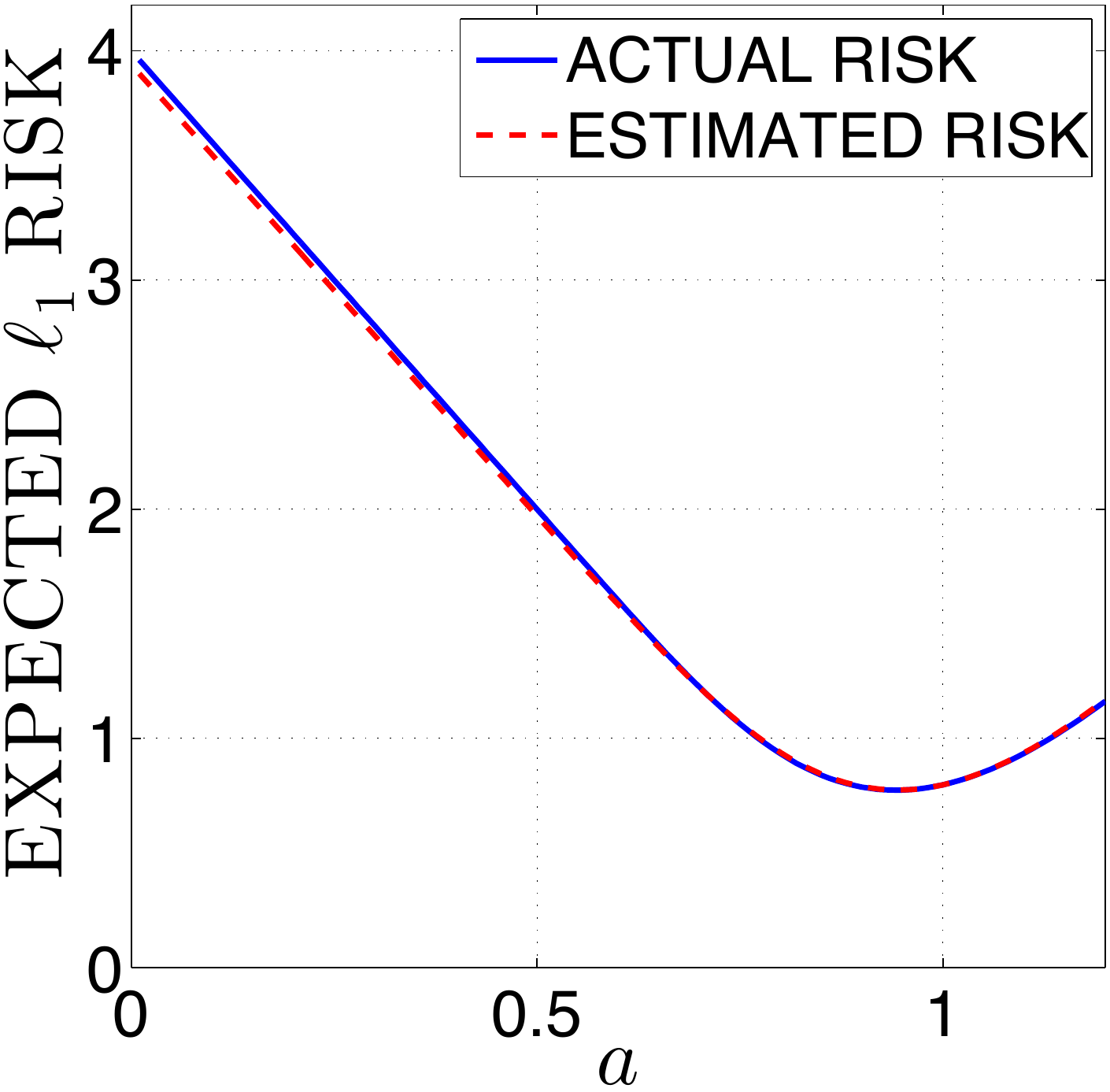}&
\includegraphics[width=1.1in]{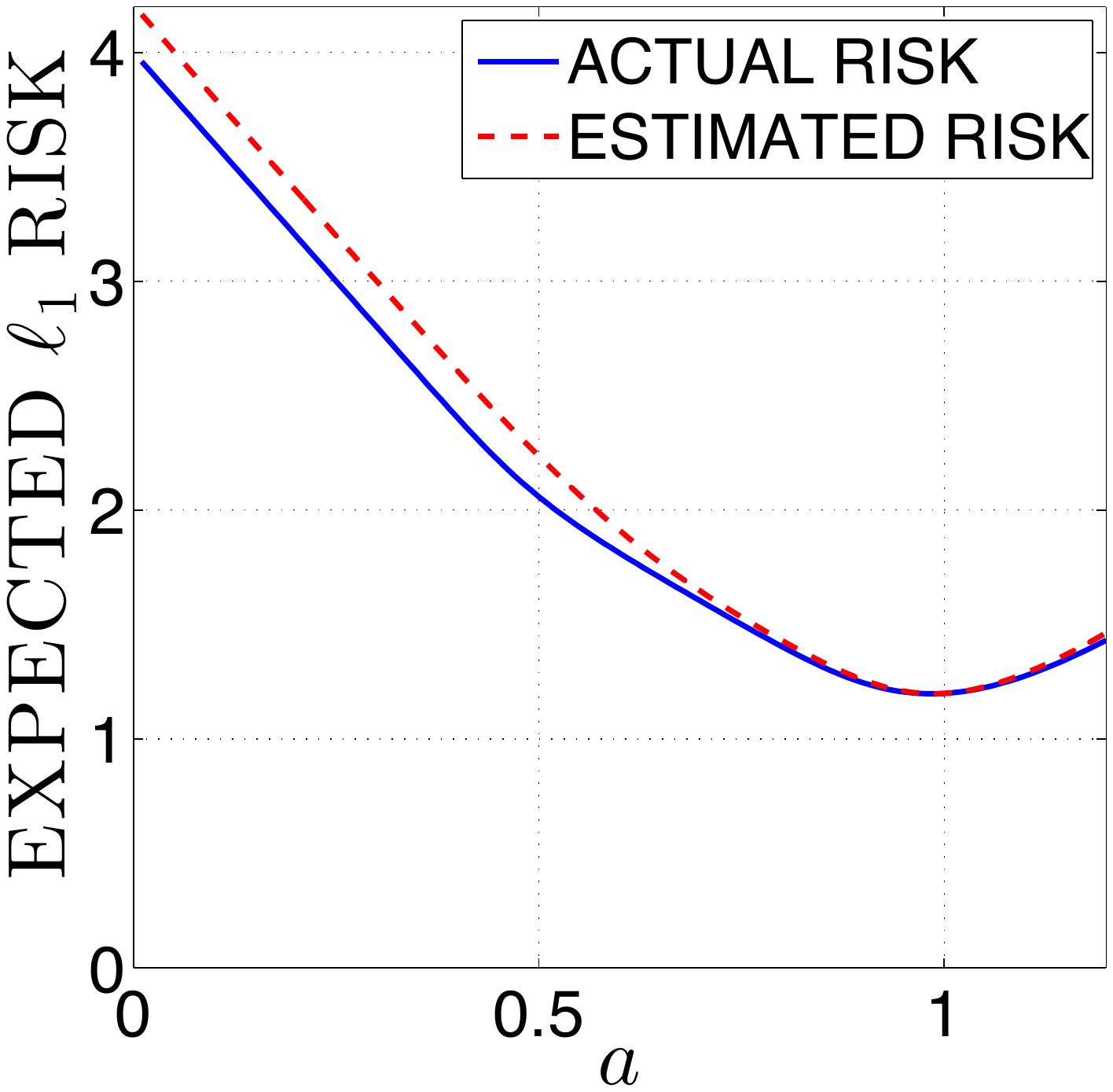}&
\includegraphics[width=1.1in]{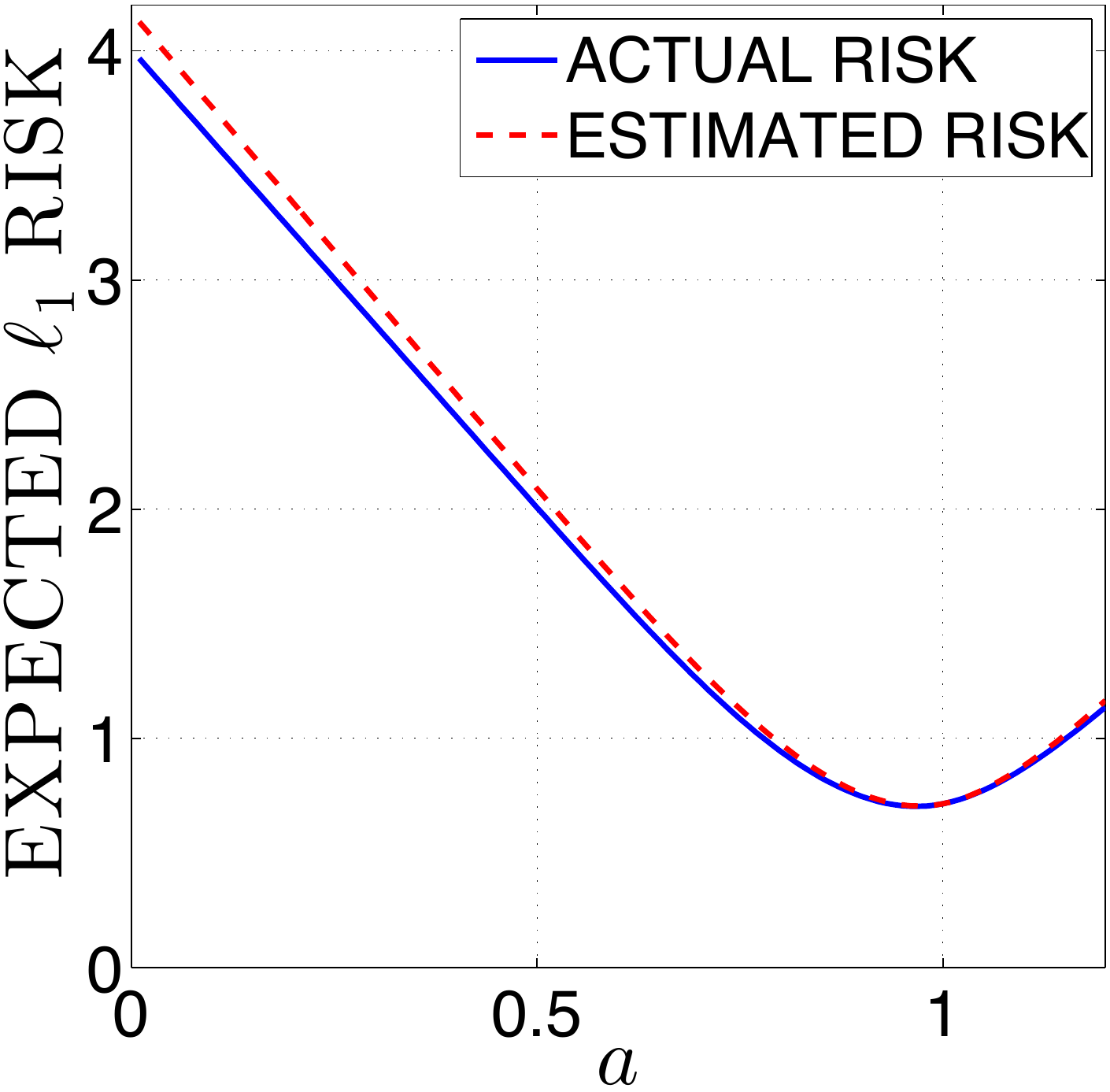}\\
\text{(a)} & \text{(b)} & \text{(c)}
\end{array}$
\caption{\small  (Color online) The expected $\ell_1$-risk and its estimate versus $a$, averaged over 100 noise realizations for: (a)  Gaussian noise with $\sigma^2=1$; (b) three-component GMM; and (c) a four-component GMM approximation to the Laplacian distribution.}
\label{riskL1_fig}
\end{figure}

\indent The MPE is parametrized by $\epsilon$, which has to be appropriately chosen in order to achieve optimal denoising performance. To suppress the direct dependence on $\epsilon$, we consider the accumulated probability of error, namely $\int_{0}^{\infty}\mathbb{P}\left( \left| \widehat{s}-s \right|>\epsilon \right)\mathrm{d} \epsilon$ as the risk to be minimized.
For a nonnegative random variable $Y$, we know that $\mathcal{E} \{Y\}=\int_{0}^{\infty}\mathbb{P}\left( Y>\epsilon \right)\mathrm{d} \epsilon.$ Therefore, the accumulated probability of error is the expected $\ell_1$ distortion:
\begin{equation}
\mathcal{E}\{ \left| \widehat{s}-s \right|\}  =  \int_{0}^{\infty}\mathbb{P}\left( \left| \widehat{s}-s \right|>\epsilon \right)\mathrm{d} \epsilon.
\label{MPE_L1_eq}
\end{equation}
For Gaussian noise distribution,
\begin{eqnarray}
\mathcal{R}_{\ell_1}\left(a,s \right)=\mathcal{E}\{ \left| \widehat{s}-s \right|\} = \int_{0}^{\infty} Q \left(  \frac{\epsilon-(a-1)s}{a\sigma} \right)\mathrm{d} \epsilon \nonumber\\ +\int_{0}^{\infty} Q \left(  \frac{\epsilon+(a-1)s}{a\sigma} \right)\mathrm{d} \epsilon.
\label{MPE_L1_gauss_eq}
\end{eqnarray}
Denoting $u=\displaystyle \frac{\epsilon-(a-1)s}{a\sigma}$ and $\mu=-\displaystyle \frac{(a-1)s}{a\sigma}$, the first integral in \eqref{MPE_L1_gauss_eq} is evaluated as\\
$\displaystyle\int_{0}^{\infty}Q \left(\frac{\epsilon-(a-1)s}{a\sigma} \right)\mathrm{d} \epsilon=a\sigma \int_{\mu}^{\infty} Q \left(  u \right)\mathrm{d}u$
\begin{eqnarray}
&& = a\sigma \left(    \int_{0}^{\infty} Q \left(  u \right)\mathrm{d}u -  \int_{0}^{\mu} Q \left(  u \right)\mathrm{d}u   \right)\nonumber \\
&& = a\sigma \left(    \frac{1}{\sqrt{2\pi}}- \mu Q \left(  \mu \right)-  \frac{1}{\sqrt{2\pi}}\left( 1-e^{-\frac{\mu^2}{2}}  \right)    \right)\nonumber \\
&& = a\sigma \left( \frac{e^{-\frac{\mu^2}{2}} }{\sqrt{2\pi}} - \mu Q \left(  \mu \right) \right).
\label{evaluate_L1}
\end{eqnarray}
The second term in (\ref{MPE_L1_gauss_eq}) can be evaluated by replacing $\mu$ with $-\mu$ in (\ref{evaluate_L1}). Combining both integrals, we obtain the expression for the \textit{expected} $\ell_1$ \textit{distortion}:\\
$\mathcal{R}_{\ell_1}\left(a,s \right) = a\sigma \left[ \sqrt{\frac{2}{\pi}}e^{-\frac{\mu^2}{2}} - \mu Q \left( \mu \right) + \mu Q \left( - \mu \right) \right]$
\begin{eqnarray}
&&= a\sigma \left[ \sqrt{\frac{2}{\pi}}\exp\left(-\frac{(a-1)^2 s^2}{2a^2 \sigma^2}\right)+ 2\frac{(a-1)s}{a\sigma}\right.\nonumber\\
& & \left. Q \left( -\frac{(a-1)s}{a\sigma} \right)  -\frac{(a-1)s}{a\sigma}  \right].
\label{L1_final_expression}
\end{eqnarray}
An estimate of the  expected $\ell_1$ distortion is calculated by replacing $s$ with an estimate $\tilde{s}$, which could also be $x$, to begin with. In Figure~\ref{riskL1_fig}(a), we show the variation of the original $\ell_1$ distortion and its estimate obtained by setting $\tilde{s}=x$, as functions of $a$, averaged over $100$ independent realizations of $\mathcal{N}\left(0,1\right)$ noise. The actual parameter value is $s=4$. The figure shows that the minimum of the expected $\ell_1$ risk is close to that of its estimate.
In principle, one could iteratively minimize the $\ell_1$ distortion by starting with $\widehat{s}=x$ and successively refining it. Such an approach is given in Algorithm $1$. An illustration of the denoising performance of the iterative algorithm is deferred to Section~\ref{L1_sec_results}.
\begin{algorithm}[t]
\caption {Iterated minimization of the expected $\ell_1$ distortion.}
\begin{algorithmic}
\STATE {\bf  1.} {\bf Initialization:} Set $j \leftarrow 1$, $\widehat{s}^{(j)}\leftarrow x$, and $N_{\text{iter}}$ = Maximum iteration count.
\STATE {\bf  2.} {\bf Iterate until $j$ exceeds $N_{\text{iter}}$:} 
\begin{itemize}
\item Find $a_{\text{opt}}^{(j)} = \text{arg\,} \underset{0 \leq a \leq 1}{\min\,} \mathcal{R}_{\ell_1}\left(a,\widehat{s}^{(j)} \right)$ by grid-search.
\item $j \leftarrow j+1.$
\item Compute $\widehat{s}^{(j)} = a_{\text{opt}}^{(j-1)}  x.$
\end{itemize}
\STATE {\bf  3.} {\bf Output:} Denoised estimate $\widehat{s}^{(j)}$.
\end{algorithmic}
\label{Algorithm1}
\end{algorithm}
\subsection{Expected $\ell_1$ risk Using GMM Approximation}
\label{GMM_l1_risk}
For the GMM p.d.f. in (\ref{GMM_pdf_eqn}), the expected $\ell_1$ distortion evaluates to (cf. Appendix~ \ref{expected_L1_derivation} for the derivation)
\begin{eqnarray}
\mathcal{R}_{\ell_1}=\sum_{m=1}^{M} a \alpha_m  \sigma_{m} \left( \sqrt{\frac{2}{\pi}}e^{-\frac{\mu_{m}^2}{2}} -2 \mu_{m} Q \left(  \mu_{m} \right) + \mu_{m}  \right),
\label{L1_final_gmm_expression}
\end{eqnarray}
where $\mu_m=-\displaystyle \frac{(a-1)s+\theta_{m}}{a\sigma_{m}}$. The expected $\ell_1$ risk and its estimate for a multimodal (cf. Figure~\ref{risk_gmm_fig}(b)) and Laplacian noise p.d.f.s are shown in Figures \ref{riskL1_fig}(b) and \ref{riskL1_fig}(c), respectively. We observe that,  in both cases, the locations of the minima of the true risk and its estimate are in good agreement.
\subsection{ Optimum Shrinkage $a_{\text{opt}}$ Versus Posterior SNR}
\label{sec_aopt}
\begin{figure}[t]
\centering
$\begin{array}{cc}
\includegraphics[width=1.75in]{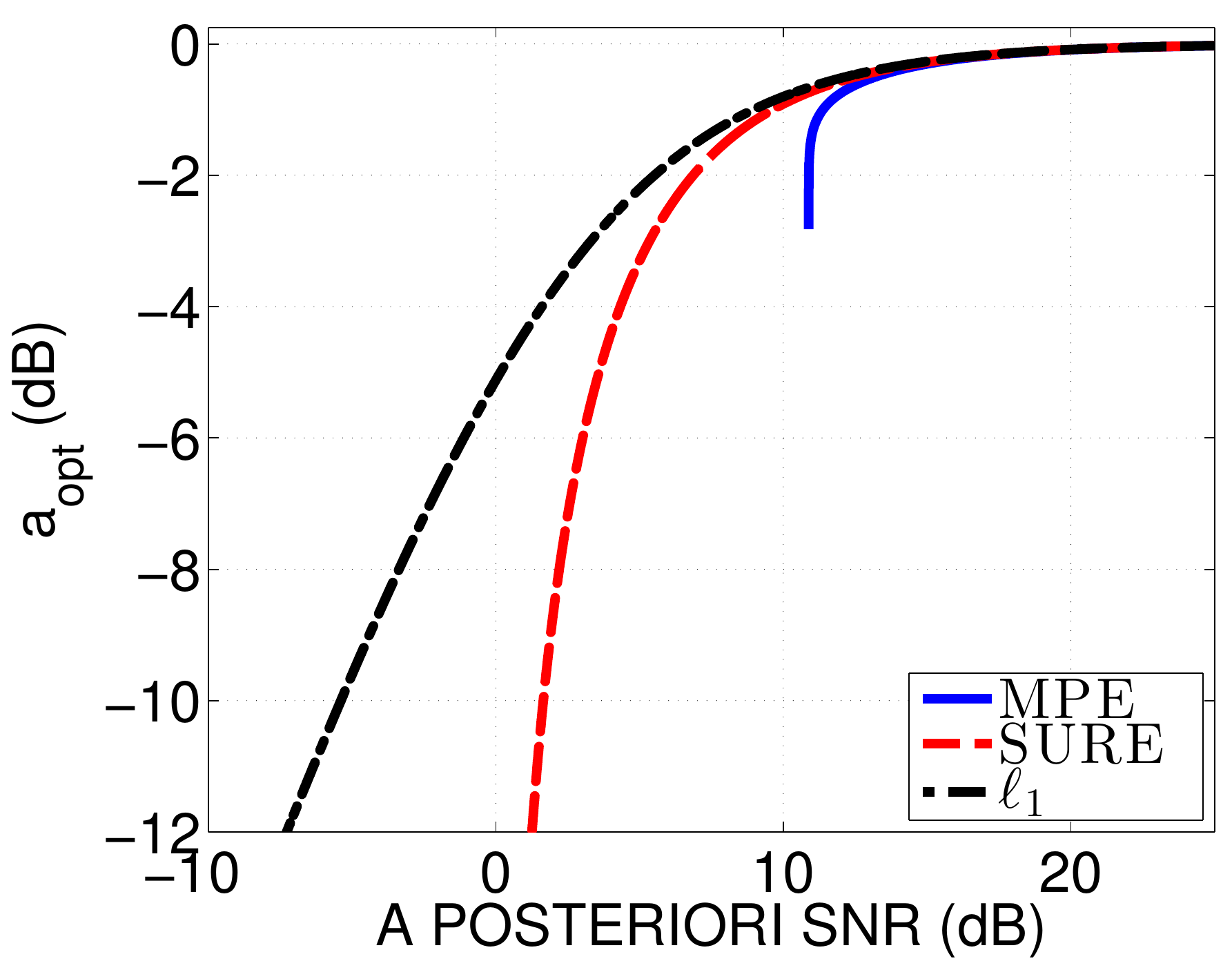}&
\includegraphics[width=1.75in]{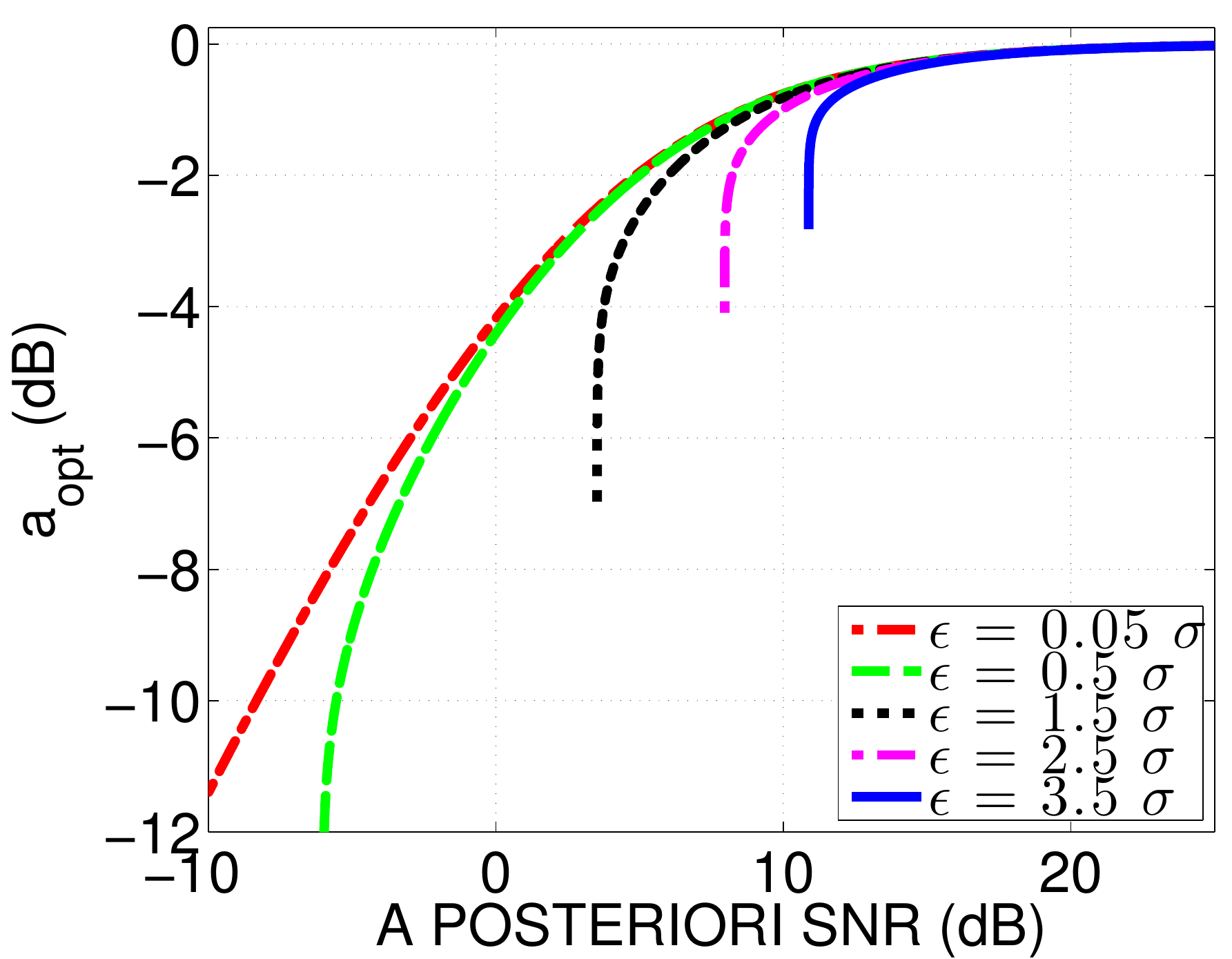}\\
\text{(a)} & \text{(b)}
\end{array}$
\caption{\small  (Color online) Shrinkage parameter profiles as a function of \textit{a posteriori} SNR, corresponding to different risk functions: (a) MPE, SURE, expected $\ell_1$ distortion; and (b) MPE for different values of $\epsilon$. The shrinkage factor $a_{\text{opt}}$ is plotted on a log scale mainly to highlight the fine differences among various attenuation profiles.}
\label{gain_apsnr}
\end{figure}

We next study the behavior of $a_{\text{opt}}$ for different input SNRs to compare the denoising capabilities of the MPE and the expected $\ell_1$-distortion-based shrinkage estimators.  The optimum pointwise shrinkage parameter $a_{\text{opt}}$  for Gaussian noise statistics, obtained by minimizing SURE, MPE risk estimate, and the estimated $\ell_{1}$ risk, for different values of the \textit{a posteriori} SNR  $\displaystyle\frac{x^2}{\sigma^2}$ is plotted in Figure~\ref{gain_apsnr}(a). To illustrate the effect of $\epsilon$, the variation of $a_{\text{opt}}$ versus \textit{a posteriori} SNR for MPE corresponding to Gaussian noise is shown in Figure~\ref{gain_apsnr}(b), for different $\epsilon$. We observe that the shrinkage profiles are characteristic of a reasonable denoising algorithm, as Figures~\ref{gain_apsnr}(a) and \ref{gain_apsnr}(b) exhibit that  the  shrinkage parameters increase as the \textit{a posteriori} SNR increases. Whereas in case of the MPE, the choice of $\epsilon$ is crucial, the expected $\ell_1$ distortion does not require tuning such a parameter. Moreover, the MPE attenuation profile for larger values of $\epsilon$ is reminiscent of a hard-thresholding function, whereas the expected $\ell_1$ distortion has an attenuation profile that resembles a soft-threshold.
\section{Performance of the Expected $\ell_1$ Distortion-Based Pointwise Shrinkage Estimator} \label{L1_sec_results}
\begin{figure}[t]
\centering
$\begin{array}{cc}
\hspace{-0.27cm}\includegraphics[width=1.847in]{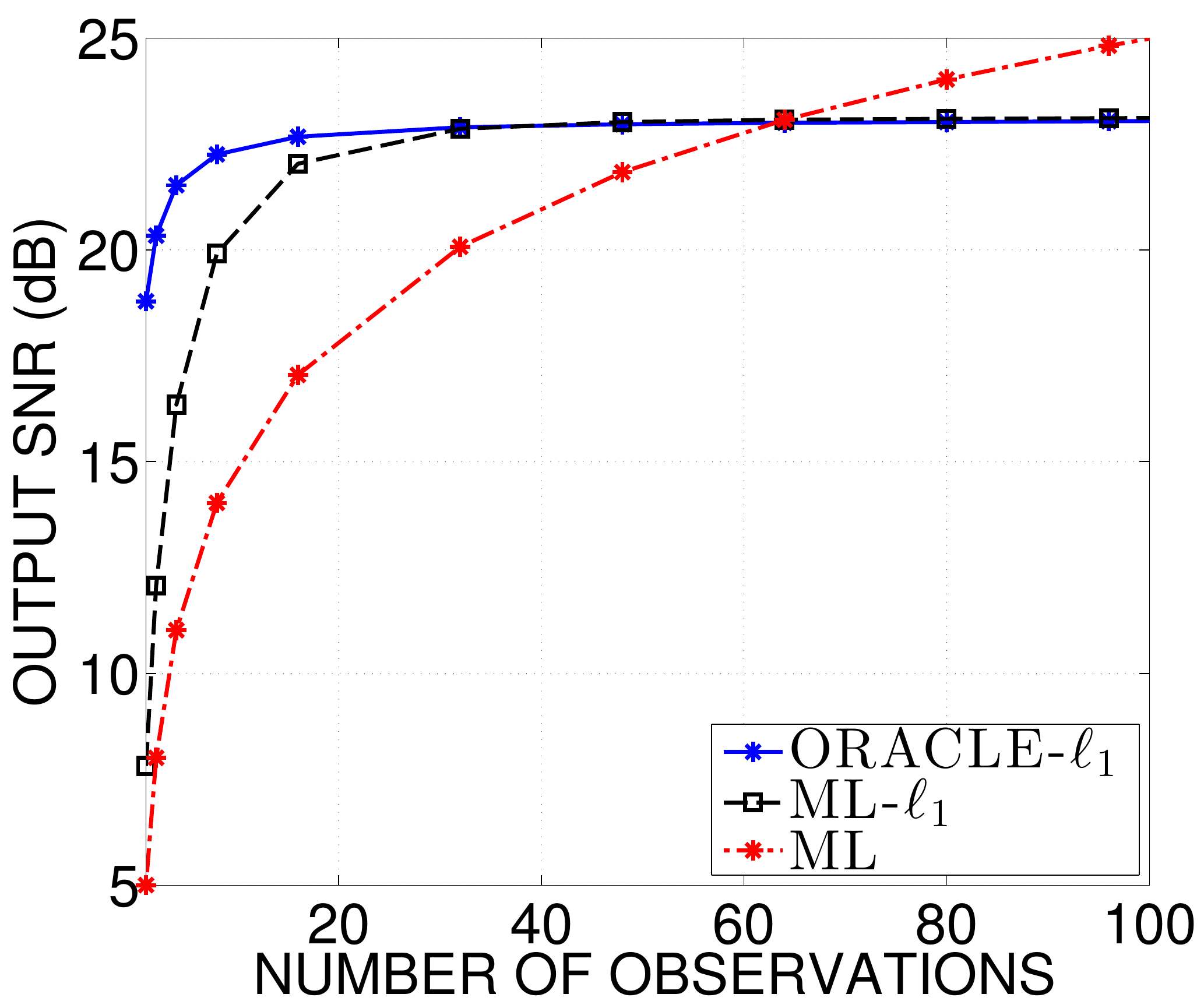}&\hspace{-0.26cm}
\includegraphics[width=1.84in]{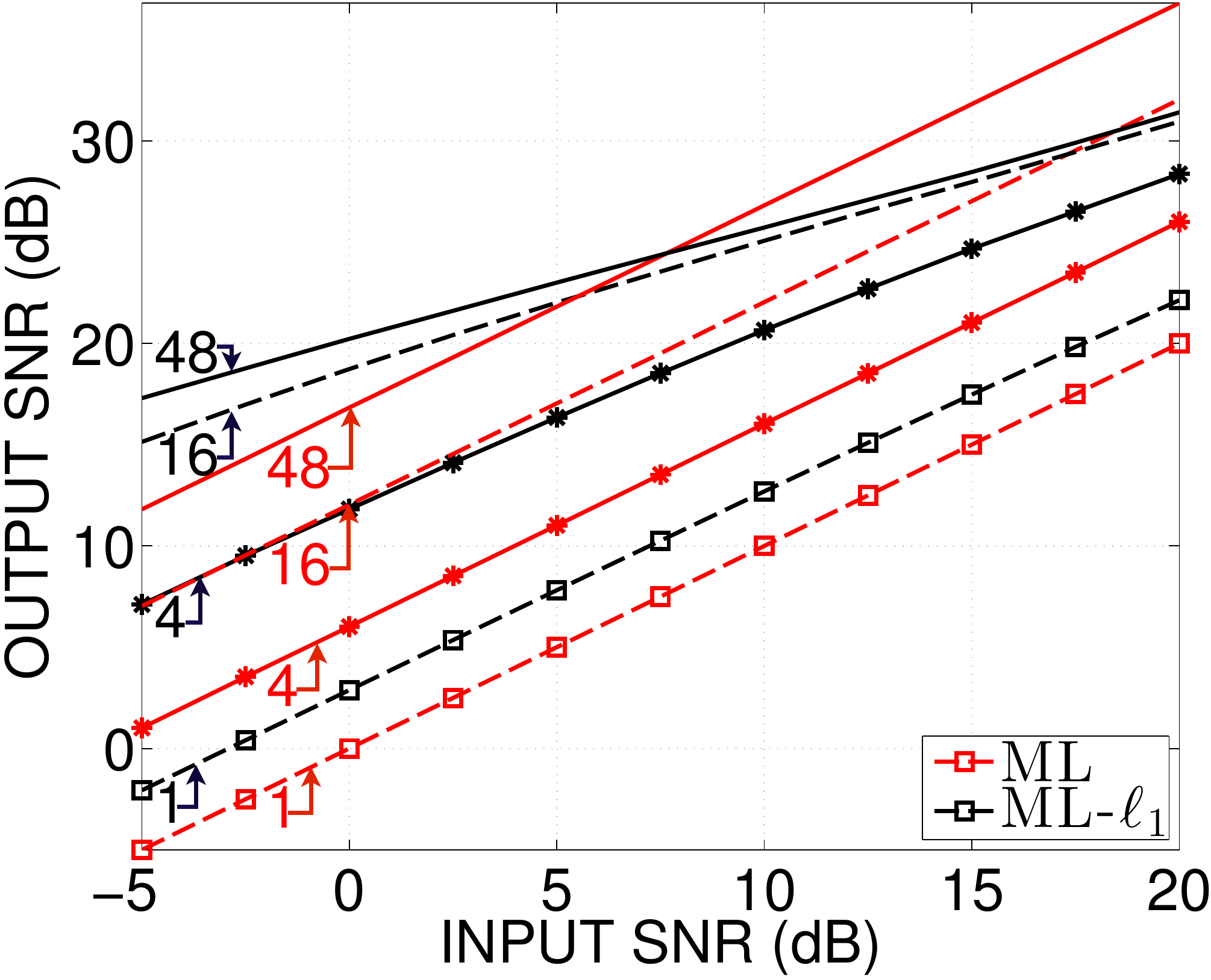}\\
\text{(a)} & \text{(b)}
\end{array}$
\caption{\small (Color online)  Comparison of denoising performance for different number of observations, for the \textit{Piece-Regular} signal in Gaussian noise: (a) variation of output SNR  for different number of observations, corresponding to input SNR $5$ dB; and (b) variation of output SNR with respect to the number of observations $M$ and the input SNR. The numerical values on the curves indicate the corresponding values of $M$. In both (a) and (b), the results are averaged over $100$ independent noise realizations.}
\label{dif_snr_obs_mll1_and_ML}
\end{figure}

In a practical denoising application, we have only one noisy realization from which the clean signal has to be estimated. However, it is instructive to consider the case of multiple realizations as it throws some light on the performance comparisons vis-\`a-vis other estimators such as the ML estimator. Consider the observation model $\bold x^{(m)}=\bold s+ \bold w^{(m)}$ in $\mathbb{R}^n$,  $1\leq m \leq M$, where one has access to $M$ noisy copies of the signal $\bold s$, and the noise vectors $\bold w^{(m)}$ are drawn independently from the $\mathcal{N}\left(\bold 0,\sigma^2 I_n\right)$ distribution. The ML estimator of the $i^{\text{th}}$ signal coefficient $s_i$ is given by $\hat{s}_{\text{ML},i}=\frac{1}{M} \sum_{m=1}^{M} x_i^{(m)}$, where  $x_i^{(m)}$ is the $i^{\text{th}}$ component of $\bold x^{(m)}$. Dropping the subscript $i$, as each  coefficient is treated independently of the others, the shrinkage estimator takes the form $\widehat{s}=a_{\text{opt}} \hat{s}_{\text{ML}}$. To study the behavior of the estimate with respect to $M$, we consider two variants: (i) where $a_{\text{opt}}$ is obtained by minimizing $\mathcal{R}_{\ell_1}\left(s,a\right)$, referred to as the oracle-$\ell_1$; and (ii) where $a_{\text{opt}}$ is chosen to minimize $\mathcal{R}_{\ell_1}\left(\hat{s}_{\text{ML}},a\right)$, referred to as ML-$\ell_1$. The output SNR as a function of $M$ for the \textit{Piece-Regular} signal, corresponding to an input SNR of $5$ dB, is shown in Figure \ref{dif_snr_obs_mll1_and_ML}(a). For all three estimators, namely, oracle-$\ell_1$, ML-$\ell_1$, and the ML estimate, the output SNR increases with $M$. However, for the oracle-$\ell_1$ and the ML-$\ell_1$ estimators, the output SNR stagnates as $M$ increases beyond $40$. For $M\leq 60$, the oracle-$\ell_1$ and the ML-$\ell_1$ shrinkage estimators exhibit  better performance  compared with the  ML estimator. As one would expect, the performance of the ML-$\ell_1$ estimator matches with that obtained using the oracle-$\ell_1$ as $M$ becomes large, because the ML estimate converges in probability to the true parameter. For $M=1$, which is often the case in practice, the ML-$\ell_1$ estimate significantly dominates the ML estimator as seen in Figure \ref{dif_snr_obs_mll1_and_ML}(a). The SNR gain over the ML estimator could be further improved by using the iterative minimization algorithm introduced  in Section \ref{sec_integrated_error} (cf. Algorithm \ref{Algorithm1}). The performance of the ML-$\ell_1$ and the ML estimators, for different values of $M$ and input SNR is shown in Figure~\ref{dif_snr_obs_mll1_and_ML}(b). The figures show that for small values of SNR and $M$, the ML-$\ell_1$ estimate outperforms the ML estimator. This is of significant importance in a practical setting where we have only one noisy realization ($M = 1$).
\subsection{Iterative Minimization of the Expected $\ell_1$-Risk} 
\label{Iterative_versus_Noniterative}
\begin{figure}[t]
\centering
$\begin{array}{cc}
\includegraphics[width=1.75in]{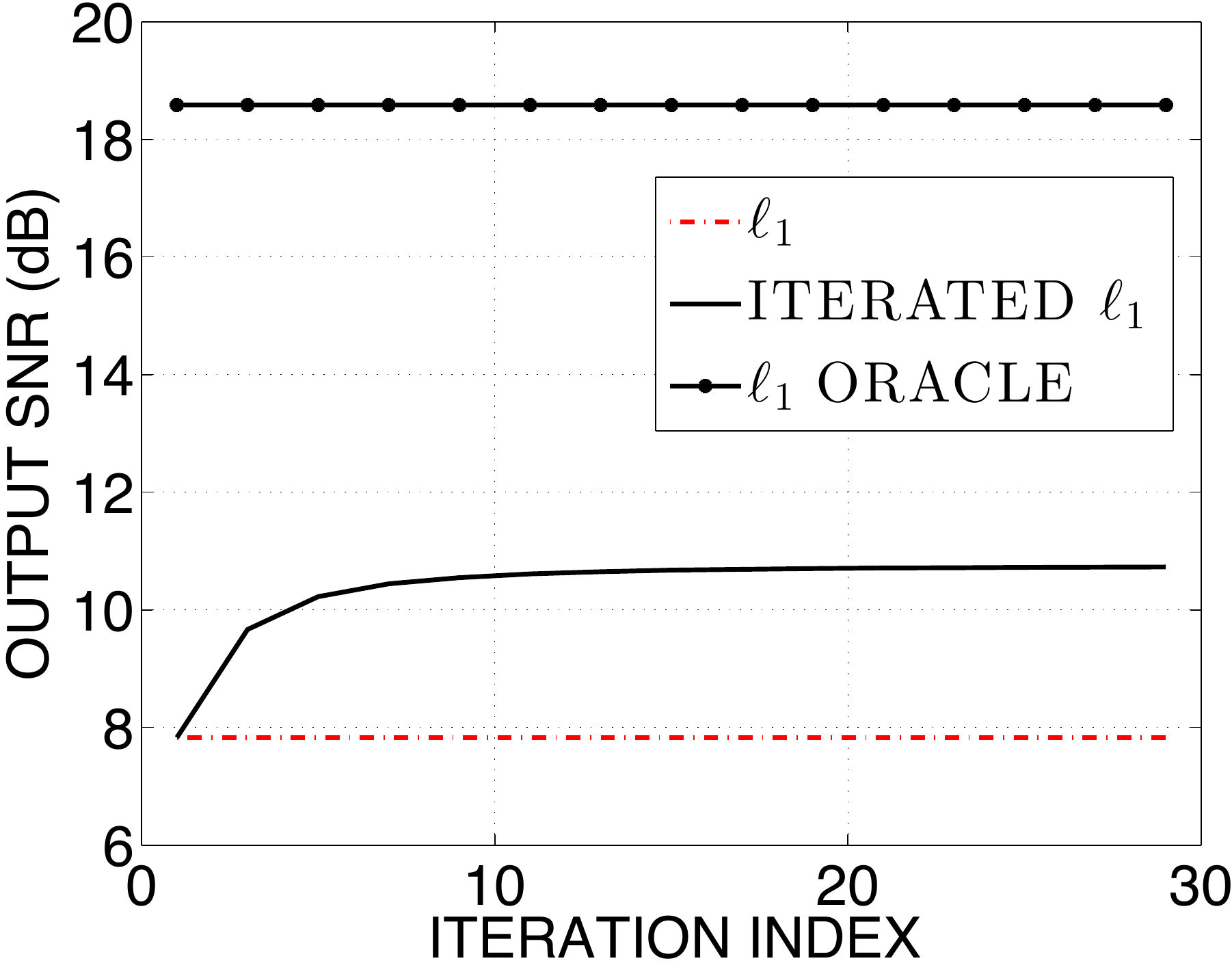}&
\includegraphics[width=1.75in]{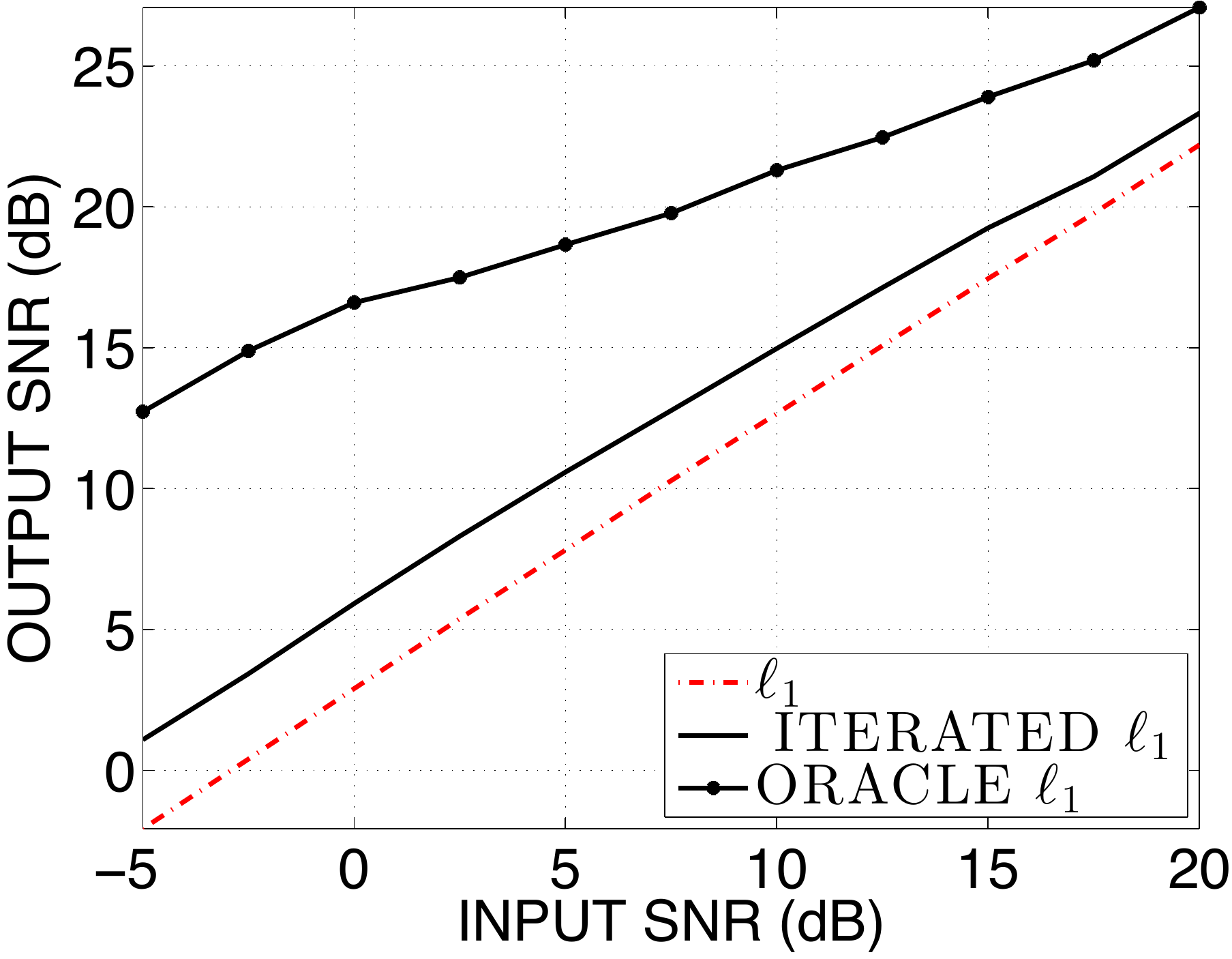}\\
\text{(a)} & \text{(b)}
\end{array}$
\caption{\small  (Color online)  Performance of $\ell_1$ risk minimization-based pointwise shrinkage estimator: (a) Variation of output SNR  versus iterations, where the signal considered is \textit{Piece-Regular} and noise is Gaussian with input SNR  $5$ dB; and (b) Variation of output SNR  versus input SNR (averaged over $100$ independent noise realizations), where the number of iterations in Algorithm $1$ is fixed at $N_{\text{iter}}=20$.}
\label{PR_in_out_snr_L1_iter}
\end{figure}

\begin{figure}[t]
\centering
$\begin{array}{cc}
\includegraphics[width=1.75in]{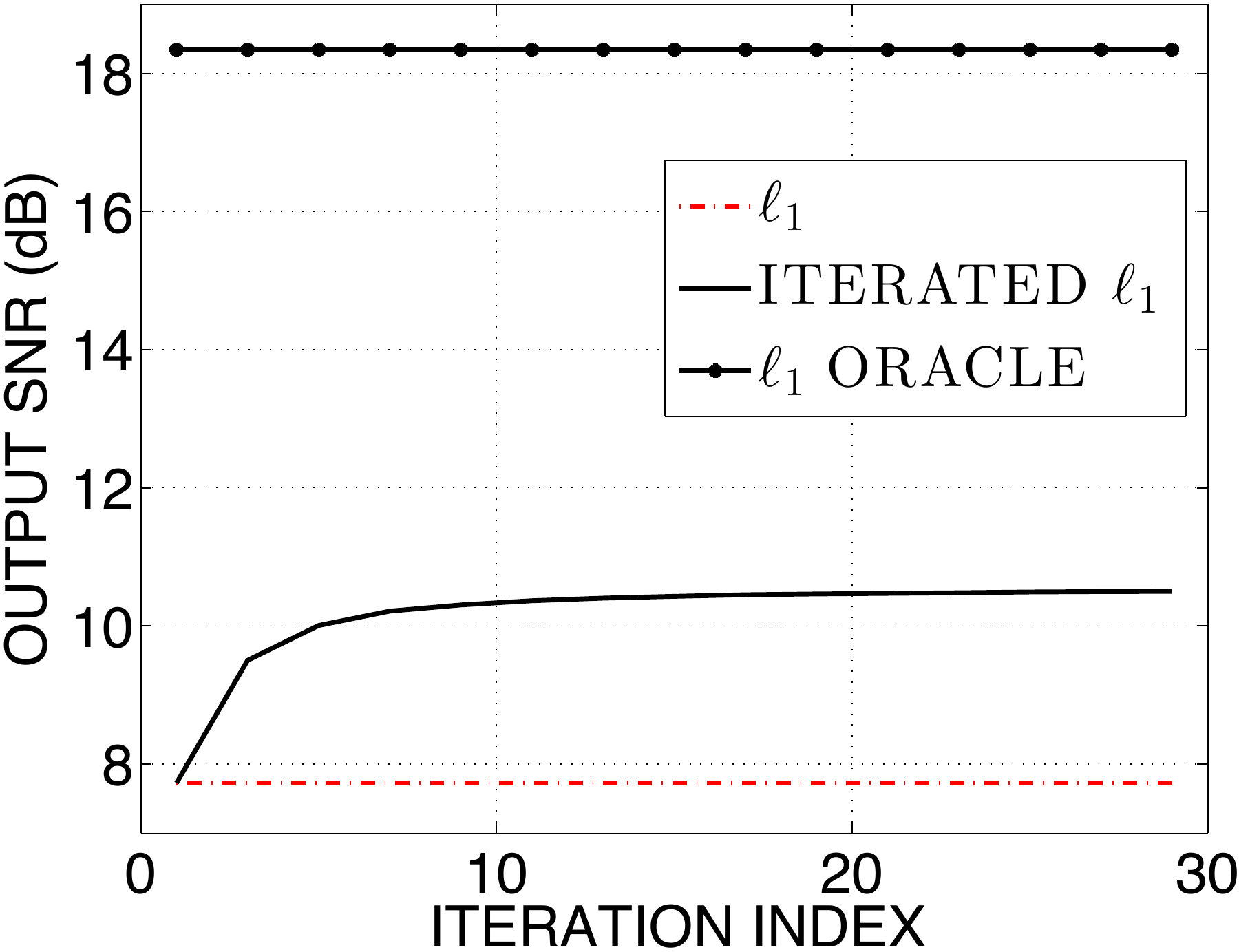}&
\includegraphics[width=1.75in]{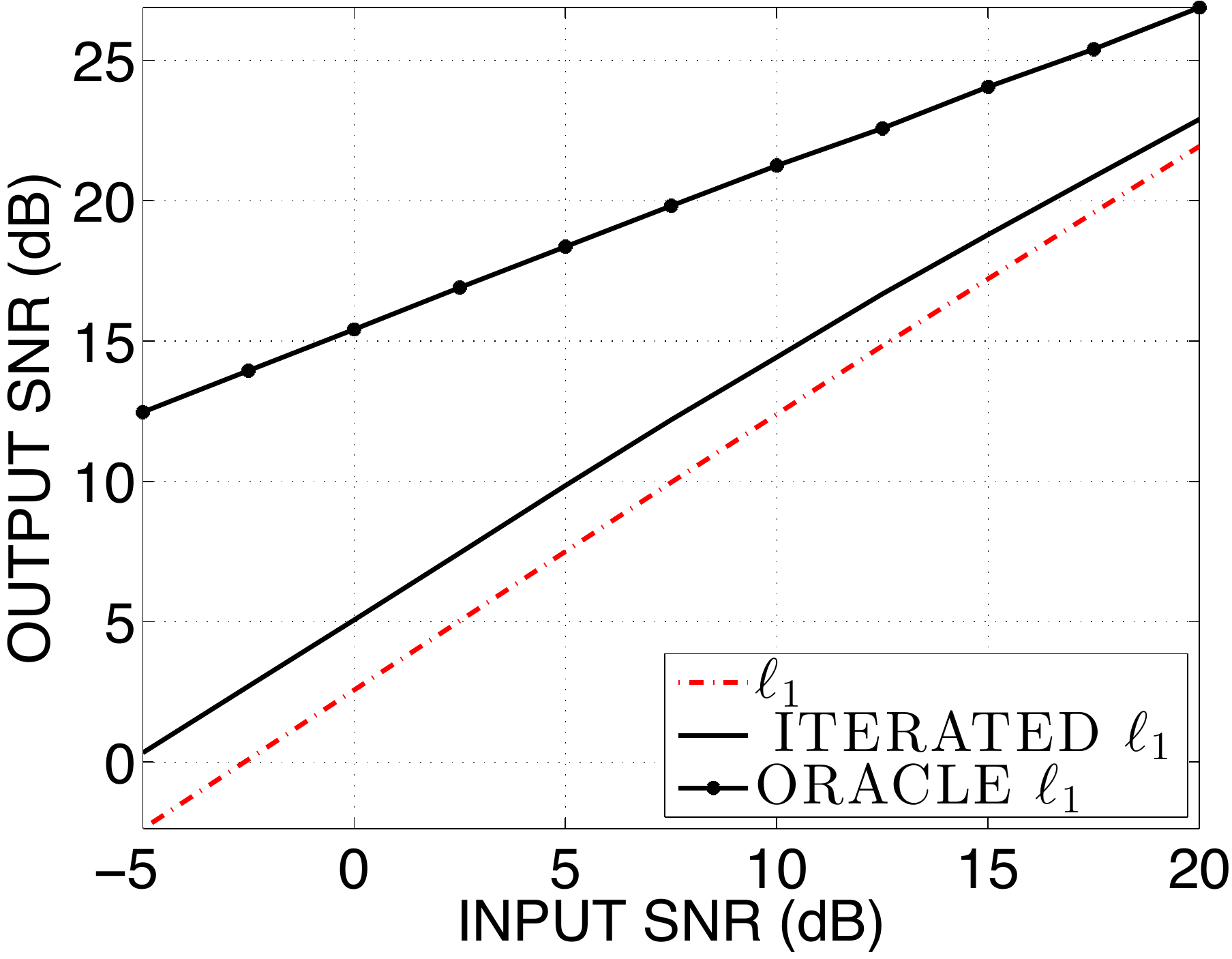}\\
\text{(a)} & \text{(b)}
\end{array}$
\caption{\small  (Color online) Performance of pointwise shrinkage estimator based on $\ell_1$ risk minimization: (a) Variation of output SNR versus iterations, corresponding to the \textit{Piece-Regular} signal contaminated by noise whose p.d.f. is given in Figure~\ref{risk_gmm_fig}(b). The input SNR is $5$ dB. (b) Output SNR versus input SNR (averaged over $100$ independent noise realizations), where the number of iterations in Algorithm $1$ is taken as $N_{\text{iter}}=20$.}
\label{PR_in_out_snr_L1_gmm_iter}
\end{figure}

\begin{figure}[t]
\centering
$\begin{array}{cc}
\includegraphics[width=1.75in]{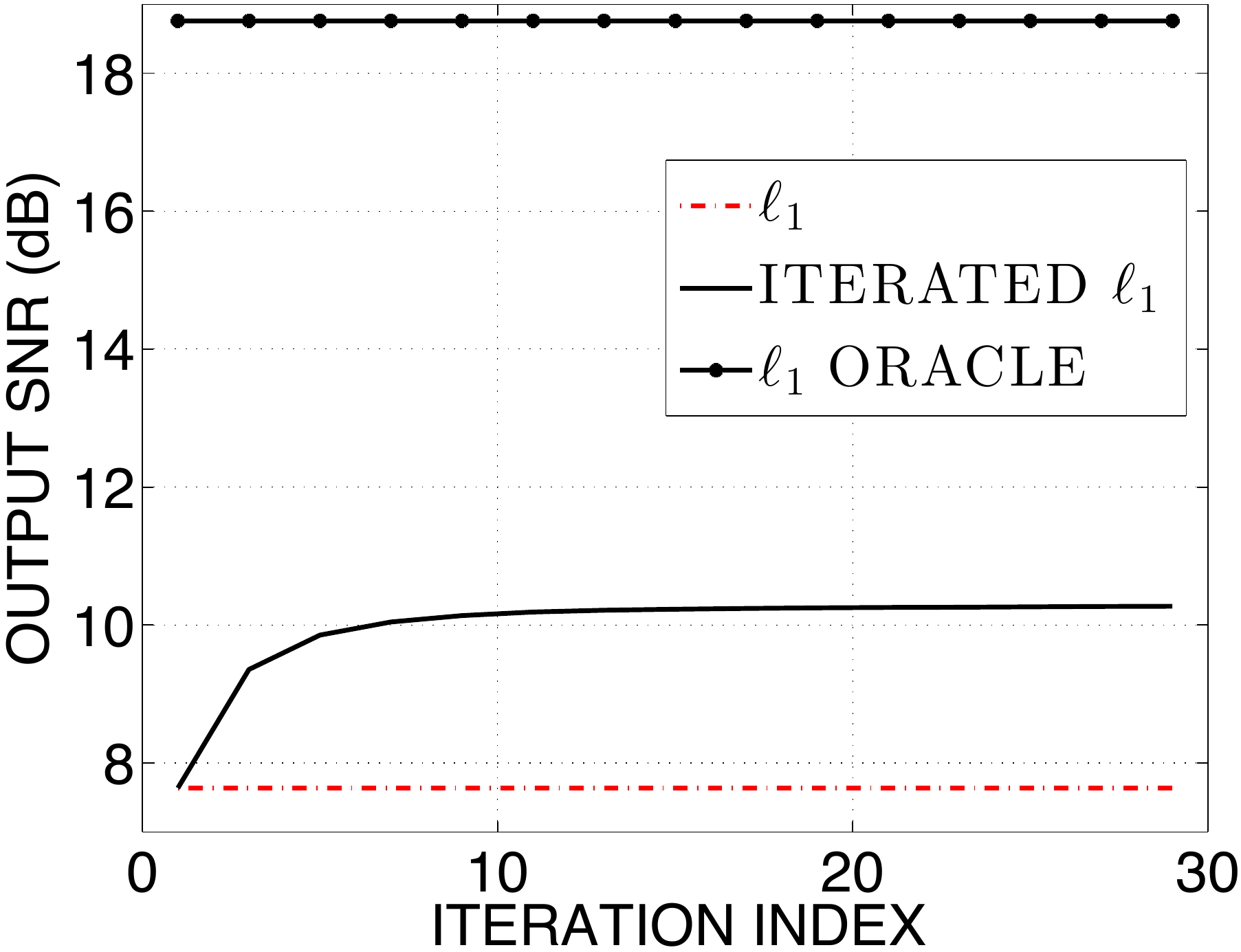}&
\includegraphics[width=1.75in]{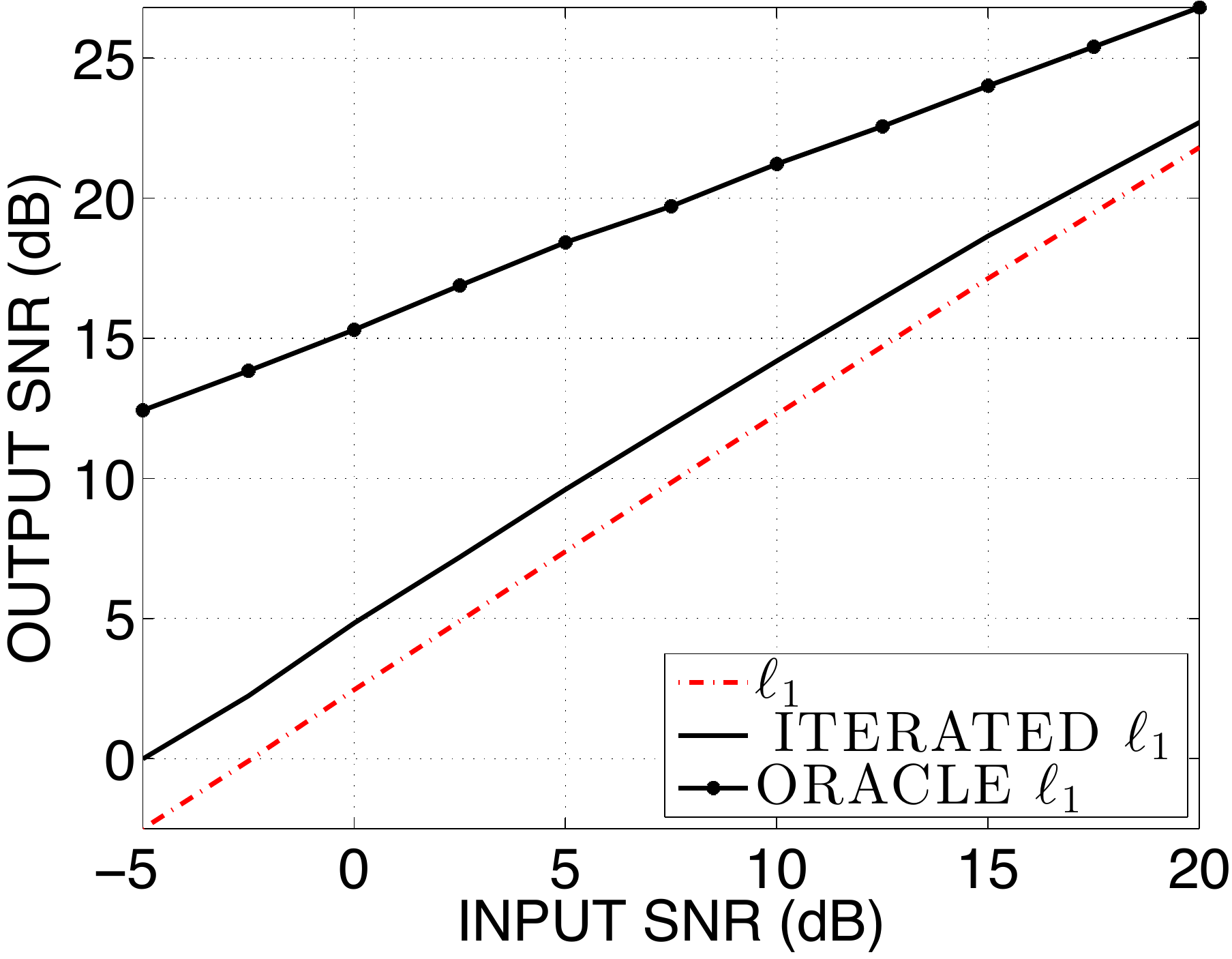}\\
\text{(a)} & \text{(b)}
\end{array}$
\caption{\small  (Color online) Performance of pointwise shrinkage estimator obtained by $\ell_1$ risk minimization: (a) Output SNR versus iterations for the \textit{Piece-Regular} signal in Laplacian noise with an input SNR of $5$ dB; and (b) Output SNR versus input SNR for $N_{\text{iter}}=20$. In (a) and (b), the Laplacian distribution is modeled using a four-component GMM to calculate the $\ell_{1}$-risk estimate. The results in (b) are averaged over $100$ realizations.}
\label{PR_in_out_snr_L1_laplacian_iter}
\end{figure}

\indent When $M=1$, the ML-$\ell_1$ estimator is obtained by minimizing $\mathcal{R}_{\ell_1}\left(x,a\right)$, where $x$ is the noisy version of $s$. We refer to this estimate as the non-iterative $\ell_1$-based shrinkage estimator. Following Algorithm 1, one could iteratively refine the estimate, starting from $x$. We compare the non-iterative $\ell_1$-based estimator with its iterative counterpart, and present the results in Figures \ref{PR_in_out_snr_L1_iter}, \ref{PR_in_out_snr_L1_gmm_iter}, and \ref{PR_in_out_snr_L1_laplacian_iter}, corresponding to Gaussian, multimodal (c.f. Figure~\ref{risk_gmm_fig}(b)), and a GMM approximation to the Laplacian noise, respectively. The output SNR obtained using the {\it oracle}-$\ell_1$ estimator, calculated by minimizing $\mathcal{R}_{\ell_1}\left(s,a\right)$, is also shown for benchmarking the performance.\\
\indent We make the following observations from the Figures \ref{PR_in_out_snr_L1_iter}, \ref{PR_in_out_snr_L1_gmm_iter}, and \ref{PR_in_out_snr_L1_laplacian_iter}: (i) the output SNR increases with iterations, albeit marginally after about 10 iterations; (ii) the iterative method consistently dominates the non-iterative one, with an overall SNR improvement of about $2$ to $3$ dB, for input SNR in the range $-5$ dB to $20$ dB; and (iii) the SNR gain of the iterative technique also reduces for higher input SNR, similar to other denoising algorithms.
\section{Performance Assessment of MPE and $\ell_1$-Risk Minimization Algorithms Versus State-Of-The-Art Denoising Algorithms}
\label{sec_State-Of-The-Art}
\begin{figure*}[t]
\centering
$\begin{array}{ccc}
\includegraphics[width=6 cm]{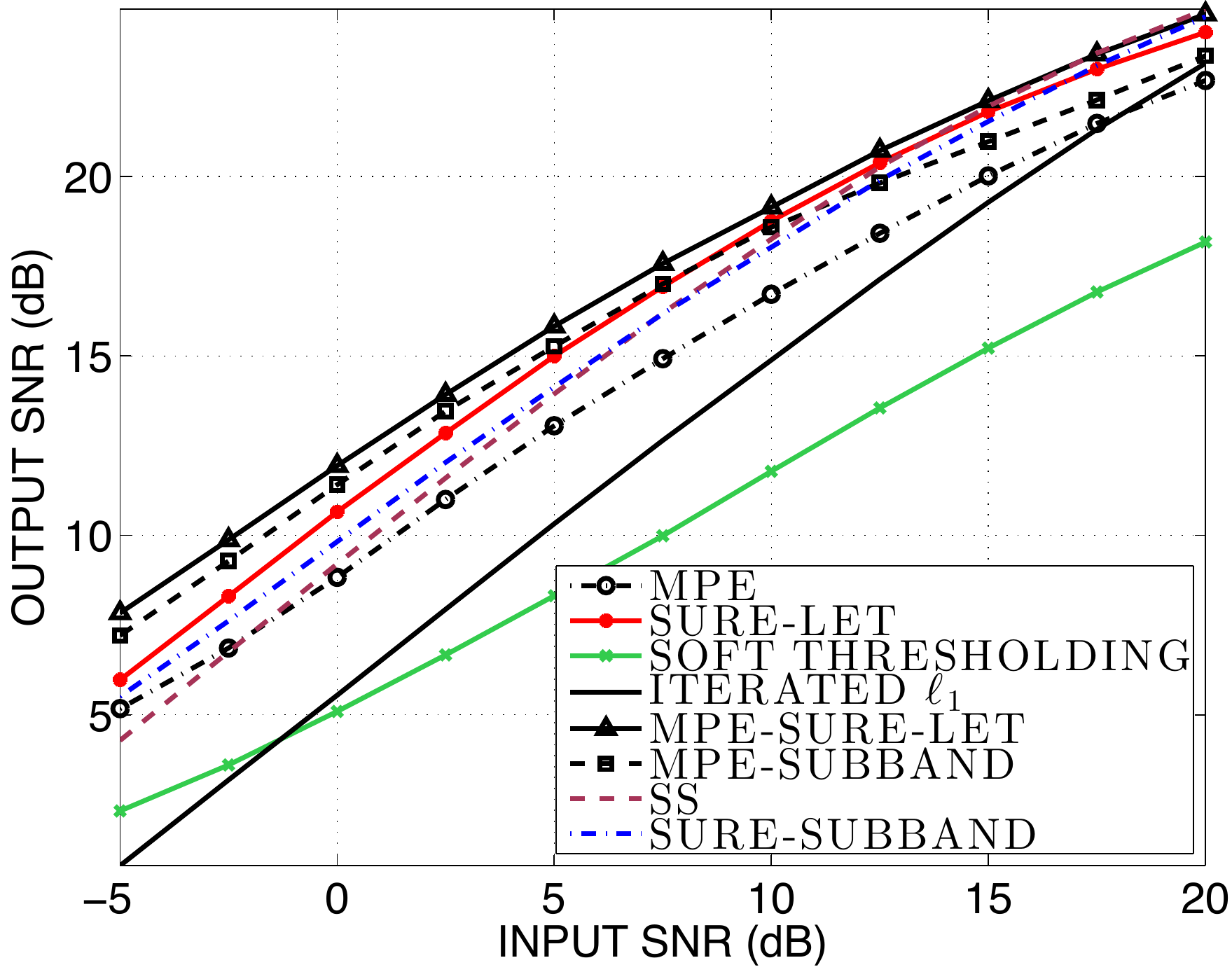}&
\includegraphics[width=6 cm]{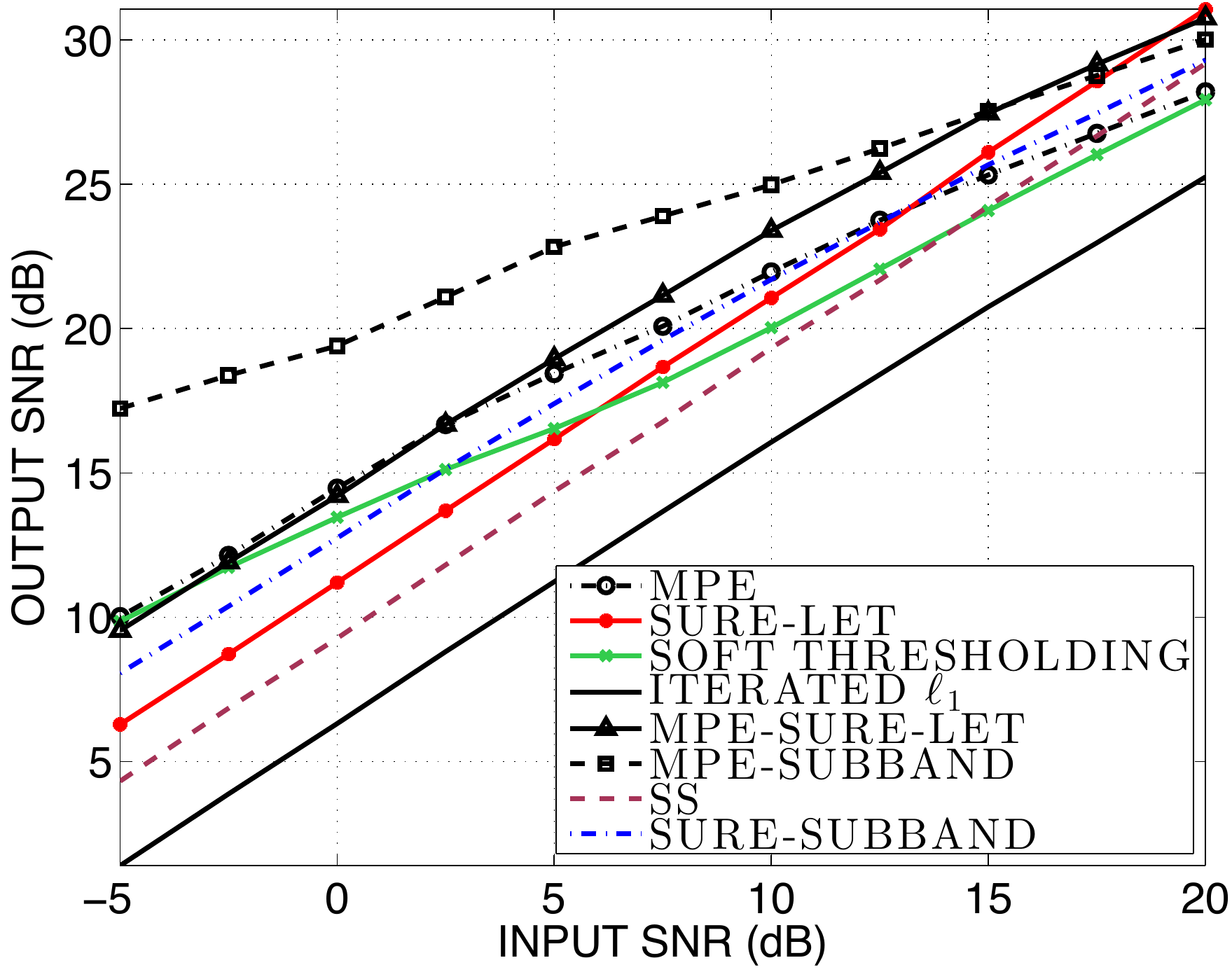}&
\includegraphics[width=6 cm]{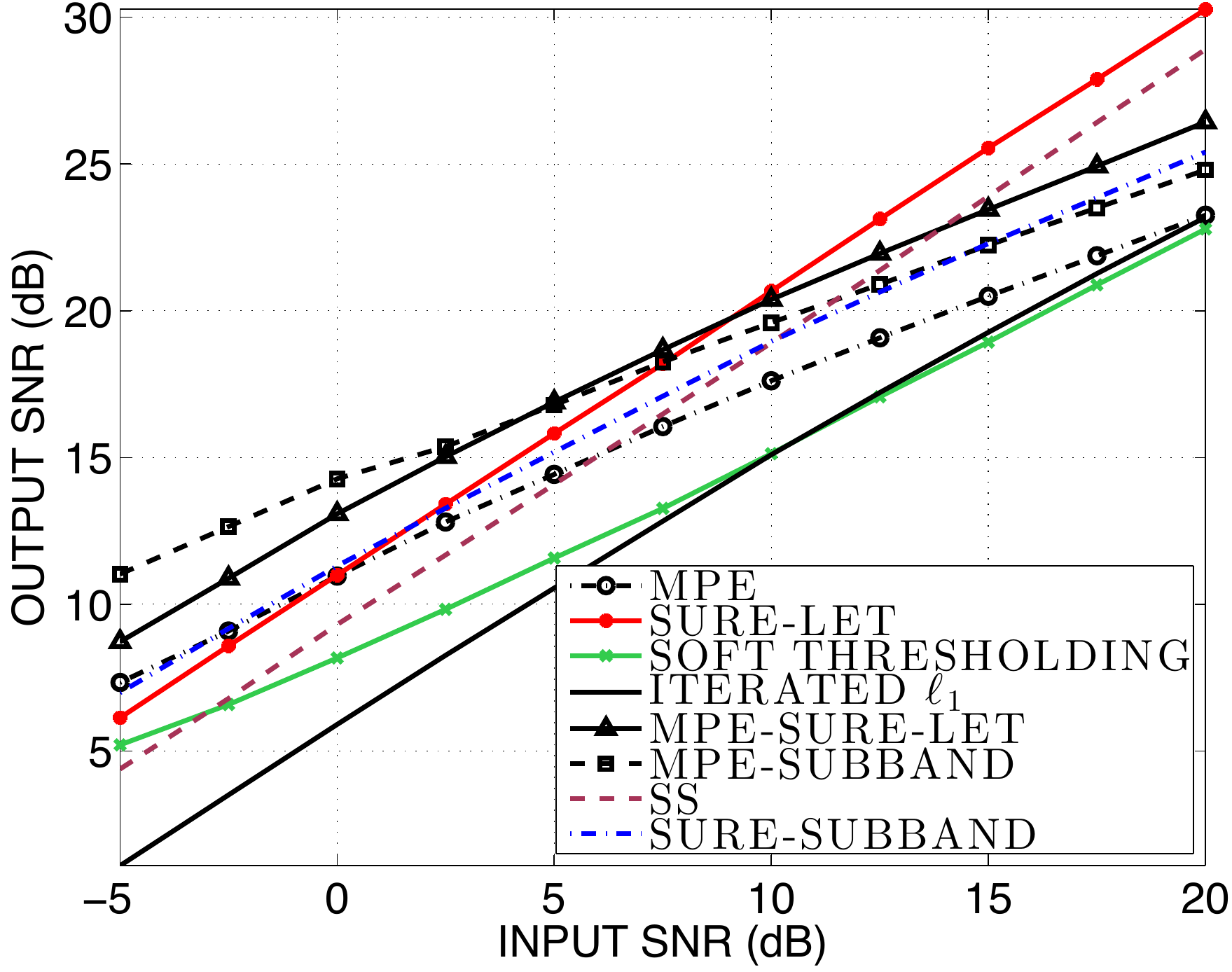}\\
\text{(a) ECG signal} &  \text{(b)} \hspace{0.15cm} \textit{HeaviSine} \text{ signal}  & \text{(c)}  \hspace{0.15cm}  \textit{Piece-Regular} \text{ signal}
\end{array}$
\caption{\small Output SNR of various denoising algorithms, averaged over $100$ noise realizations, corresponding to different input SNRs. The figures also demonstrate how the MPE-based estimator could  be used as an add-on to the SURE-LET algorithm to boost the overall denoising performance.}
\label{den_comparison_diff_algorithms_fig}
\end{figure*}

\indent We compare the MPE and the $\ell_1$-based shrinkage estimators with three state-of-the-art denoising algorithms:  (i) wavelet soft-thresholding\footnote{\label{first_foot}A Matlab implementation is included in the \textit{Wavelab} toolbox available at: \\{\url{http://statweb.stanford.edu/~wavelab/}}.} \cite{donoho}; (ii) the SURE-LET  denoising algorithm\footnote{\label{second_foot}A MATLAB implementation of the SURE-LET algorithm is available at: \\ \url{http://bigwww.epfl.ch/demo/suredenoising}.} \cite{Luisier}; and (iii) smooth sigmoid shrinkage (SS) \cite{Atto1} in the wavelet domain \footnote{Pastor et al. kindly provided the MATLAB implementation of their denoising technique \cite{Atto1}, which facilitated the comparisons reported in this paper.}. In \cite{donoho}, a wavelet-based soft-thresholding scheme is used for denoising, with the threshold selected as $\tau=\sigma \sqrt[]{2 \log(N)}$ for an $N$ length signal. The SURE-LET technique employs a \textit{linear expansion of thresholds} (LET), which is a linear combination of elementary denoising functions and optimizes for the coefficients by minimizing the SURE criterion. In \cite{Atto1}, a smooth sigmoid shrinkage is applied on the wavelet coefficients to achieve denoising, and the parameters of the sigmoid, which control the degree of attenuation, are obtained by minimizing the SURE objective. We consider ECG signals taken from the \textit{PhysioBank} database, and the \textit{HeaviSine} and \textit{Piece-Regular} signals taken from \textit{Wavelab} toolbox for performance evaluation.\\
\indent  The  noise is assumed  to follow a  Gaussian distribution and the output SNR values are averaged over $100$ independent realizations. The noise variance is estimated using a median-based estimator \cite{mallat}, which is also used by Luisier et al.\footref{first_foot} and Donoho\footref{second_foot}. In  SURE-LET, SS,  and wavelet thresholding techniques, denoising is performed using Symmlet-4, with three levels of decomposition, as these settings were found to be the best for the ECG signal (following \cite{Skrishnan}). In case of MPE and $\ell_1$-based shrinkage estimators, denoising is performed in the DCT domain. We use the shorthand notations MPE and MPE-subband to denote the pointwise and subband shrinkage estimators, respectively. The corresponding SURE-based subband shrinkage estimator is denoted as  SURE-subband. We set $k=16$ and $\epsilon=1.75\sqrt{k}\sigma$ for computing the subband shrinkage parameters. These parameters have not been specifically optimized; however, they were found to work well in practice. The output SNR as a function of the input SNR, obtained using various algorithms, is shown in Figure~\ref{den_comparison_diff_algorithms_fig}.\\
\indent From the ECG signal denoising performance shown in Figure~\ref{den_comparison_diff_algorithms_fig}(a),  we observe that the MPE  estimate consistently dominates the soft-thresholding-based denoising  for   input SNRs ranging from $-5$ dB to $20$ dB. The iterative $\ell_1$-distortion-based shrinkage estimator (20 iterations) yields lower output SNR compared with the MPE-based estimate for input SNR values in the range $-5$ to $17.5$ dB, but surpasses it for relatively higher values of input SNR ($17.5$ to $20$ dB). The SURE-LET and the SS algorithms dominate both MPE and the $\ell_1$-based shrinkage estimators, because they use more sophisticated denoising functions in the transform domain, thereby offering greater flexibility. For  input SNR range of $0$ dB to $20$ dB, the expected  $\ell_1$-distortion-based shrinkage estimator consistently outperforms the soft-thresholding-based techniques.\\
\indent We have also found that it is possible to boost the denoising performance of an algorithm in the low-SNR regime by adding the MPE denoiser in tandem, that is, by replacing $\tilde{s}$ in the expression for the MPE risk estimate in (\ref{MPE_gauss_eqn}) with the estimate obtained using a denoising technique, for example, the SURE-LET. We refer to this tandem approach as MPE-SURE-LET in Figure~\ref{den_comparison_diff_algorithms_fig}. This approach results in $1$ to $2$ dB gain in output SNR over SURE-LET for low and medium values of input SNR. We observe in Figure \ref{den_comparison_diff_algorithms_fig} that the MPE-subband estimator outperforms the competing algorithms (except for MPE-SURE-LET in Figure \ref{den_comparison_diff_algorithms_fig}(a)),  at low input SNR. 

\section{Conclusions}
\label{sec_conclusion}
\indent We have proposed a new framework for signal denoising based on a novel criterion, namely the probability of error. Our framework is applicable to scenarios where the noise samples are independent and  additively  distort the signal. Denoising is performed by transform-domain shrinkage and  the optimum shrinkage parameter is obtained by minimizing an estimate of the MPE risk. We have considered both pointwise and subband shrinkage estimators within the MPE paradigm. The performance of the proposed MPE estimators depends on the choice of the error-tolerance parameter $\epsilon$. In pointwise shrinkage, to deal with the issue of selecting an appropriate $\epsilon$, we have proposed two approaches. In the first one, we experimentally determined an $\epsilon$ value that results in maximum SNR gain for a particular signal by evaluating the output SNR for different $\epsilon$. In the second approach, we  computed the accumulated probability of error, which is the expected $\ell_1$ distortion, and developed an iterative algorithm for minimization. We demonstrated that  the shrinkage estimator obtained using the expected $\ell_1$ risk outperforms the classical ML estimator, when the number of observations is small or the input SNR is low. We  also showed that the  shrinkage estimator obtained by iteratively minimizing the $\ell_1$ risk dominates the non-iterative approach in terms of the output SNR.\\
\indent  Extensive performance comparison of the proposed MPE and the $\ell_1$ distortion-based approaches with state-of-the-art denoising algorithms is carried out on real ECG signals and \textit{Wavelab} signals. Experimental results demonstrate that the  shrinkage estimator based on the  MPE-risk estimate outperforms the SURE-based  estimator in terms of  SNR gain,  particularly in the regime of low SNR and smaller subband size. The proposed MPE-framework could be used as an add-on over an  existing denoising technique, leading to an estimator that has a higher  output SNR, particularly in the low input SNR regime.\\
\indent For deriving the expression and validating the performance of the MPE-based subband shrinkage estimator, we considered denoising of signals corrupted with Gaussian noise. Experimentally, we have found that increase in the subband size leads to an increase in output SNR, and saturates beyond a point. We have also observed that, when the subband size or the input SNR is low, the MPE-based estimate has superior performance compared with the SURE-based estimator.\\
\indent We demonstrated that the optimum shrinkage parameter obtained by minimizing  estimates of the MPE/$\ell_1$ distortions  increases monotonically with the  increase in a posteriori SNR. Such behavior of the shrinkage parameter is essential for denoising.  A theoretical characterization of this behavior is needed and may lead to interesting inferences, which could potentially lead to a rigorous convergence proof for the iterative expected $\ell_1$ distortion minimization technique. Another important observation is that, for lower input SNRs, the proposed denoising framework yields a higher output SNR compared with the MSE-based techniques. The improvement in performance in terms of SNR of the denoised output may be attributed to the fact that the MPE framework incorporates knowledge of the distribution of the observations, which goes beyond the second-order statistics considered in {\it de facto} MSE-based optimization. We also believe that this is the first attempt at demonstrating competitive denoising performance with probability of error chosen as the distortion metric, in a non-Bayesian estimation framework.

\appendices
\section{Perturbation of SURE-Based Pointwise Shrinkage}
\label{sure_pert}
To analyze the perturbation in the location of the minimum of the SURE cost function, in comparison with the true MSE, one needs to evaluate  
\begin{equation*}
P_e^{\text{SURE}}=\mathbb{P}\left\{ \left|  a_{\text{opt}}\left(s\right) - a_{\text{opt}}\left(x\right)  \right| \geq \delta \right\},
\end{equation*}
where $a_{\text{opt}}\left(s\right)=\frac{s^2}{s^2+\sigma^2}$ and  $a_{\text{opt}}\left(x\right)=1-\frac{\sigma^2}{x^2}$. Let 
\begin{equation*}
h\left(x\right)=a_{\text{opt}}\left(s\right) - a_{\text{opt}}\left(x\right)=\left(\frac{s^2}{s^2+\sigma^2}- 1+\frac{\sigma^2}{x^2}\right).
 \end{equation*}
 The Taylor-series expansion of $h(x)$ about  $s$ yields 
 \begin{equation*}
h(x)=\frac{\sigma^4}{s^2\left(s^2+\sigma^2\right)}-2 \frac{w \sigma^2}{s^3}+\sum_{n=2}^{\infty}\frac{d^{\left(n\right)}\left(s\right) }{n!} w^n,
\end{equation*}
where $h^{\left(n\right)}$ is the $n^{\text{th}}$ derivative $h$. Using the first-order Taylor series approximation $h(x)\approx h(s)+w h^{\left(1\right)}\left(s\right)$, we obtain
\begin{equation*}
h(x)\approx \frac{\sigma^4}{s^2\left(s^2+\sigma^2\right)}-2 \frac{w \sigma^2}{s^3},
\end{equation*}
which, in turn, leads to an approximation of the perturbation probability $P_e^{\text{SURE}}$:
\begin{equation*}
P_e^{\text{SURE}}=\mathbb{P}\left\{ \left|  h\left(x\right)  \right| \geq \delta \right\} \\
\approx \mathbb{P}\left\{ \left|  \frac{\sigma^4}{s^2\left(s^2+\sigma^2\right)}-2 \frac{w \sigma^2}{s^3}\right| \geq \delta \right\}.
\end{equation*}
Invoking $w\sim \mathcal{N}\left(0,\sigma^2\right)$, and using  the Chernoff bound\cite{Mitzenmacher}, we obtain
\begin{equation*}
P_e^{\text{SURE}} \leq 2\exp\left( -\frac{s^6}{8\sigma^6} \left( \delta - \frac{\sigma^4}{\left(  s^2+\sigma^2 \right)s^2}   \right)^2    \right).
\end{equation*}
Consequently, to satisfy an upper bound on the deviation probability of the form $P_e^{\text{SURE}}  \leq  \alpha$, for a given $\delta>0$, one must ensure that
\begin{equation}
  \frac{s^6}{8\sigma^6} \left( \delta - \frac{\sigma^4}{\left(  s^2+\sigma^2 \right)s^2}   \right)^2  \geq \log \left(\frac{2}{\alpha}\right).
\label{SURE_snr_requirement1}
\end{equation}
The condition in \eqref{SURE_snr_requirement1} translates to an equivalent condition on the minimum required  SNR  $\displaystyle{\frac{s^2}{\sigma^2}}$ to achieve a certain $P_e^{\text{SURE}}$. 
\section{ Expected $\ell_1$ Risk for GMM}
\label{expected_L1_derivation}
For  additive noise with the p.d.f. given in  (\ref{GMM_pdf_eqn}), we have
\begin{eqnarray}
\mathcal{E}\{ \left| \widehat{s}-s \right|\}&=& \sum_{m=1}^{M}\alpha_m \left(  \int_{0}^{\infty} Q \left(  \frac{\epsilon-(a-1)s-\theta_m}{a\sigma_m} \right) \right. \mathrm{d} \epsilon\nonumber\\ &+&  \int_{0}^{\infty}\left. Q \left(  \frac{\epsilon+(a-1)s+\theta_m}{a\sigma_m} \right)\mathrm{d} \epsilon \right),
\label{gmm_l1_integrate}
\end{eqnarray}
using (\ref{gmm_mpe_expr}) and (\ref{MPE_L1_eq}). Letting  $\mu_m=-\displaystyle \frac{(a-1)s+\theta_{m}}{a\sigma_{m}}$ and  $ u_m=\displaystyle \frac{\epsilon -(a-1)s-\theta_{m}}{a\sigma_{m}}$, we get
\small
\begin{eqnarray}
 \displaystyle\int_{0}^{\infty} Q \left( \frac{\epsilon -(a-1)s-\theta_{m}}{a\sigma_{m}} \right)\mathrm{d} \epsilon
=  a\sigma_m \left( \frac{e^{-\frac{\mu_{m}^2}{2}} }{\sqrt{2\pi}} - \mu_{m} Q \left(  \mu_{m} \right) \right).
\label{L1_gmm_qfunc_integral}
\end{eqnarray}
\normalsize
Subsequently, substituting (\ref{L1_gmm_qfunc_integral}) in (\ref{gmm_l1_integrate}) yields
\begin{eqnarray}
\mathcal{E}\{ \left| \widehat{s}-s \right|\}=\sum_{m=1}^{M} a \alpha_m  \sigma_{m} \left( \sqrt{\frac{2}{\pi}}e^{-\frac{\mu_{m}^2}{2}} -2 \mu_{m} Q \left(  \mu_{m} \right) + \mu_{m}  \right),\nonumber
\end{eqnarray}
which is the expression for the expected $\ell_1$ distortion for noise following a GMM distribution.



\begin{thebibliography}{99}

\bibitem{poor}
H. V. Poor, \emph{An Introduction to Signal Detection and Estimation}, Springer, 1994.

\bibitem{kay}
{S. Kay and Y. C. Eldar, ``Rethinking  biased estimation," \emph{IEEE Signal Process. Mag.}, vol. 25, no. 3, pp. 133--136, May 2008.}

\bibitem{Stein}
C. M. Stein, ``Estimation of the mean of a multivariate normal distribution,'' \emph{Ann. Stat.}, vol.~9, no.~6, pp.~1135--1151, Nov. 1981.

\bibitem{Jamesproof}
{W. James and C. Stein, ``Estimation with quadratic loss,'' \emph{Berkeley Symp. Math. Statist. and Prob.}, Berkeley, CA, vol.1, pp.~361--379, 1961.}

\bibitem{Luisier}
{F.~Luisier, T.~Blu, and M.~Unser, ``A new SURE approach to image denoising: Interscale orthonormal wavelet thresholding,'' \emph{IEEE Trans. Image Process.}, vol. 16, no. 3, pp. 593--606, Mar. 2007.}

\bibitem{Bluv}
{F. Luisier, T. Blu,  and M. Unser, ``SURE-LET for orthonormal wavelet-domain video denoising,'' \emph{IEEE Trans.  Circ.  Syst.   Video Tech.}, vol. 20, no. 6, pp. 913--919, Jun. 2010.}

\bibitem{Blu3}
{F. Luisier and T. Blu, ``The SURE-LET multichannel  image denoising: Interscale orthonormal wavelet thresholding,'' \emph{IEEE Trans. Image Process.}, vol. 17, no. 4, pp. 482$-$492, Apr. 2008.}

\bibitem{Blu}
{T. Blu and F. Luisier, ``The SURE-LET approach to image denoising,'' \emph{IEEE Trans. Image Process.}, vol. 16, no. 11, pp. 2778--2786, Nov. 2007.}

\bibitem{Zheng}
{N. Zheng, X. Li, T. Blu, and T. Lee, ``SURE-MSE speech enhancement for robust speech recognition,'' in \emph{Proc. $7^{\text{th}}$ Int. Symp. Chinese Spoken Language Process.}, pp. 271--274, Nov. 2010.}

\bibitem{Donoho}
D. L. Donoho and I. M. Johnstone, ``Adapting to unknown smoothness via wavelet shrinkage,'' \emph{J. Amer. Stat. Assoc.}, vol. 90, no. 432, pp.1200--1224, Dec. 1995.

\bibitem{Benazza}
{A.~Benazza-Benyahia and J. C. Pesquet, ``Building robust wavelet estimators for multicomponent images using Stein's principle,''  \emph{IEEE Trans. Image Process}, vol. 14, no. 11, pp. 1814--1830, Nov. 2005.}

\bibitem{Zhang}
{X. Zhang and M. D. Desai, ``Adapting denoising based on SURE risk,'' \emph{IEEE  Signal Process. Lett.}, vol. 5, no. 10, pp. 265$-$267, Oct. 1998.}

\bibitem{Atto1}
{A. M. Atto, D. Pastor, and G. Mercier, ``Smooth adaptation by sigmoid shrinkage,'' \emph{EURASIP J. Image and Video Process.}, article ID 532312,  2009.}

\bibitem{Atto2}
{A. M. Atto, D. Pastor, and G. Mercier, ``Optimal SURE parameter for sigmoidal wavelet shrinkage,'' in \emph{Proc. Eur. Signal Process. Conf.}, pp. 40-43, Aug. 2009.}

\bibitem{Pesquet}
{J. C. Pesquet, A. Benazza-Benyahia, and C. Chaux, ``A SURE approach for digital signal/image deconvolution problems,'' \emph{IEEE Trans. Signal Process.}, vol. 57, no. 12, pp. 4616--4632, Dec. 2009.}

\bibitem{Ramani}
S. Ramani,  T. Blu, and M. Unser, ``Monte Carlo SURE: A black-box optimization of regularization parameters for general denoising algorithms,'' \emph{IEEE Trans. Image Process}, vol. 17, no. 9, pp. 1540$-$1544, Sep. 2008.

\bibitem{Dimitri2}
{D. V. De  Ville and M.  Kocher, ``Nonlocal means with dimensionality reduction and SURE-based parameter selection,'' \emph{IEEE  Trans. Image Process.}, vol. 20, no. 9, pp. 2683$-$2690, Sep. 2011.}

\bibitem{Hsung}
{T. Hsung, D. Lun, and  K. C. Ho, ``Optimizing the multiwavelet shrinkage denoising,'' \emph{IEEE  Trans. Signal Process.}, vol. 53, no. 1, pp. 240$-$251, Jan. 2005.}

\bibitem{Qiu}
{T. Qiu, A. Wang, N. Yu, and A. Song, ``LLSURE: Local linear SURE-based edge-preserving image filtering,'' \emph{IEEE Trans. Image Process.}, vol. 22, no. 1, pp. 80--90, Jan. 2013.}


\bibitem{Krim}
H. Krim, D. Tucker, S. Mallat, and D. Donoho, ``On denoising and best signal representation,'' \emph{IEEE Trans. Info. Theory}, vol. 45, no. 7, pp. 2225$-$2238, Nov. 1999.

\bibitem{Harini}
{H. Kishan and C. S. Seelamantula, ``SURE-fast bilateral filters,'' in \emph{Proc. IEEE Int. Conf. Acoust., Speech, Signal Process.},  pp. 1129--1132, Mar. 2012.}

\bibitem{Skrishnan}
{S. R. Krishnan and C. S. Seelamantula, ``On the selection of optimum Savitzky-Golay filters,'' \emph{IEEE Trans. Signal Process.}, vol. 61, no. 2, pp. 380--391, Jan. 2013.}

\bibitem{Hudson}
{H. M. Hudson, ``A natural identity for exponential families with applications in multi-parameter estimation,'' \emph{Ann. Stat.}, vol. 6, no. 3, pp. 473--484, 1978.}

\bibitem{Hwang}
{J. T. Hwang, ``Improving upon standard estimators in discrete exponential families with applications to Poisson and negative binomial cases," \emph{Ann. Stat.}, vol. 10, no. 3, pp. 857--867, 1982.}

\bibitem{Eldar}
{Y. C. Eldar, ``Generalized SURE for exponential families: Applications to regularization,'' \emph{IEEE Trans. Signal Process.}, vol. 57, no. 2, pp. 471--481, Feb. 2009.}

\bibitem{Giryes}
{R. Giryes, M. Elad, and Y. C. Eldar, ``The projected GSURE for automatic parameter tuning in iterative shrinkage methods,'' \emph{Appl. Comput. Harmon. Anal.}, vol. 30, no. 3, pp. 407--422, May 2011.}


\bibitem{Nag}
{N. R. Muraka and C. S. Seelamantula, ``A risk-estimation-based comparison of mean-square error and Itakura-Saito distortion measures for speech enhancement,'' in \emph{Proc. Interspeech}, pp. 349--352, Aug. 2011.}

\bibitem{Krishnan}
{S. R. Krishnan and C. S. Seelamantula, ``A generalized Stein's estimation approach for speech enhancement based on perceptual criteria,'' in \emph{Proc. Workshop on Statistical and Perceptual Audition (SAPA)--Speech Communication with Adaptive Learning (SCALE)}, Sep. 2012.}

\bibitem{Johnstone}{I. M. Johnstone, ``Gaussian estimation: Sequence and wavelet models,'' Jun. 2013, Available at : \textit{http://statweb.stanford.edu/~imj/GE06-11-13.pdf}.} 

\bibitem{icassp2014}
J. Sadasivan, S. Mukherjee, and C. S. Seelamantula, ``An optimum shrinkage estimator based on minimum-probability-of-error criterion and application to signal denoising,'' in \emph{Proc. IEEE Intl. Conf. on Acoust. Speech and Signal Process.}, pp. 4249--4253, 2014.

\bibitem{Plataniotis}
K. N. Plataniotis and D. Hatzinakos,  \emph{Gaussian Mixtures and Their Applications to Signal Processing}, CRC Press, Dec. 2000. 

\bibitem{Sorenson}
H. W. Sorenson and D. L. Alspach, ``Recursive Bayesian estimation using Gaussian sums,'' \emph{Automatica}, Vol.  7, pp. 465--479, 1971.

\bibitem{wavelab}
{WAVELAB toolbox  [Online].  Available: {\url{http://statweb.stanford.edu/~wavelab/ }}}

\bibitem{donoho}
{D. L. Donoho, ``Denoising by soft thresholding,'' \emph{IEEE Trans. Info. Theory.}, vol. 41, no. 3, pp. 613--627, May 1995.}

\bibitem{Redner}
R. Redner and H. Walker. ``Mixture densities, maximum likelihood and the em algorithm,'' \emph{SIAM Review}, Vol.  26, no. 2, pp. 195--239,  Apr. 1984.

\bibitem{roth} 
B. M. G. Kibria and A. H. Joarder, ``A short review of multivariate $t$-distribution," \emph{J.   Stat. Res.}, vol. 40, no. 1, pp. 59--72, 2006.

\bibitem{mallat}
S. Mallat, \emph{A Wavelet Tour of Signal Processing}, 3rd edition, Academic Press, 2009.

\bibitem{Socheleau}
{D. Pastor  and F. Socheleau, ``Robust estimation of noise standard deviation in presence of signals with unknown distributions and occurrences,'' \emph{IEEE Trans. Signal Process.}, vol. 60, no. 4, pp. 1545--1555,  Apr.  2012. }

\bibitem{KRRao}
 K. R. Rao and P. Yip, \emph{Discrete Cosine Transform: Algorithms, Advantages, Applications}. Boston, MA: Academic, 1990


\bibitem{Boll}
{S.~Boll, ``Suppression of acoustic noise in speech using spectral subtraction,'' \emph{ IEEE Trans.~Acoust.~Speech, Signal Process.}, vol. 27, no. 2, pp.~113--120, Apr. 1979.} 
\bibitem{PLoizou}
{P.~Loizou, \emph{Speech Enhancement --- Theory and Practice}, CRC Press, 2007.}


\bibitem{database}
{PhysioBank database [Online]. Available: http://www.physionet.org/ physiobank/database/aami-ec13/}

\bibitem{Mitzenmacher}
M. Mitzenmacher and E. Upfal, \emph{ Probability and Computing:  Randomized Algorithms and Probabilistic Analysis}, Cambridge University Press, 2005.


\end{thebibliography}
\end{document}